\documentclass[11pt,hyper,letter]{JHEP3}
\usepackage{graphicx,amssymb,amsmath,amsfonts,cite,subfigure}

\def\N{{\mathcal{N}}}

\def\<{\langle}
\def\th{\theta}
\def\>{\rangle}
\def\({\left (}
\def\){\right )}
\def\[{\left[}
\def\]{\right]}

\def\beq{\begin{equation}}
\def\eeq{\end{equation}}

\def\jz{\langle J^z \rangle}

\def\p{\pi}                
\def\a{\alpha}
\def\s{\sigma}

\newcommand{\bea}{\begin{eqnarray}}
\newcommand{\eea}{\end{eqnarray}}

\def\Om{{\cal{O}}_m}
\def\Omv{\langle {\cal{O}}_m \rangle}
\def\Op{{\cal O}_{\phi}}
\def\Opv{\langle {\cal O}_{\phi}\rangle}

\def\lagr{{\cal L}}

\def\nn{\nonumber}

\def\TT{\langle T_{tt} \rangle}
\def\psibar{\overline{\psi}}

\def\a{\alpha}
\def\b{\beta}

\def\e{\epsilon}           
  
\def\g{\gamma}

\def\k{\kappa}             

\def\o{\omega}  
\def\p{\phi}              
\def\th{\theta}                   
\def\s{\sigma}                                   

\def\th{\theta}

\def\ie{\textit{i.e.} }

\title{\LARGE A Chiral Magnetic Effect from AdS/CFT with Flavor}

\author{Carlos Hoyos,$^1$\footnotemark[1]\, Tatsuma Nishioka,$^2$\footnotemark[2] and Andy O'Bannon,$^3$\footnotemark[3]\,
\\
$^1$Department of Physics, University of Washington \\ Seattle, WA 98195-1560, United States
\\
\\
$^2$Department of Physics, Princeton University \\Princeton, NJ 08544, United States
\\
\\
$^3$Department of Applied Mathematics and Theoretical Physics\\ University of Cambridge\\ Cambridge CB3 0WA, United Kingdom}

\footnotetext[1]{E-mail address: \email{choyos@phys.washington.edu}}
\footnotetext[2]{E-mail address: \email{nishioka@princeton.edu}}
\footnotetext[3]{E-mail address: \email{A.OBannon@damtp.cam.ac.uk}}

\abstract{For (3+1)-dimensional fermions, a net axial charge and external magnetic field can lead to a current parallel to the magnetic field. This is the chiral magnetic effect. We use gauge-gravity duality to study the chiral magnetic effect in large-$N_c$, strongly-coupled $\N=4$ supersymmetric $SU(N_c)$ Yang-Mills theory coupled to a number $N_f \ll N_c$ of $\N=2$ hypermultiplets in the $N_c$ representation of $SU(N_c)$, \textit{i.e.} flavor fields. Specifically, we introduce an external magnetic field and a time-dependent phase for the mass of the flavor fields, which is equivalent to an axial chemical potential for the flavor fermions, and we compute holographically the resulting chiral magnetic current. For massless flavors we find that the current takes the value determined by the axial anomaly. For massive flavors the current appears only in the presence of a condensate of pseudo-scalar mesons, and has a smaller value than for massless flavors, dropping to zero for sufficiently large mass or magnetic field. The axial symmetry in our system is part of the R-symmetry, and the states we study involve a net flow of axial charge to the adjoint sector from an external source coupled to the flavors. We compute the time rate of change of axial charge and of energy both in field theory and from holography, with perfect agreement. In contrast to previous holographic models of the chiral magnetic effect, in our system the vector current is conserved and gauge-invariant without any special counterterms.}


\preprint{DAMTP-2011-44\\PUPT-2369}

\begin{document}

\section{Introduction}\label{ss:intro}

Consider a (3+1)-dimensional system of free, massless Dirac fermions $\psi$. The Lagrangian of such a system has two $U(1)$ symmetries, the vector one $U(1)_V$, with conserved current $\psibar \gamma^{\mu} \psi$, and the axial one $U(1)_A$, with conserved current $\psibar \gamma^{\mu} \g^5 \psi$. $U(1)_A$ is anomalous, and can be explicitly broken by a nonzero Dirac mass. If we introduce an axial chemical potential $\mu_5$ then we expect an imbalance in the number of left- and right-handed fermions. If we further introduce an external $U(1)_V$ magnetic field $B$ then, assuming the fermions have positive charge, we expect their spins to align with $B$, and since they are massless their momenta will also align or anti-align depending on their chirality. Given the imbalance in chirality, we expect a net $U(1)_V$ current parallel to $B$. This is the simplest example of the chiral magnetic effect (CME)\cite{1998PhRvL..81.3503A,Kharzeev:2007jp,Fukushima:2008xe}.

For free fermions, the axial anomaly determines the size of the chiral magnetic current as follows. An axial chemical potential is equivalent to a background $U(1)_A$ gauge field with constant time component, $A^5_t=\mu_5$, or to a time-dependent phase $\psi \rightarrow e^{i \g^5 \mu_5 t} \psi$. Via the axial anomaly such a phase shift can be traded for a $\theta$-term of the form $a(t,\vec{x}) F \wedge F$ with $a(t,\vec{x}) = \mu_5 t$ and $F$ the $U(1)_V$ field strength. The spacetime-dependent source $a(t,\vec{x})$ can be regarded as a background, non-dynamical axion field. Writing $F=dA$ and integrating by parts, we obtain an interaction of the form $da \wedge A \wedge F$. Varying the action with respect to $A$ we find the chiral magnetic current, which is parallel to $B$ and has magnitude
\beq\label{eq:CME}
J=\frac{\mu_5}{2\pi^2}B.
\eeq
The quantity $\sigma \equiv J/B$ is called the chiral magnetic conductivity. When the system has nontrivial time evolution $\s$ becomes a function of time, or in Fourier space a function of frequency $\sigma(\o)$. For free fermions, the DC limit $\sigma(0)$ is fixed by the axial anomaly. Generically, interactions can modify $\sigma(\o)$, including $\sigma(0)$ \cite{Fukushima:2010zza}. Notice that the CME occurs only when parity P and charge conjugation times parity CP symmetries are broken.

A CME may occur in heavy-ion collisions such as those produced at the Relativistic Heavy-Ion Collider (RHIC) and the Large Hadron Collider (LHC)~\cite{Kharzeev:2007jp,Kharzeev:2004ey,Kharzeev:2007tn}. The dominant interaction in the early stages of collisions is the strong nuclear force, as described by Quantum Chromodynamics (QCD). Indeed, the quark-gluon plasma (QGP) created at RHIC appears to involve strongly-interacting degrees of freedom far from equilibrium: the plasma appears to thermalize quickly and have a short mean free path (both signs of strong interactions \cite{Shuryak:2003xe, Shuryak:2004cy}) but also expands and cools rapidly until hadronization occurs.

Two conditions must hold to produce a CME in a heavy-ion collision. First, the QCD vacuum apparently preserves P and CP, so some event-by-event violation of these is required. In a medium such as the QGP, one possible mechanism for such violations are fluctuations of the topological charge density. Second, the collision must be non-central, \textit{i.e.} the nuclei must not perfectly overlap upon impact. In that case, the net charge combined with the net angular momentum can produce large magnetic fields, although these may die quickly as the QGP expands \cite{Kharzeev:2007jp}.

Assuming that P and CP are broken and a magnetic field is present, we know of two mechanisms to produce a CME in finite-temperature QCD. The first occurs for sufficiently large temperatures, where the QCD plasma is deconfined and chiral symmetry is restored. In that case we may invoke a na\"{i}ve picture of quarks as freely propagating fermions in a magnetic field, and apply the arguments above.

A second, more subtle, mechanism, discussed for example in ref.~\cite{Asakawa:2010bu}, may occur at lower temperatures, when the QCD plasma is in a confined state with chiral symmetry broken. Here we expect a gas of hadrons rather than a QGP. The key observation is that an external electromagnetic field can convert a neutral pseudo-scalar meson, such as the $\pi^0$, $\eta$, or $\eta'$, into a neutral vector meson, such as the $\rho$. More precisely, any effective action describing QCD and electromagnetic interactions will include for example a vertex of the form $B \pi^0 \rho$. The vector meson so produced will be polarized in the direction of the magnetic field, and via interactions with charged mesons can induce a current parallel to $B$, thus producing a CME even in a confined phase. The same process may also occur in the late stages of QGP evolution, during hadronization when metastable domains with spontaneous breaking of P and CP could be formed \cite{Kharzeev:1998kz}.

To our knowledge, analysis of RHIC data appears to favor the presence of a CME in the QGP, although a better understanding of systematic errors and backgrounds is still needed before a firm conclusion can be made \cite{Selyuzhenkov:2005xa,Voloshin:2008jx,Abelev:2009txa,Wang:2009kd}. The strong interactions and far-from-equilibrium evolution of the QGP in a heavy-ion collision make a clean theoretical prediction for $\sigma$ very difficult. Lattice simulations suggest that the CME occurs in thermal equilibrium \cite{Buividovich:2009zzb,Buividovich:2009zj,Abramczyk:2009gb,Yamamoto:2011gk}. Lattice simulations cannot yet reliably determine the time evolution of $\sigma$, or equivalently $\sigma(\o)$, which is crucial for estimating the size of the chiral magnetic current in a heavy-ion collision.

An alternative approach to the CME is the anti-de Sitter/Conformal Field Theory correspondence (AdS/CFT) \cite{Maldacena:1997re,Gubser:1998bc,Witten:1998qj}, or more generally gauge-gravity duality. Gauge-gravity duality equates a strongly-coupled non-Abelian gauge theory with a weakly-coupled theory of gravity on some background spacetime, such that the field theory lives on the boundary of the spacetime, hence the duality is holographic. In particular, a black hole spacetime is dual to a thermal equilibrium state in which the center symmetry is spontaneously broken, such as the high-temperature, deconfined phase of a confining theory, where the temperature of the field theory coincides with the Hawking temperature of the black hole \cite{Witten:1998zw}.

Gauge-gravity duality has been most successful at describing out-of-equilibrium physics, especially near-equilibrium physics, \textit{i.e.} hydrodynamics. Most importantly, \textit{all} gauge theories with a gravity dual (in states with $SO(2)$ rotational symmetry \cite{Erdmenger:2010xm}) have the same, very small, ratio of shear viscosity $\eta$ to entropy density $s$, namely $\eta/s=1/4\pi$ \cite{Kovtun:2004de}, which is surprisingly close to the value estimated for the QGP at RHIC \cite{Romatschke:2007mq,Luzum:2008cw}. We take such universality, and indeed the universality of hydrodynamics in general, as encouragement to study the CME in many holographic systems, following refs.~\cite{Rebhan:2009vc,Yee:2009vw,Gorsky:2010xu,Rubakov:2010qi,Gynther:2010ed,Brits:2010pw,Kalaydzhyan:2011vx}, including systems without confinement or chiral symmetry breaking in vacuum.

A conserved $U(1)$ current in the field theory is dual to a $U(1)$ gauge field in the bulk and, roughly speaking, an anomaly for the current is dual to a (4+1)-dimensional Chern-Simons term for the bulk gauge field. The latter is thus typically a key ingredient in holographic descriptions of the CME \cite{Rebhan:2009vc,Yee:2009vw,Gorsky:2010xu,Rubakov:2010qi,Gynther:2010ed,Brits:2010pw,Kalaydzhyan:2011vx}. More generally, holographic models dual to fluids with anomalous currents have been constructed for example in refs.~\cite{Kalaydzhyan:2011vx,Erdmenger:2008rm,Banerjee:2008th,Eling:2010hu}. These holographic studies are complementary to field theory studies of the effects of triangle anomalies on hydrodynamics \cite{Son:2009tf,Sadofyev:2010pr,Neiman:2010zi,Kharzeev:2010gd}, which themselves have been applied to study the CME in heavy-ion collisions~\cite{KerenZur:2010zw,Kharzeev:2010gr,Burnier:2011bf}.

One holographic model of QCD, the Sakai-Sugimoto model \cite{Sakai:2004cn,Sakai:2005yt} includes a bulk Chern-Simons term, although some confusion has arisen as to whether the CME occurs in this model at all. The problem in this model is that the vector current is anomalous under $U(1)_V\times U(1)_A$ transformations and therefore is not conserved in the presence of arbitrary external sources. Modifying the vector current such that it is conserved, which in the gravity dual requires adding certain boundary counterterms, causes the chiral magnetic current $J$ to vanish \cite{Rebhan:2009vc}. To our knowledge, no consensus has emerged on whether a CME occurs in the Sakai-Sugimoto model.\footnote{An alternative way to fix the normalization of the currents in the Sakai-Sugimoto model is to demand that the bulk action be invariant under gauge transformations that are non-vanishing at spatial infinity (in field theory directions), which leads to different bulk counterterms \cite{Bergman:2008qv} and produces a non-vanishing chiral magnetic current. For the sake of argument, here we are taking the phenomenological point of view that a $U(1)_A$ current in the presence of a $U(1)_V$ chemical potential should coincide with the weak-coupling result, which occurs with the bulk counterterms of ref.~\cite{Rebhan:2009vc} but not those of ref.~\cite{Bergman:2008qv}.}

The authors of refs.~\cite{Gynther:2010ed,Brits:2010pw} argued that for any bulk theory with a Chern-Simons term both a conserved vector current and nonzero $J$ are possible, but at a price: the bulk gravity solution becomes non-regular. More precisely, in Euclidean signature the bulk gauge field will not vanish at the horizon and hence will not be a regular one-form~\cite{Kobayashi:2006sb}. In Lorentzian signature the gauge field solution will be regular only on the future horizon. One conclusion is that no reliable holographic description of the CME in thermal equilibrium (regular on the past and future horizons) exists for a conserved vector current. Effectively, a source for the gauge field must be introduced at the black hole horizon, the meaning of which is unclear from a field theory point of view. We were thus motivated to study other models where such issues could be avoided or at least clarified.

We consider $\N=4$ supersymmetric $SU(N_c)$ Yang-Mills (SYM) theory, in the 't Hooft large-$N_c$ limit and with large 't Hooft coupling, coupled to a number $N_f\ll N_c$ of $\N=2$ supersymmetric hypermultiplets in the fundamental representation of the gauge group, \ie flavor fields. We introduce a complex mass $m = |m| e^{i \phi}$ for the flavor fields into the superpotential with a time-dependent phase $\phi = \omega t$, following refs.~\cite{Evans:2008zs,O'Bannon:2008bz,Evans:2008nf,Das:2010yw}. For the fermions that effectively introduces an axial chemical potential $\mu_5 = \frac{1}{2} \omega$. The theory also has a $U(1)_V$ symmetry that we will call baryon number. We introduce a baryon number magnetic field $B$ and compute (holographically) the resulting chiral magnetic current.

$\N=4$ SYM at large $N_c$ and large 't Hooft coupling is holographically dual to type IIB supergravity on $AdS_5 \times S^5$ \cite{Maldacena:1997re}. The $N_f \ll N_c$ hypermultiplets are dual to a number $N_f\ll N_c$ of probe D7-branes extended along $AdS_5 \times S^3$ \cite{Karch:2002sh}. The phase of the mass corresponds to the position of the D7-branes in one of the transverse directions on the $S^5$, hence $\omega$ corresponds to the angular frequency of the D7-branes in that direction, and the axial charge density corresponds to the angular momentum of the D7-branes. The axial anomaly is realized holographically via the Wess-Zumino (WZ) coupling of D7-branes to the background Ramond-Ramond (RR) flux on the $S^5$. The $U(1)_V$ current is dual to the $U(1)$ gauge field on the worldvolume of the D7-branes, which thus encodes both the $U(1)_V$ magnetic field and the chiral magnetic current.

In short, our system is a D7-brane in $AdS_5 \times S^5$, rotating with angular frequency $\omega$ on the $S^5$ and with a worldvolume magnetic field $B$. At finite temperature we replace $AdS_5$ with AdS-Schwarzschild. When $|m|=0$ the value of the chiral magnetic current agrees with the result from the calculation using the anomaly, eq.~\eqref{eq:CME}. When $|m|$ is nonzero we find that a chiral magnetic current appears only when a certain $U(1)_V$-neutral pseudo-scalar operator has a nonzero expectation value, signaling the breaking of C times time reversal, CT. We interpret this as a neutral pseudo-scalar condensate being converted into a vector condensate by the magnetic field, in a manner somewhat similar to the CME in the low-temperature phase of QCD \cite{Asakawa:2010bu}. For nonzero $|m|$ the value of the chiral magnetic current is less that that in eq.~\eqref{eq:CME}, and indeed both the current and the expectation value of the pseudo-scalar drop to zero for sufficiently large $|m|$ or $B$.

Although we were motivated to find a model describing a CME in equilibrium with regular bulk solutions, in states with a nonzero $|m|$ and a CME we can demonstrate that our system is out of equilibrium in two ways. First, we simply observe that the scalars in the $\N=2$ hypermultiplet have the same mass, with the same phase, as the fermions, so when $|m|$ is nonzero the Lagrangian has explicit time dependence and hence the system cannot be in equilibrium. The explicit time dependence disappears in the limit $|m| \rightarrow 0$. Second, we observe that in our system the axial symmetry is part of the R-symmetry under which the adjoint fields of $\N=4$ SYM are also charged, hence axial charge in the flavor sector can ``leak'' into the adjoint sector, also taking energy with it. We compute, both in the field theory and from holography, the rate at which that occurs, with perfect agreement. We find that the rate is nonzero only when the pseudo-scalar has nonzero expectation value. Our solutions are stationary because we inject an equal amount of charge from an external source coupled to the flavor fields. The corresponding supergravity statement is that we pump angular momentum into the D7-branes which then flows into a bulk horizon.

This paper is organized as follows. In section \ref{ss:fieldtheory} we describe the main characteristics of the field theory with a flavor mass that has a time-dependent phase. In section \ref{ss:sols} we describe the gravity dual and perform the holographic computation of the chiral magnetic current. In section \ref{ss:anomaly} we compare with other holographic models and explain some of the issues related to current conservation, the $N_c$ dependence of anomalies, and to thermal equilibrium, and we compute the axial charge loss rate for our system. In section \ref{ss:discuss} we summarize and discuss our results. We collect some technical results in three appendices.

\section{The Theory in Question}\label{ss:fieldtheory}

We study $\N=4$ SYM theory in the 't Hooft limit of $N_c \rightarrow \infty$ with Yang-Mills coupling squared $g_{YM}^2 \rightarrow 0$ keeping the 't Hooft coupling $\lambda \equiv g_{YM}^2 N_c$ fixed, followed by the limit $\lambda \gg 1$. The theory has an $SO(6)_R$ R-symmetry. The field content of $\N=4$ SYM theory is the gauge field, four Weyl fermions, and three complex scalars. The former are in the \textbf{4} representation and the latter in the \textbf{6} representation of $SO(6)_R\simeq SU(4)_R$. We will also consider an $\N=4$ SYM plasma with equilibrium temperature $T$.

We next introduce a number $N_f$ of $\N=2$ supersymmetric hypermultiplets in the fundamental representation of $SU(N_c)$, which in analogy with QCD we call flavor fields. In $\N=1$ notation the field content of the hypermultiplet is two chiral superfields of opposite chirality, $Q$ and $\tilde{Q}$ in the $N_c$ and $\bar{N_c}$ representations, respectively. Each chiral superfield consists of a complex scalar and a Weyl fermion. We denote the scalars, the squarks, as $q$ and $\tilde{q}$ and combine the Weyl fermions into a Dirac fermion $\psi$. The flavor fields' couplings break the $SO(6)_R$ symmetry down to $SO(4) \times U(1)_R$, of which an $SU(2)_R \times U(1)_R$ subgroup is the $\N=2$ R-symmetry. The $U(1)_R$ does not affect the squarks but acts as an axial symmetry for the quarks. Given that our theory has only this Abelian chiral symmetry, we will use ``axial symmetry'' and ``chiral symmetry'' interchangeably. As in QCD, the axial $U(1)_R$ symmetry is anomalous. The flavor fields also have a $U(1)_V$ symmetry that simply rotates $Q$ and $\tilde{Q}^{\dagger}$ by the same phase.

We will work in the probe limit, which consists of keeping $N_f$ fixed when we take the 't Hooft $N_c \rightarrow \infty$ limit, and then working to leading order in the small parameter $N_f/N_c$. Physically that corresponds to neglecting quantum effects due to the flavor fields, such as the running of the coupling. For instance, when the 't Hooft coupling is small the probe limit consists of discarding diagrams with (s)quark loops.

In the probe limit some part of the $U(1)_R$ anomaly survives, as we now explain. Three types of triangle diagram contribute to the anomaly, each with a $U(1)_R$ current at one vertex and two other currents at the other vertices. For example one diagram has the $U(1)_R$ current and two gauge currents. We will denote that as the $U(1)_R \times SU(N_c) \times SU(N_c) \equiv U(1)_R SU(N_c)^2$ anomaly, with similar notation for other anomalies. Both adjoint and flavor fields will appear in the loop, hence that diagram will have an order $N_c^2$ contribution and an order $N_f N_c$ contribution. The next diagram is the $U(1)_R^3$ anomaly, which will similarly receive order $N_c^2$ and $N_f N_c$ contributions. The third diagram is the $U(1)_RU(1)_V^2$ anomaly. Only flavor fields carry the $U(1)_V$ charge, hence that diagram will be order $N_f N_c$, with no $N_c^2$ contribution. In the probe limit we neglect the order $N_f N_c$ contribution to the first two diagrams, since that is sub-leading. For the third diagram, however, the order $N_f N_c$ term is leading, hence we retain it.\footnote{On the supergravity side the first two anomalies would appear in the type IIB supergravity sector, while the third will be associated with a WZ term on a probe D7-brane.} That anomaly will give rise to the CME in our system.

$\N=2$ supersymmetry allows for a constant, complex mass $m=|m|e^{i\phi}$ for the flavor fields. A nonzero $|m|$ explicitly breaks $U(1)_R$. We will introduce a mass with a time dependent phase: $|m| e^{i\phi} = |m| e^{i \omega t}$. Let us recall not only how $\phi = \o t$ is equivalent to an axial chemical potential for the quarks, but also how $\phi = \o t$ introduces explicit time dependence in the potential terms for $q$ and $\tilde{q}$. Of the adjoint scalars, only one is charged under $U(1)_R$. We denote this scalar as $\Phi$. The flavor couplings in the $\N=1$ superpotential $W$ are then
\beq
W\supset{\tilde Q}\Phi Q+|m|e^{i\phi} {\tilde Q}Q.
\eeq
Integrating over superspace, we find the (normal) potential, from which we extract terms involving the squarks and terms involving the quarks. The terms involving $q$ are \cite{Das:2010yw}
\beq\label{qpot}
V_q= q^\dagger |\Phi|^2 q-|m| e^{i\phi} q^\dagger \Phi^\dagger q-|m| e^{-i\phi} q^\dagger\Phi q+|m|^2 q^\dagger q.
\eeq
The potential includes identical terms for $\tilde{q}$. The quark contribution is simply\footnote{Our $\gamma^5$ is Hermitian and squares to the identity.}
\beq
V_{\psi}= |m| \, \overline{\psi}e^{i\phi \g^5}\psi.
\eeq
If we now perform a chiral rotation
\beq
\psi\to e^{-i\gamma^5 \phi/2} \psi,
\eeq
then the derivative in $\psi$'s kinetic term will act on $\phi$, producing a new term that we may include in the potential,
\beq
V_{\psi}=  |m| \overline{\psi}\psi-\frac{\partial_{\mu}\phi}{2}\,  \overline{\psi}\gamma^\mu \gamma^5 \psi.
\eeq
If we introduce $\phi = \o t$ then clearly $\o$ is equivalent to twice the axial chemical potential,
\beq\label{muomega}
\omega=2 \mu_5.
\eeq
Crucially, when $|m|$ is nonzero the squark terms of the form in eq.~\eqref{qpot} explicitly depend on $t$, so the Hamiltonian depends explicitly on time, energy is not conserved, and the system cannot be in equilibrium.

For the CME in low-temperature QCD the central players are the light pseudo-scalar and vector fields, the $\pi^0$, $\eta$, $\eta'$ and the $\rho$, respectively. Excitations of these fields, the mesons, produce poles in the corresponding retarded two-point functions. Our theory has operators analogous to these which will play a role in our realization of the CME, so let us describe them in detail.

Our theory has gauge-invariant (s)quark bilinears, \textit{i.e.} gauge-invariant operators built from two fields in the $N_c$ and $\bar{N_c}$ representations. The two-point functions of these operators exhibit poles which we will call mesons in analogy with QCD. Unlike mesons in QCD, these modes are not associated with chiral symmetry breaking or confinement, rather they are deeply bound states with masses on the order of $|m|/\sqrt{\lambda}$ \cite{Kruczenski:2003be}. When $|m|$ is nonzero these are the lightest flavor degrees of freedom in our system.

To determine the operators relevant for the CME, we treat $|m|$ and $\phi$ as external sources. We denote the associated operators as ${\cal O}_m$ and $\mathcal{O}_{\phi}$, respectively. For instance, varying minus the action with respect to $|m|$, we find a dimension three operator
\beq
{\cal O}_m =-\frac{\delta S}{\delta |m|}= \overline{\psi}e^{i\phi \gamma^5}\psi - e^{i\phi} q^\dagger \Phi^\dagger q-e^{-i\phi} q^\dagger\Phi q - e^{i\phi} \tilde{q}^\dagger \Phi^\dagger \tilde{q}-e^{-i\phi} \tilde{q}^\dagger\Phi \tilde{q}+2|m| \left( q^\dagger q + \tilde{q}^\dagger \tilde{q} \right).
\eeq
When $\phi$ is constant, $\Om$ is just the $\N=2$ supersymmetric completion of the standard quark mass operator. Notice that $\Om$ is charged under $U(1)_R$, and hence may serve as an order parameter for chiral symmetry breaking when $|m|=0$. Notice also that if $\phi = \o t$ then ${\cal O}_m$ depends explicitly on time. Varying minus the action with respect to $\phi$ we find a dimension four operator,
\beq\label{vevphiFT}
{\cal O}_\phi=-\frac{\delta S}{\delta \phi}=|m| \, i\overline{\psi}e^{i \phi \gamma^5}\gamma^5 \psi + |m|\, q^\dagger \, i\left(e^{-i\phi}\Phi - e^{i\phi}\Phi^\dagger\right) q + |m|\, \tilde{q}^\dagger \, i\left(e^{-i\phi}\Phi - e^{i\phi}\Phi^\dagger\right) \tilde{q}.
\eeq
Notice that ${\cal O}_{\phi} \propto |m|$, and again if $\phi = \o t$ then ${\cal O}_{\phi}$ depends explicitly on time.

The $U(1)_V$ baryon number and $U(1)_R$ currents will also be involved in the CME. We denote the conserved $U(1)_V$ current as $J^{\mu}$,
\beq
J^{\mu} = \overline{\psi} \g^{\mu} \psi - i \left( q^{\dagger} D^{\mu} q - \left( D^{\mu} q\right)^{\dagger} q \right) - i \left( \tilde{q} \left( D^{\mu} \tilde{q}\right)^{\dagger} - \left(D^{\mu} \tilde{q} \right)\tilde{q}^{\dagger} \right).
\eeq
The contribution to the R-current from flavor fields is the same as half the axial current $J_R^\mu=\frac{1}{2}\overline{\psi} \g^{\mu} \g^5 \psi$.\footnote{We are identifying R-charge transformations with shifts $\phi\to \phi+\delta \phi$, which for the quarks imply the $U(1)_A$ transformation $\psi\to e^{i\phi \gamma^5/2}\psi$. With this convention the R-charge of the quarks is $1/2$.} As mentioned in the introduction, adjoint fields also contribute to the R-current, hence the axial current will not be conserved even in the absence of anomalies and when $|m|=0$. We discuss the non-conservation of quark axial charge in detail in section \ref{ss:equilcme}.

Since discrete spacetime symmetries play a central role in the CME, we will also present the transformation properties of various operators under C, P and T, when $\phi = \o t$:
\begin{equation}
\begin{array}{c|ccc}
 & {\rm C} & {\rm P} & {\rm T}\\ \hline
V_q & \omega\to-\omega & {\rm even} & \omega\to-\omega \\
V_{\psi} & {\rm even} & \omega\to-\omega & {\rm even}\\
 i\overline{\psi}e^{i\omega t \gamma^5}\gamma^5 \psi & {\rm even} & \omega\to -\omega +\,{\rm odd} & {\rm odd}\\
 q^\dagger \, i\left(e^{i\omega t}\Phi^\dagger-e^{-i\omega t}\Phi\right) q &  \omega\to -\omega +\,{\rm odd} & {\rm even} & {\omega\to -\omega}
\end{array}\nonumber
\end{equation}
where $\o \rightarrow -\o$ means that a sign flip of $\o$ is the only change, and $\omega\to -\omega +\,{\rm odd}$ means a sign flip of $\o$ plus an overall sign flip are the only changes. The potential is not invariant under CPT, which is compatible with the breaking of Lorentz symmetry by the explicit time dependence. The only discrete spacetime symmetry under which the potential is invariant is CT. Notice that ${\cal O}_\phi$ is CT odd, so an expectation value $\Opv$ may serve as an order parameter for spontaneous CT breaking.

Finally, let us explain the analogy between the CME in our system and in QCD. Our system has no dynamical electromagnetic $U(1)$, so to obtain a CME we will introduce a non-dynamical external $U(1)_V$ magnetic field $F_{xy} = B$. The phase $\phi$ will play the role of a non-dynamical external axion field $a(t,\vec{x})$. The operator $\Op$ will play the role of a light, neutral pseudo-scalar, such as the $\pi^0$. At zero temperature and finite mass, we can then think of an expectation value $\Opv$ as a condensate of pseudo-scalar mesons. The $U(1)_V$ current $J^{\mu}$ will play the role of a vector meson field, like the $\rho$, so we can think of the chiral magnetic current $\jz$ as a condensate of vector mesons. Our holographic calculations will show that $\jz$ is nonzero only when $\Opv$ is nonzero, except in the chirally symmetric case $|m|=0$. Our interpretation is that the mechanism for the CME in our system is similar to that of low-temperature QCD: the magnetic field converts pseudo-scalar mesons into vector mesons polarized in the direction of $B$. Notice that away from the chiral limit the CME occurs in our system only when CT is spontaneously broken, in contrast to the free fermion case which required P and CP breaking.

\section{Chiral Magnetic Effect from Spinning Probe Branes}\label{ss:sols}

We begin in type IIB string theory with a supersymmetric intersection of $N_c$ D3-branes and $N_f$ D7-branes:
\beq
\begin{array}{c|cccccccccc}
   & x_0 & x_1 & x_2 & x_3 & x_4 & x_5 & x_6 & x_7 & x_8 & x_9\\ \hline
\mbox{D3} & \times & \times & \times & \times & & &  &  & & \\
\mbox{D7} & \times & \times & \times & \times & \times  & \times
& \times & \times &  &   \\
\end{array}
\eeq

Open strings with both ends on the D3-branes give rise at low energies to $\N=4$ $SU(N_c)$ SYM theory, while open strings with one end on the D3-branes and one on the D7-branes give rise to $\N=2$ hypermultiplets in the fundamental representation. The $SO(6)$ isometry in the directions $(x_4,\ldots,x_9)$ corresponds to the $SO(6)_R$ symmetry of $\N=4$ SYM theory. Clearly the D7-branes break that to $SO(4) \times U(1)_R$, corresponding to rotations in $(x_4,\ldots,x_7)$ and $(x_8,x_9)$ respectively. If we separate the D3- and D7-branes in the overall transverse directions, $x_8$ and $x_9$, then the 3-7 and 7-3 strings acquire a finite length, giving the hypermultiplets a mass. The complex mass $|m| e^{i \phi}$ thus corresponds simply to the relative positions of the D3- and D7-branes in that plane, with $|m|$ the separation distance and $\phi$ the angle in the plane. The breaking of $U(1)_R$ by a nonzero $|m|$ appears simply as the breaking of rotational symmetry in the $(x_8,x_9)$-plane. A time-dependent phase $\phi = \omega t$ corresponds to D7-branes spinning in the $(x_8,x_9)$-plane.

We take the usual limits for the D3-branes, $N_c \rightarrow \infty$ with $g_s N_c$ fixed, followed by taking $g_s N_c \gg 1$, where $g_s$ is the string coupling and $\alpha'$ is the string length squared. We thus obtain type IIB supergravity in the near-horizon geometry of the D3-branes, $AdS_5 \times S^5$ where each factor has radius of curvature $L^4/\a'^2 = 4 \pi g_s N_c \gg 1$. The solution includes $N_c$ units of RR five-form flux on the $S^5$. AdS/CFT equates this theory with the low-energy theory on the D3-branes, $\N=4$ SYM theory, with Yang-Mills coupling $g_{YM}^2 = 4\pi g_s$ and 't Hooft coupling $\lambda = g_{YM}^2 N_c$, so the theory is in the 't Hooft large-$N_c$ limit with $\lambda \gg 1$.

We will use an $AdS_5 \times S^5$ metric of the form
\bea\label{eq:metric}
ds^2 &  = & -|g_{tt}| \, dt^2 + g_{xx} \, d\vec{x}^2 + g_{rr} \, dr^2 + g_{SS} \, ds^2_{S^3} +g_{RR} \, dR^2 + g_{\phi\phi} \, d\phi^2, \\
& = & \frac{\rho^2}{L^2} \left( -dt^2 + d\vec{x}^2 \right) + \frac{L^2}{\rho^2} \left( dr^2 + r^2 ds^2_{S^3} + dR^2 + R^2 d\phi^2 \right)
\eea
where $\rho$ is the $AdS_5$ radial coordinate with the boundary at $\rho \rightarrow \infty$. The field theory lives in Minkowski space with coordinates $(t,\vec{x})$. We have split the six directions transverse to the D3-branes into $\mathbb{R}^4 \times \mathbb{R}^2$ where the latter $\mathbb{R}^2$ represents the $(x_8,x_9)$-plane. We have written the metric of both the $\mathbb{R}^4$ and the $\mathbb{R}^2$ in spherical coordinates. The former has radial coordinate $r$ with $ds_{S^3}^2$ the metric of a unit $S^3$, while the latter has radial coordinate $R$ and circle coordinate $\phi$. Notice that $\rho^2 = r^2 + R^2$. The self-dual RR five-form can be derived from a four-form potential
\beq\label{eq:fourform}
C_4 = g_{xx}^2 \, \textrm{vol}_{\mathbb{R}^{3,1}} - g_{SS}^2 \, d\phi \wedge \textrm{vol}_{S^3}.
\eeq
Starting now we will use units in which $L\equiv 1$.  We can convert between string theory and supergravity quantities using $\a'^{-2} = \lambda$.

The $\N=4$ SYM theory at finite temperature $T$ is dual to supergravity in an AdS-Schwarzschild spacetime. In that case only $|g_{tt}|$ and $g_{xx}$ change, becoming
\beq
|g_{tt}| = \rho^2 \frac{\gamma^2}{2} \frac{f^2(\rho)}{H(\rho)}, \qquad g_{xx} = \rho^2 \frac{\gamma^2}{2} H(\rho).
\eeq
with
\beq
f(\rho) = 1 - \frac{1}{\rho^4}, \qquad H(\rho) = 1 + \frac{1}{\rho^4}.
\eeq
In these coordinates the horizon is always at $\rho=1$, but the Hawking temperature, which we identify with the $\N=4$ SYM temperature, is $T=\gamma/\pi$, which we can vary by changing the parameter $\gamma$. We recover the $T=0$ limit by first rescaling $\rho \rightarrow \sqrt{2}\rho/\g$, and the same for $r$ and $R$, and then taking $\g \rightarrow 0$.

If we keep $N_f$ finite as $N_c \rightarrow \infty$ we may treat the D7-branes as probes. The action describing the D7-brane's dynamics, $S_{D7}$, consists of two types of terms, a Dirac-Born-Infeld (DBI) term and Wess-Zumino (WZ) terms. We will consider only the $U(1)$ worldvolume theory of coincident D7-branes, so we will need only the Abelian D7-brane action,
\beq\label{D7Action}
S_{D7} = S_{DBI} + S_{WZ},
\eeq
\bea\label{Action}
S_{DBI} & = &- N_f T_{D7} \int d^8\xi \sqrt{-\textrm{det}\left( g_{ab}^{D7} + \left( 2\pi \a'\right) \tilde{F}_{ab}\right)}, \\
S_{WZ} & = & + \frac{1}{2} N_f T_{D7} \left( 2\pi\a'\right)^2\int P[C_4] \wedge \tilde{F} \wedge \tilde{F},
\eea
where $T_{D7} = \frac{g_s^{-1} \a'^{-4}}{(2\pi)^7}$ is the D7-brane tension, $\xi^a$ are the worldvolume coordinates, $g_{ab}^{D7}$ is the induced metric on the brane, $\tilde{F}_{ab}$ is the $U(1)$ worldvolume field strength, and $P[C_4]$ is the pullback of the RR four-form to the D7-branes.

Let us introduce some convenient notation. First, we will absorb a factor of $(2\pi\a')$ into the field strength $(2\pi\a') \tilde{F}_{ab} \equiv F_{ab}$. Our D7-branes will be extended along $AdS_5 \times S^3$ inside $AdS_5 \times S^5$, that is, along the Minkowski coordinates $(t,\vec{x})$, the radial direction $r$, and the $S^3 \subset S^5$. In what follows we consider solutions for which the D7-brane Lagrangian will depend only on $r$, so we may trivially perform the integrations over the Minkowski and $S^3$ directions, producing factors of their respective volumes, $V_{\mathbb{R}^{3,1}}$ and $2\pi^2$. We will absorb the factor of the infinite volume of Minkowski space into the action, $S_{D7}/V_{\mathbb{R}^{3,1}} \rightarrow S_{D7}$. From now on we will refer to this rescaled action as the D7-brane action. We will absorb the factor of the $S^3$ volume into an overall factor
\beq
\label{eq:prefactordefinition}
\N \equiv N_f T_{D7} \, 2\pi^2 = \frac{\lambda N_f N_c}{(2\pi)^4},
\eeq
where in the second equality we converted to field theory quantities.

We now need an appropriate ansatz for the worldvolume fields to describe a CME in the field theory. The two scalars on the D7-brane worldvolume are $R$ and $\phi$. The former is dual to the operator $\Om$ while the latter is dual to $\Op$. More specifically, the asymptotic values of $R$ and $\phi$ will be (proportional to) the modulus and phase of the complex mass, as is obvious from the initial D3/D7 intersection. We thus introduce $R(r)$ and $\phi(t,r) = \omega t + \varphi(r)$, which produces a time-dependent phase for the mass and allows for nonzero $|m|$, $\Omv $ and $\Opv$.

We can motivate the $r$-dependence in $\phi(t,r)$ from previous experience with probe branes in holographic spacetimes, following ref.~\cite{Karch:2007pd}. Suppose we introduce only $\phi(t) = \omega t$ and then perform a T-duality in the $\phi$ direction.\footnote{Of course, T-duality is not a well-defined operation for an angular direction that can shrink to zero size, but here we are simply illustrating the similarities between spinning branes and branes with a worldvolume electric field.} The background solution changes, and the D7-brane becomes a D8-brane extended in $\phi$ with worldvolume gauge field, $A_{\phi}(t) \propto \omega t$. The angular frequency $\omega$ becomes a constant electric field on the D8-brane, $F_{t\phi} \propto \omega$. A probe brane in a gravitational potential well, such as AdS, and with a constant worldvolume electric field will generally have a tachyonic instability: the gravitational potential reduces the effective tension of open strings, so at some point the constant electric field can rip strings apart. This tachyonic instability causes the Lorentzian action to become imaginary. Turning that around, an imaginary action signals a tachyon, since whenever a Lorentzian action $S$ becomes a complex number, the weight factor $e^{iS}$ in a path integral will have either an exponentially growing or exponentially decaying mode. The cure for the D8-brane's instability is to introduce $r$-dependence \cite{Karch:2007pd}, \textit{i.e.} $A_{\phi}(t,r) \propto \omega t + \varphi(r)$, producing a new constant of integration, associated with $\varphi(r)$, that we can adjust to maintain reality of the action and hence avoid the instability. T-duality back to a D7-brane produces the $\phi(t,r)$ above, although now the physical interpretation of the instability is very different. Now the instability occurs because in a gravitational potential well the local speed of light decreases, while the probe brane rotates at a constant angular frequency, so at some point the probe brane may have linear velocity faster than the local speed of light. The D7-brane cures the problem by ``twisting'' in $\phi$ as a function of $r$~\cite{Evans:2008zs,O'Bannon:2008bz,Evans:2008nf}.

For the CME we need a $U(1)_V$ magnetic field and we expect a current $\jz$. The $U(1)_V$ current $J^{\mu}$ is dual to the $U(1)$ gauge field on the D7-branes, hence we include in our ansatz $F_{xy} = B$ and $A_z(r)$. In total, then, our ansatz includes\footnote{We work in $A_r=0$ gauge. Recall also that our $B$ includes a factor of $(2\pi\a') \propto \lambda^{-1/2}$.} $R(r)$, $\phi(t,r) = \omega t + \varphi(r)$, $F_{xy}=B$ and $A_z(r)$.

We can argue that the probe D7-brane action must depend only on derivatives of $\phi$ and $A_z$ as follows. The background solution has an isometry in $\phi$: the metric and four-form in eqs.~\eqref{eq:metric} and \eqref{eq:fourform} are invariant under constant shifts of $\phi$. Recalling that the scalars on the worldvolume of a D-brane are Goldstone bosons associated with the breaking of translation invariance in the transverse directions, and that Goldstone bosons can have only derivative interactions, we can conclude that the action $S_{D7}$ will involve only derivatives of $\phi$. For $A_z$ we argue simply that the action depends only on the field strength and not $A_z$ itself. Inserting our ansatz into the action, we find
\begin{align}\label{actionden}
S_{DBI} &= -\N \int dr \, g_{SS}^{3/2} g_{xx}^{3/2} \, \sqrt{1 + \frac{B^2}{g_{xx}^2}} \sqrt{\left(|g_{tt}|-g_{\p\p} \dot{\p}^2\right)\left(g_{rr}+g_{RR}\,R'^2 + g^{xx} A_z'^2\right)+|g_{tt}|g_{\p\p}\p'^2} \ , \nonumber\\
S_{WZ} &= -\N B \omega \int dr \, g_{SS}^2 A_z' \ .
\end{align}
where primes denote $\frac{\partial}{\partial r}$ and dots denote $\frac{\partial}{\partial t}$. As advertised, the action depends only on derivatives of $\phi$ and $A_z$ and hence produces two ``constants of motion,'' $\frac{\delta S_{D7}}{\delta \phi'}$ and $\frac{\delta S_{D7}}{\delta A_z'}$. We may thus solve for $\phi'$ and $A_z'$ and obtain an action for $R(r)$ only. We may do so in several ways. One is to solve for $\phi'$ and $A_z'$, derive $R(r)$'s equation of motion from $S_{D7}$, and then plug the solutions for $\phi'$ and $A_z'$ into that. Alternatively, we can plug the solutions directly into $S_{D7}$, perform a Legendre transform with respect to both $\phi'$ and $A_z'$, and then derive $R(r)$'s equation of motion. We may also proceed in stages, for example by solving for and Legendre-transforming with respect to one only and then repeating the process for the second. The simplest approach turns out to be solving for $\phi'$ first and then for $A_z'$.

The equation of motion for $\phi'$ is
\beq\label{eq:pirho}
\frac{\delta S_{D7}}{\delta \phi'} = -\N g_{SS}^{3/2} g_{xx}^{3/2} \sqrt{1 + \frac{B^2}{g_{xx}^2}} \frac{|g_{tt}|g_{\p\p} \phi'}{\sqrt{\left(|g_{tt}|-g_{\p\p} \dot{\p}^2\right)\left(g_{rr}+g_{RR}\,R'^2 + g^{xx} A_z'^2\right)+|g_{tt}|g_{\p\p}\p'^2}}\equiv \alpha,
\eeq
where $\a$ is the first constant of motion. Solving for $\p'$ we get
\beq
\label{phisol}
\phi'^2 = \frac{\alpha^2}{|g_{tt}|g_{\p\p}} \frac{ \left(|g_{tt}|-g_{\p\p} \dot{\p}^2\right)\left(g_{rr}+g_{RR}\,R'^2 + g^{xx} A_z'^2\right)}{\N^2 g_{xx}^3 g_{SS}^3 |g_{tt}|g_{\p\p} \left(1+\frac{B^2}{g_{xx}^2}\right) - \alpha^2}.
\eeq
Next we Legendre transform with respect to $\phi'$,
\beq
\label{eq:legendretransformphi}
\hat{S}_{D7} = \hat{S}_{DBI} + \hat{S}_{WZ} = S_{D7} - \int dr \phi' \, \frac{\delta S_{D7}}{\delta \phi'}.
\eeq
Notice that $S_{WZ}$ does not participate here, $\hat{S}_{WZ} = S_{WZ}$, so we focus on $S_{DBI}$,
\bea
\hat{S}_{DBI} & = & S_{DBI} - \int dr \phi' \, \frac{\delta S_{DBI}}{\delta \phi'} \\
& = & -\N \int dr \, g_{SS}^{3/2} g_{xx}^{3/2} \sqrt{|g_{tt}|-g_{\p\p} \dot{\p}^2}\sqrt{g_{rr}+g_{RR}\,R'^2 + g^{xx} A_z'^2}\sqrt{1 + \frac{B^2}{g_{xx}^2}- \frac{\a^2/\N^2}{|g_{tt}|g_{\p\p}g_{xx}^3g_{SS}^3}}. \nn
\eea
The equation of motion for $A_z'$ is then
\beq\label{eq:azeq}
\frac{\delta \hat{S}_{D7}}{\delta A_z'} = \frac{\delta \hat{S}_{DBI}}{\delta A_z'} + \frac{\delta \hat{S}_{WZ}}{\delta A_z'} \equiv \beta,
\eeq
where $\beta$ is the second constant of motion. The two terms in $\beta$ are
\beq
\label{eq:betadef}
\frac{\delta \hat{S}_{DBI}}{\delta A_z'}= -\N g_{SS}^{3/2} g_{xx}^{3/2} \sqrt{|g_{tt}|-g_{\p\p} \dot{\p}^2}\frac{g^{xx} A_z'}{\sqrt{g_{rr}+g_{RR}\,R'^2 + g^{xx} A_z'^2}}\sqrt{1 + \frac{B^2}{g_{xx}^2}- \frac{\a^2 / \N^2}{|g_{tt}|g_{\p\p}g_{xx}^3g_{SS}^3}},
\eeq
\beq
\frac{\delta \hat{S}_{WZ}}{\delta A_z'} = - \N B \o g_{SS}^2.
\eeq
We now solve for $A_z'$,
\beq
\label{azsol}
A_z' = \frac{\left(\b + \N B \o g_{SS}^2\right) g_{xx}\sqrt{g_{rr}+g_{RR}R'^2}}{\sqrt{\N^2 g_{xx}^3 g_{SS}^3 \left(|g_{tt}|-g_{\p\p} \dot{\p}^2\right)\left(1 + \frac{B^2}{g_{xx}^2}- \frac{\a^2/\N^2}{|g_{tt}|g_{\p\p}g_{xx}^3g_{SS}^3}\right)-g_{xx} \left(\b+\N B \o g_{SS}^2\right)^2  }}.
\eeq
Finally, we Legendre transform with respect to $A_z'$,
\bea
\label{eq:shathat}
\hat{\hat{S}}_{D7} & = & \hat{S}_{D7} - \int dr A_z' \frac{\delta \hat{S}_{D7}}{\delta A_z'} \\
& = & -\N \int dr \sqrt{g_{rr} + g_{RR} R'^2} \nn \\
& \times & \sqrt{g_{xx}^3 g_{SS}^3 \left(|g_{tt}|-g_{\p\p} \dot{\p}^2\right)\left(1 + \frac{B^2}{g_{xx}^2}- \frac{\a^2/\N^2}{|g_{tt}|g_{\p\p} g_{xx}^3g_{SS}^3}\right)- g_{xx} \left( \frac{\b}{\N} + B \o g_{SS}^2\right)^2}. \nn
\eea
We can derive $R(r)$'s equation of motion from this final form of the action, although is it cumbersome and unilluminating, so we will not present it.

We can now explain how to extract field theory information from bulk solutions. The fields have the following near-boundary asymptotic expansions:
\bea\label{eq:Rasymptoticexpansion}
R(r) & = & c_0 + \frac{c_2}{r^2} + \frac{1}{2} c_0 \, \omega^2 \frac{\log r}{r^2} + O\left(\frac{\log r}{r^4}\right),\\
\label{eq:phiasymptoticexpansion}
\phi(t,r) & = & \omega t + \frac{\a}{2\N c_0^3} \left( + \frac{c_0}{r^2} - \frac{c_2 + \frac{1}{8} c_0 \, \omega^2}{r^4} - \frac{1}{2} c_0 \, \omega^2 \frac{\log r}{r^4}  \right) + O\left(\frac{\log r}{r^6}\right),\\
\label{eq:azasymptoticexpansion}
A_z(r) & = & c_z +\frac{1}{2} \frac{\frac{\b}{\N} + B \omega}{r^2} - \frac{1}{2} \frac{ c_0^2 B \omega}{r^4} + O\left(\frac{1}{r^6}\right),
\eea
where $c_0$, $c_2$, and $c_z$ are constants.

In each case the leading term acts as a source for the dual operator. $c_0$ is the asymptotic separation between the original D3-branes and the D7-branes, so the magnitude of the mass is $c_0$ times the string tension, $|m| = \frac{c_0}{2\pi\a'}$. The leading term in $\phi(t,r)$, in our case $\omega t$, is the phase of the mass. $c_z$ is a source for $J^z$, equivalent to the $A_z$ component of an external gauge field. In our case we may safely set $c_z=0$.

The coefficients of the sub-leading terms determine the expectation values of the dual operators. The exact relations follow from the holographic dictionary, which equates the on-shell bulk action with minus the generating functional of the field theory. In appendix A we calculate the expectation values using holographic renormalization. For $\Omv$ we find
\beq
\label{eq:omvev}
\Omv =  (2\pi\a')\N \left( -2 c_2 - \frac{1}{2} \o^2 c_0 - \frac{1}{2} \o^2 c_0 \ln c_0^2 \right).
\eeq
Factors of the AdS radius, which we have set to one, make the argument of the logarithm dimensionless.\footnote{At first glance, the $\Omv$ in eq.~\eqref{eq:omvev} appears to diverge in the flavor decoupling limit $c_0 \propto |m| \to \infty$, which is counter-intuitive. Both the analytic argument in the appendix of ref.~\cite{O'Bannon:2008bz} and numerical calculations confirm that in fact $\Omv \to 0$ in that limit.} For the other operators we find
\beq
\label{eq:opvev}
\Opv = \a, \qquad \jz = -(2\pi\a')\beta.
\eeq

As shown in the last section, when $|m|$ is nonzero the operators $\Om$ and $\Op$ depend explicitly on time. In our solutions $c_0$, $c_2$, and $\a$ will be time-independent constants, however, so the expressions above are only consistent for nonzero $|m|$ if the state in which we evaluate $\Omv$ and $\Opv$ has time dependence that cancels the time-dependence of the operators. Our configurations correspond then to a steady state.

Notice that $\Omv$, $\Opv$, and $\jz$ are all of different orders in the large-$N_c$ and large-$\lambda$ counting. $\Omv$ is of order $(2\pi\a')\N \propto \sqrt{\lambda} N_f N_c$ times factors of order one in the large-$N_c$ and large-$\lambda$ counting, such as $c_0$ and $c_2$. In contrast, $\Opv = \a$, where from eq.~\eqref{eq:pirho} we see that $\a$ is of order $\N \propto \lambda N_f N_c$ times factors of order one, so $\Opv$ is bigger than $\Omv$ by a factor of $\sqrt{\lambda}$. Recall, however, that $\Op$ is $|m|$ times a dimension three operator. Using $|m| = \frac{c_0}{2\pi\a'} \propto \sqrt{\lambda} \, c_0$, we see that if $\Opv$ scales as $\lambda N_f N_c$ then the expectation value of the dimension three operator must scale as $\Opv/|m| \propto \sqrt{\lambda} N_f N_c$. The expectation values of the two dimension three operators, $\Omv$ and $\Opv/|m|$, thus have the same scaling. On the other hand we have $\jz \propto \a' \b$. From eq.~\eqref{eq:betadef} we see that in terms of large-$N_c$ and large-$\lambda$ counting $\b \propto \N A_z'$. Recall that we have absorbed a factor of $(2\pi\a')$ into $A_z'$. If we extract that factor, then we find $\b \propto (2\pi\a') \N \propto \sqrt{\lambda} N_f N_c$ and hence $\jz = - (2\pi\a') \b \propto N_f N_c$. We thus find that $\jz$'s normalization is independent of the coupling $\lambda$. That is not surprising. $\jz$ is our chiral magnetic current, whose normalization is fixed by the $U(1)_A U(1)_V^2$ anomaly, and so is determined by the $U(1)_A$ and $U(1)_V$ charges, not the 't Hooft coupling.

The massless limit $|m| \to 0$, or equivalently $c_0 \to 0$, is subtle. For one thing, in that limit the phase of the mass becomes ill-defined. Moreover, in that limit we see from $\phi(t,r)$'s asymptotic expansion in eq.~\eqref{eq:phiasymptoticexpansion} that $\a$ must also vanish, since otherwise the coefficients of the $r$-dependent terms in $\phi(t,r)$'s expansion would diverge. The vanishing of $\a$ in that limit makes sense, since $\a = \Opv$ and we know from section~\ref{ss:fieldtheory} that $\Opv \propto |m|$. As mentioned in section~\ref{ss:fieldtheory}, in our system $\Opv$ is an order parameter for spontaneous CT breaking, so the vanishing of $\Opv$ as $|m| \to 0$ suggests that CT will always be restored in that limit. We must be cautious, however, since the expectation value of the dimension three operator $\Opv/|m|$ need not vanish as $|m| \to 0$, so CT could still be broken. We will argue in section \ref{ss:solsfiniteT} that it actually is restored in the states we consider.

Lastly, the R-charge density appears in the bulk as the angular momentum of the D7-brane. In what follows we will not compute the R-charge density or $\Omv$, rather our focus will be on $\Opv$ and $\jz$. We will only care whether the R-charge density is zero or not. To determine that we only need to know whether the embedding $R(r)$ is nonzero or not: if $R(r)=0$ then the D7-brane has no angular momentum, while if $R(r)$ is nonzero then the D7-brane has angular momentum. That is clear from the original D3/D7 construction, since if $R(r)=0$ then the D7-brane is not extended at all in the $(x_8,x_9)$-plane and so cannot have angular momentum.

\subsection{Solutions at Zero Temperature}\label{ss:solszeroT}

Our goal now is to solve $R(r)$'s equation of motion, derived from eq.~\eqref{eq:shathat}, numerically in the pure AdS background dual to the zero-temperature vacuum of $\N=4$ SYM.

Let us first quickly review what happens when $B$ and $\o$ are zero. Here we have no time-dependent phase for the flavor mass, and we expect no current, so we also set $\phi$ and $A_z$ to zero. The induced metric on the D7-brane is then\footnote{Starting now we suppress the $r$ dependence in $R(r)$ for notational clarity, $R(r) \rightarrow R$, unless stated otherwise.}
\beq
\label{eq:inducedmetric}
ds^2_{D7} = \rho^2 (-dt^2 + d\vec{x}^2) + \frac{1}{\rho^2} (dr^2 (1+R'^2) + r^2 ds^2_{S^3}),
\eeq
where $\rho^2 = r^2 + R^2$, and the action becomes
\beq
S_{D7} = - \N \int dr \, r^3 \sqrt{1+R'^2}.
\eeq
Inside the square root factor appearing in the action is a sum of squares, hence the action will be extremized only when $R'=0$, or in other words when the solution is constant $R=c_0$. These solutions describe flavor fields with an $\N=2$ supersymmetry-preserving constant mass. $\N=2$ supersymmetry demands that $\Omv =0$, which is indeed the case for these solutions, which have $c_2=0$ and hence via eq.~\eqref{eq:omvev} $\Omv=0$.

The D7-brane is always extended along $r$ from the boundary $r=\infty$ to $r=0$, however for these constant solutions the D7-brane does not fill all of $AdS_5$. At the boundary the D7-brane wraps the maximum-volume equatorial $S^3$ inside the $S^5$, but as it extends into $AdS_5$, to smaller $r$, the $S^3$ shrinks and eventually collapses to zero size at the ``North pole'' of the $S^5$, which occurs when $r=0$. Recalling that the radial coordinate of $AdS_5$ is not $r$ but $\rho = \sqrt{r^2 + R^2}$, we see that at $r=0$ the D7-brane has only reached $\rho =c_0$: from the perspective of an observer in $AdS_5$ the D7-brane simply ends at that point. The trivial solution $R=0$ describes massless flavors. In that case the D7-brane fills all of $AdS_5$.

Notice that $R=c_0$ is a smooth solution because $R'(0)=0$, that is, the slope of $R$ is zero when the $S^3$ collapses at $r=0$. If that does not occur then we see from eq.~\eqref{eq:inducedmetric} that the D7-brane will have a conical singularity at $r=0$. The regularity condition $R'(0)=0$ remains true when $B$ and $\o$ are nonzero. In what follows we will find solutions for which $R'(0)$ is nonzero and hence the D7-brane develops a conical singularity at $r=0$.

Now let us introduce a nonzero $B$, with $\phi$ and $A_z$ still zero \cite{Filev:2007gb,Filev:2007qu,Albash:2007bk,Erdmenger:2007bn}. Roughly speaking a nonzero $B$ ``pushes'' the D7-brane toward the boundary. More precisely, suppose we fix $c_0 \propto |m| =0$. Here an infinite number of solutions appear, of which only one is the trivial solution $R(r)=0$. The key question then is which solution has the smallest on-shell action and hence is physically preferred? A numerical analysis reveals that the trivial solution is \textit{not} the preferred one: that honor is reserved for a nontrivial $R(r)$ \cite{Filev:2007gb,Filev:2007qu,Albash:2007bk,Erdmenger:2007bn}. In fact, as we increase $B$ the position where the D7-brane ends, $\rho = R(0)$, increases. Physically, the nonzero $B$ causes the D7-brane to ``bend,'' and increasing $B$ pushes the endpoint of the D7-brane closer to the boundary. Notice what that means in the field theory: the preferred solution, being nontrivial, necessarily has a nonzero $c_2$, which from eq.~\eqref{eq:omvev} indicates a nonzero $\Omv$, hence chiral symmetry is spontaneously broken. The general lesson is that in our system a nonzero $B$ promotes D7-brane bending, or in field theory language chiral symmetry breaking. The same remains true at nonzero $c_0$, although in that case $c_0$ explicitly breaks chiral symmetry.

Now let us return to nonzero $B$, $\phi$, and $A_z$, and follow the arguments of refs.~\cite{Evans:2008zs,Evans:2008nf,O'Bannon:2008bz,Das:2010yw}. The induced D7-brane metric is now
\beq\label{eq:zerotempinducedmetric}
ds^2_{D7} = g_{tt}^{D7} dt^2 + 2 g_{tr}^{D7} dt dr + g_{rr}^{D7} dr^2 + \rho^2 d\vec{x}^2 + \frac{r^2}{\rho^2} ds^2_{S^3},
\eeq
\beq
\label{eq:zerotempmetric}
g_{tt}^{D7} = \rho^2 \left(-1 + \frac{\o^2R^2}{(r^2 + R^2)^2} \right), \qquad g_{tr}^{D7} = \frac{R^2 \o \phi'}{r^2 + R^2}, \qquad g_{rr}^{D7} =  \frac{1}{\rho^2} (1+R'^2 + R^2 \p'^2),
\eeq
while the Legendre-transformed action in eq.~\eqref{eq:shathat} is
\bea
\hat{\hat{S}}_{D7} & = & - \N \int dr \, r^2 \sqrt{1+R^2} \\ & \times & \sqrt{\left(1-\frac{\o^2 R^2}{(r^2 + R^2)^2}\right)\left(1+\frac{B^2}{(r^2+R^2)^2}-\frac{\a^2/\N^2}{R^2 r^6}\right)-\frac{1}{r^6}\left(\frac{\b}{\N}+\frac{B\o r^4}{(r^2+R^2)^2}\right)^2}. \nonumber
\eea
The square root in the second line is of the form\footnote{Notice that the Legendre-transformed action here has the same generic form as the Legendre-tranformed D7-brane action with worldvolume electric and magnetic fields used in ref.~\cite{O'Bannon:2007in} for a holographic calculation of a Hall conductivity associated with the $U(1)_V$ symmetry. The similarity is not surprising, given the similarity between rotation and worldvolume electric fields due to T-duality, as explained above. Many of our arguments below are similar to those made in ref.~\cite{O'Bannon:2007in}.} $\sqrt{a(r)b(r)-c(r)^2}$ where
\beq
a(r) = 1-\frac{\o^2 R^2}{(r^2 + R^2)^2}, \quad b(r) = 1+\frac{B^2}{(r^2+R^2)^2}-\frac{\a^2/\N^2}{R^2 r^6}, \quad c(r) = \frac{1}{r^3}\left(\frac{\b}{\N}+\frac{B\o r^4}{(r^2+R^2)^2}\right).
\eeq
Notice that $a(r)$ may change sign between $r \rightarrow \infty$ and $r\rightarrow 0$, but does not necessarily. More specifically, $a(r)$ is always positive at $r \rightarrow \infty$ and may become negative as $r \rightarrow 0$, depending on the behavior of $R(r)$. For the moment let us suppose that $a(r)$ does change sign. We will denote the value of $r$ where $a(r)$ vanishes as $r_*$. Upon taking $a(r_*)=0$ and doing some algebra we find the equation for a semicircle,
\beq
\label{eq:hor}
\left( R(r_*) - \frac{\omega}{2} \right)^2 + r_*^2 = \frac{\omega^2}{4},
\eeq
where the radius\footnote{Recall that we are using units in which the $AdS_5$ radius is $L\equiv 1$.} is $\o/2$ and the center is at $(r_*,R(r_*)) = (0,\o/2)$.

In fact, this semicircle is a horizon on the worldvolume of the D7-brane. If we change coordinates
\beq
d\hat t = dt + \frac{g_{tr}^{D7}}{g_{tt}^{D7}}dr,
\eeq
then the induced metric becomes
\beq
ds_{D7}^2 = \hat{g}_{\hat t \hat t}^{D7} d\hat t^2 + \hat{g}_{rr}^{D7}dr^2 + g_{xx}d{\vec x}^2 + g_{SS}ds_{S^3}^2,
\eeq
with
\beq
\hat{g}_{\hat t \hat t}^{D7} = g_{t t}^{D7} \ , \qquad \hat{g}_{rr}^{D7} = g_{rr}^{D7} - \frac{(g_{tr}^{D7})^2}{g_{t t}^{D7} }.
\eeq
We then have $\hat{g}_{\hat t \hat t}^{D7} = - \rho^2 a(r)$ and hence $a(r_*)=0$ implies $g_{\hat t \hat t}^{D7} (r_*)= 0$. To understand the appearance of this horizon on the D7-brane, consider a light ray moving in the $\phi$ direction, at fixed values of all other coordinates. The line element for a light ray is null, hence $g_{tt} dt^2 + g_{\phi\phi} d\phi^2 = 0,$ which gives us the local speed of light in the $\phi$ direction
\beq
\frac{d \phi}{dt} = \sqrt{\frac{|g_{tt}|}{g_{\p\p}}} = \frac{r^2 + R^2}{R}.
\eeq
Clearly when $\o$ is large enough to make $a(r)<0$ the D7-brane is moving faster than the local speed of light at that value of $r$, and a worldvolume horizon appears. Formally we can associate a temperature with the worldvolume horizon. We will discuss the meaning of this temperature in what follows, especially in appendix~\ref{ss:heatcap}.

If $a(r)$ changes sign but $b(r)$ does not, then $a(r) b(r) - c(r)^2 < 0$ for some $r<r_*$ and because of the square root $\hat{\hat{S}}_{D7}$ becomes imaginary, signaling a tachyonic instability as explained above. To avoid the instability we demand that $b(r_*)=0$ also. Furthermore, as $a(r)$ and $b(r)$ approach zero, $c(r)$ must approach zero more quickly, otherwise we again encounter an instability. We thus also impose $c(r_*)=0$.

The condition $a(r_*)=0$ fixes the worldvolume horizon while the conditions $b(r_*)=0$ and $c(r_*)=0$ fix the two integration constants $\a$ and $\b$, or equivalently via eq.~\eqref{eq:opvev} $\Opv$ and $\jz$. In other words, for given values of $|m|$, $\o$, and $B$, regularity of the bulk solution determines unique values of the one-point functions $\Omv$, $\Opv$, and $\jz$, as is standard in AdS/CFT. Explicitly, we find
\beq
\label{eq:alpha}
\a = -\N R(r_*) r_*^3 \sqrt{1+ \frac{B^2}{\omega^2 R^2(r_*)}}, \qquad \b = -\N \frac{ B \omega r_*^4}{(R^2(r_*) + r_*^2)^2}.
\eeq
Using $a(r_*)=0$ we can also express $\a$ and $\b$ in terms of $R(r_\ast)$ alone,
\beq
\label{eq:zerotempabvalues}
\a = -\N R(r_\ast)^{3/2}|R(r_\ast )-\omega |^{3/2} \sqrt{R(r_\ast)^2 + \frac{B^2}{\omega^2}}, \qquad \b = -\N \frac{B}{\omega} (R(r_\ast) - \omega)^2.
\eeq

The D7-brane does not always develop a worldvolume horizon. What happens when it does not? In that case $a(r)>0$ for all values of $r$. If $\a$ is nonzero then $b(r)$ will change sign, rendering the action imaginary, so we demand $\a=0$. In addition if $\b$ is nonzero then again the action becomes imaginary because $a(r)b(r)$ goes as $1/r^4$ as $r \rightarrow 0$ while $c(r)^2$ goes like $\beta^2/r^6$, so clearly $\sqrt{a(r)b(r)-c(r)^2}$ will become imaginary at sufficiently small $r$. We thus also demand $\b=0$ for these solutions.

We thus have two classes of D7-brane embeddings, those with a worldvolume horizon and those without. The former have nonzero $\a$ and $\b$, or in the field theory nonzero $\Opv$ and $\jz$, hence these solutions describe a CME with CT spontaneously broken. The latter have $\a=0$ and $\b=0$ and hence no CME.

An exceptional case is the trivial solution $R(r)=0$ and $\phi(t,r) = \o t$, corresponding to a chirally-symmetric state with $|m|=0$ and $\Omv=0$. That solution has no worldvolume horizon, yet from eq.~\eqref{eq:zerotempabvalues} we see that although $\a$ vanishes, $\b$ must be nonzero to maintain reality of the action. Recalling that $B = (2\pi\a') \tilde{B}$, where $\tilde{B}$ is the value of the magnetic field in the field theory (see above eq.~\eqref{eq:prefactordefinition}), we find for the trivial solution
\beq\label{eq:trivialsolbeta}
\beta = -(2\pi \a')\N \tilde{B} \omega.
\eeq
Using eqs.~\eqref{muomega},~\eqref{eq:prefactordefinition}, and~\eqref{eq:opvev} to translate to field theory quantities, we find
\beq
\label{CMEcurrent}
\langle J^z \rangle = \frac{N_f N_c}{2\pi^2} \mu_5 \tilde{B},
\eeq
which is the value fixed by the anomaly, as expected in the chirally-symmetric case. We hasten to add three things. First, we will see that eq.~\eqref{CMEcurrent} is unchanged at finite temperature, where the trivial solution remains a valid solution. Second, the trivial D7-brane has no angular momentum, so the corresponding field theory state has zero axial charge density, despite having nonzero $\mu_5$ with zero mass gap, $|m|=0$. Third, because the anomaly's contribution to $\jz$ does not contain much dynamical information, will will isolate the more interesting dynamical contributions by writing the $\jz$ in eq.~\eqref{eq:opvev} as
\beq
\jz = -(2\pi\a') \b = (2\pi\a')^2 \N \tilde{B} \o \left( \frac{-\beta}{ (2\pi\a')\N \tilde{B}\omega}\right) = \frac{N_f N_c}{(2\pi)^2} \tilde{B}\omega \left( \frac{-\beta}{ (2\pi\a')\N \tilde{B}\omega} \right).
\eeq
The factor in parentheses in the final equality contains the non-trivial dynamical information. From eq.~\eqref{eq:alpha} we see that $\b \propto (2\pi\a')\N \tilde{B}\omega$, so the factor in parentheses also does not depend explicitly on the magnetic field, although it will depend implicitly through the embedding. Notice that the current is always proportional to $B$, as we expect for the CME.

To produce numerical solutions for the two classes of D7-brane embeddings, we must specify boundary conditions. The equation of motion for $R(r)$ is a second-order non-linear ordinary differential equation, for which we need two boundary conditions on $R(r)$. Solutions without worldvolume horizons are the simplest to produce. For these we set $\a=0$ and $\b=0$, choose a value of $R(0)$ greater than $\o/2$ to avoid a worldvolume horizon, and then impose $R'(0)=0$ to guarantee regularity. Solutions with worldvolume horizons are trickier to obtain,\footnote{Solutions with nonzero $B$ and worldvolume horizons but with $A_z(r)=0$ were obtained in ref.~\cite{O'Bannon:2008bz}. These solutions are in fact unphysical, since an ansatz with $A_z(r)=0$ is inconsistent: in ref.~\cite{O'Bannon:2008bz} the WZ term in eq.~\eqref{actionden} was omitted, but the presence of that term necessitates the introduction of $A_z(r)$. All other solutions in ref.~\cite{O'Bannon:2008bz} besides these are consistent.} since the equation of motion itself depends on the values of $\a$ and $\b$, or equivalently on $r_*$ and $R(r_*)$, so we must choose these before we can solve the equation of motion. For these solutions we first choose a point on the semicircle in eq.~\eqref{eq:hor}, which fixes the values of $\a$ and $\b$ via eq.~\eqref{eq:zerotempabvalues}. We then obtain a condition on the first derivative at that point, $R'(r_*)$, from the equation of motion itself. We omit the explicit form, which is unilluminating. With these boundary conditions we can solve the equation of motion both inside the worldvolume horizon and outside. Notice that in these cases the value of $R'(0)$, which determines whether the D7-brane has a concial singularity at $r=0$, is an \textit{output} of the calculation. Figure~\ref{Fig1} shows numerical solutions for $R(r)$ for various values of $|m|$, $\o$, and $B$.

Our first observation is that for all solutions with a worldvolume horizon $R'(0)$ is nonzero, so in our system at zero temperature all non-trivial solutions describing the CME have a conical singularity. In fact, for these solutions the on-shell action exhibits a divergence at $r=0$, taking us outside of both the probe and supergravity limits, so strictly speaking we should not trust these solutions. Nevertheless, in section \ref{ss:anomaly} we will argue that the singularities are physically sensible, being intimately related with the time rates of change of axial charge and energy in the field theory.

\FIGURE[t]{
	\includegraphics[width=6cm]{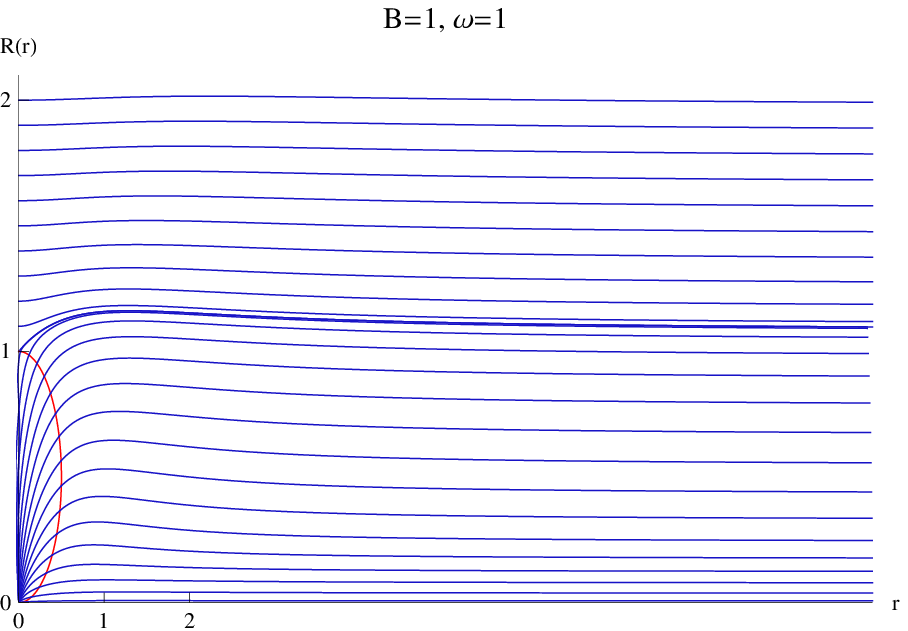}\qquad
	\includegraphics[width=6cm]{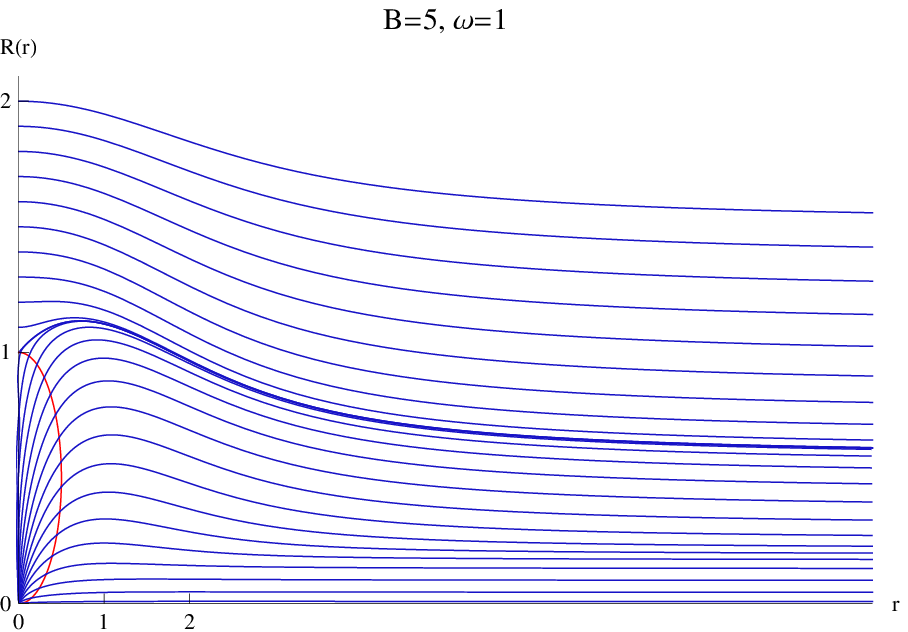}\\
	\vspace{0.4cm}
	\hspace{1cm} (a.) \hspace{6cm} (b.)\\
	\includegraphics[width=6cm]{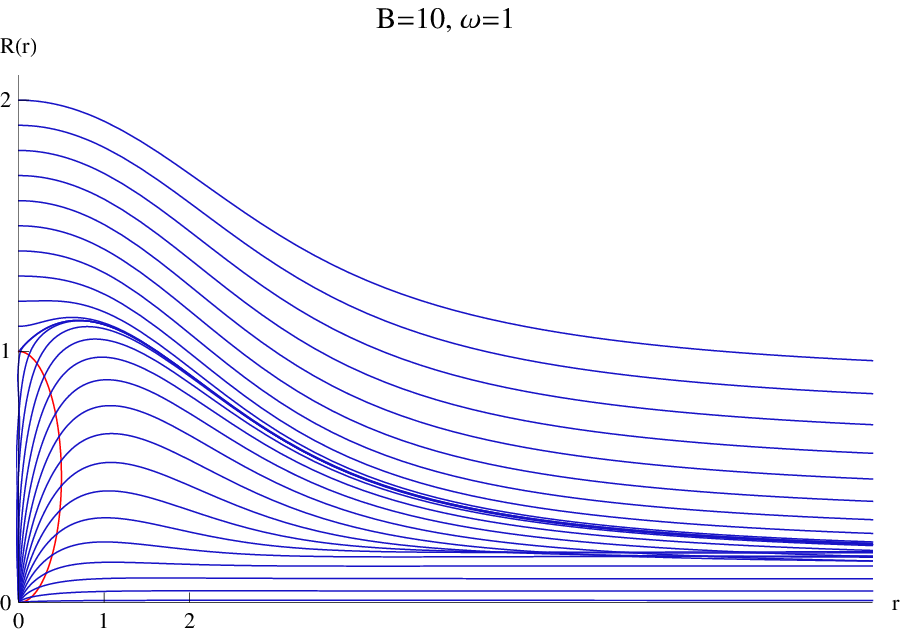}\qquad
	\includegraphics[width=6cm]{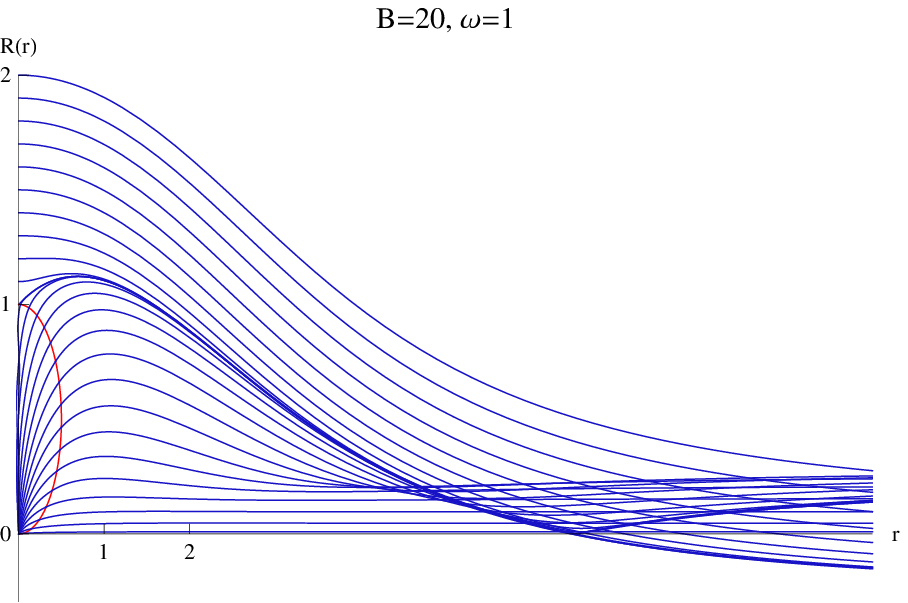}\\
	\vspace{0.1cm}
	\hspace{-0.4cm} (c.) \hspace{6cm} (d.)
	\caption{Numerical D7-brane embeddings $R(r)$ for $T=0$ and $\o=1$ for various values of $B$, in units of the $AdS_5$ radius. The red semi-circle denotes the worldvolume horizon of eq.~\eqref{eq:hor}. (a.) $B=1$. (b.) $B=5$. (c.) $B=10$. (d.) $B=20$. The asymptotic value of $R(r)$ as $r \to \infty$ (the far right in each plot) is the coefficient $c_0$ in eq.~\eqref{eq:Rasymptoticexpansion}, which is proportional to the flavor mass $|m|$. The different classes of solutions, and their behavior as functions of $B$ and $|m|$, are discussed in the accompanying text.
	}
	\label{Fig1}
}

Our second observation is that all solutions with a worldvolume horizon have nonzero $|m|$. That makes sense, since these solutions have nonzero $\a$, and hence must have nonzero $|m|$, as described above. Only solutions with $\a=0$ can describe $|m|=0$. That class of solutions includes the trivial one $R(r)=0$ as well as non-trivial solutions without worldvolume horizons.

From fig.~\ref{Fig1} we can deduce the general behavior of solutions as we increase $|m|$ or $B$, as follows. Suppose we fix $\o$ and $B$, \textit{i.e.} we choose one of figs.~\ref{Fig1} (a.) through (d.), and then begin with some $c_0 \propto |m|$ that is nonzero but much smaller than $\o$ or $\sqrt{B}$, such that the solution is very close to the trivial $R(r)=0$ solution. As we increase $|m|$, clearly the endpoint $R(0)$ also increases. For sufficiently large $|m|$ the worldvolume horizon will disappear, at which point $\a$ and $\b$ vanish. Alternatively, suppose we fix $\o$ and $c_0 \propto |m|$, and then increase $B$. Now in fig.~\ref{Fig1} we are choosing the value of a curve at the far right and then moving through the figures from (a.) to (d.). Again we see that for sufficiently large $B$ the worldvolume horizon will disappear.\footnote{From fig.~\ref{Fig1} (d.) we also see that for some D7-branes without a worldvolume horizon, $R(r)$ passes through zero for sufficiently large $B$. Such behavior has been observed many times for D7-branes with worldvolume magnetic field: see for example refs.~\cite{Filev:2007gb,Erdmenger:2007bn}. As argued in ref.~\cite{Erdmenger:2007bn}, these solutions have a sensible interpretation in the field theory as a renormalization group flow, although in equilibrium they are not always the lowest-energy solutions. These D7-branes do not describe a CME and so are of less interest to us than D7-branes with worldvolume horizons.} The corresponding field theory statements are that increasing $|m|$ or $B$ eventually restores CT and extinguishes the CME, since eventually $\Opv=0$ and $\jz=0$.  The general lesson is that chiral symmetry breaking, whether explicit via $|m|$ or spontaneous via $B$, acts against the CME in our system.

Our main result in this section is fig.~\ref{Fig2}. The green solid curve in fig~\ref{Fig2} (a.) shows the exact behavior of $\jz$, normalized to the value in eq.~\eqref{CMEcurrent}, as we increase $|m|/\mu_5$, and and the green curve in fig.~\ref{Fig2} (b.) shows the same for $\Opv/\N$. At $|m|=0$, $\jz$ takes the value determined by the anomaly, while $\Opv=0$. Increasing $|m|/\mu_5$, $\jz$ decreases monotonically and eventually reaches zero, while $\Opv$ increases, reaches a maximum, and then drops to zero. We omit the curves for $\jz$ and $\Opv$ versus $\sqrt{B}/\mu_5$, which are qualitatively similar to those in fig.~\ref{Fig2}.


\FIGURE[t]{
	\includegraphics[width=7cm]{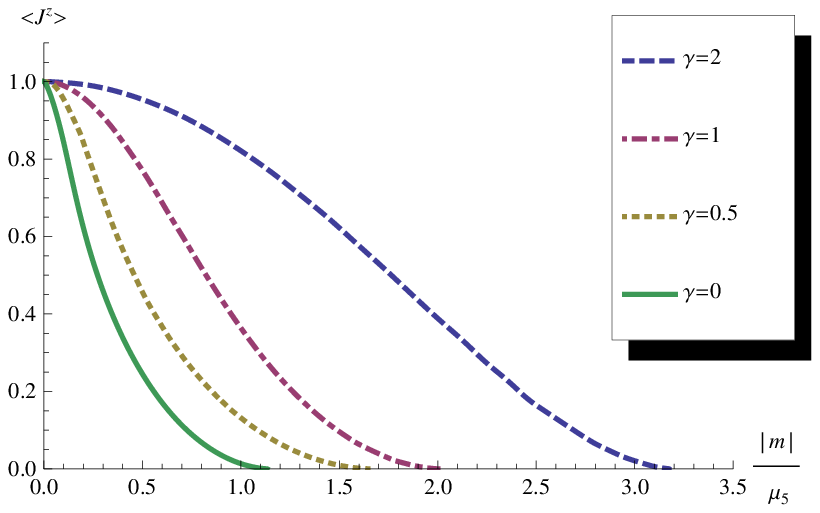}\qquad
	\includegraphics[width=7cm]{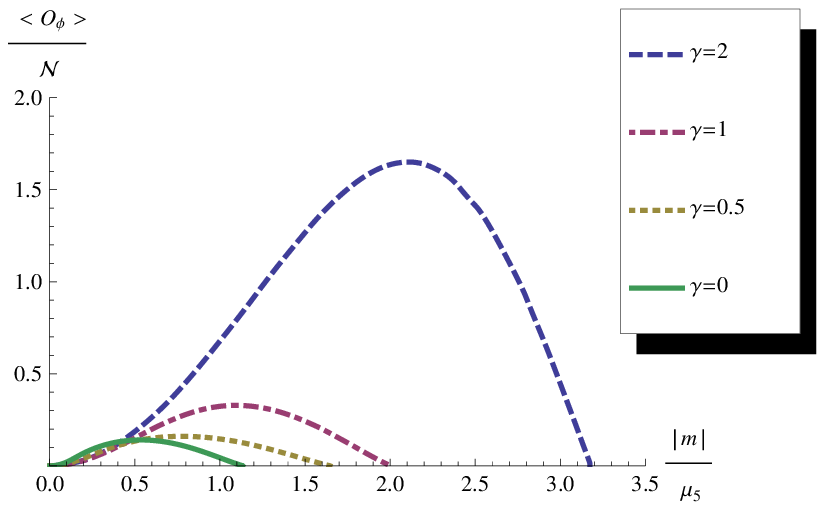}\\
	\vspace{0.1cm}
	\hspace{0.7cm} (a.) \hspace{7cm} (b.)
	\caption{(a.) The value of $\jz$, divided by the anomaly-determined value in eq.~\eqref{CMEcurrent}, as a function of the flavor mass divided by the axial chemical potential, $|m|/\mu_5$. Here we set the magnetic field $B=1$ and $\omega = 2 \mu_5 = 1$ (in units of the $AdS_5$ radius). The different curves correspond to different temperatures $T=\g/\pi$, with green solid, yellow dotted, red dot-dashed, and blue dashed corresponding to $\g=0,0.5,1,2$, respectively. (b.) The pseudo-scalar condensate $\Opv/\N$ versus $|m|/\mu_5$, with $B=1$ and $\omega=2\mu_5=1$, for the same temperatures as in (a.).
}
	\label{Fig2}
}

\subsection{Solutions at Finite Temperature}\label{ss:solsfiniteT}

We now want to solve $R(r)$'s equation of motion numerically in the AdS-Schwarzschild background, corresponding to an $\N=4$ SYM plasma at temperature $T$.

Before doing so, let us briefly review what occurs when $\o=0$ and $B=0$, \ie when the worldvolume gauge field and $\phi$ vanish, summarizing refs.~\cite{Babington:2003vm,Kirsch:2004km,Ghoroku:2005tf,Mateos:2006nu,Albash:2006ew,Mateos:2007vn}. The main difference from the zero-temperature, pure $AdS_5$ case is the presence of the AdS-Schwarzschild horizon, which divides D7-brane solutions into two categories. The first are similar to those in pure $AdS_5$, namely D7-branes for which the $S^3$ shrinks and eventually collapses to zero size at some value of $\rho$ outside the AdS-Schwarzschild horizon. The second category consists of D7-branes for which the $S^3$ shrinks but does not reach zero size by the time the D7-brane intersects the AdS-Schwzrzschild horizon. In the current context, solutions in the first category are called ``Minkowski'' embeddings while solutions in the second category are called ``black hole'' embeddings. In Euclidean signature, with compact time direction of period $1/T$, the time circle collapses to zero size at the horizon. The two types of D7-brane solution thus have distinct topology: Minkowski embeddings have a collapsing three-cycle, the $S^3$, while black hole embeddings have a collapsing one-cycle, the time circle. For Minkowski emeddings the condition to avoid a conical singularity when the $S^3$ collapses is $R'(0)=0$, while the condition for black hole embeddings to avoid a singularity is that in the $(r,R)$ plane the D7-brane must be perpendicular to the AdS-Schwarzschild horizon.

When $\o$ and $B$ both vanish, the theory has one physically meaningful dimensionless parameter, $T/|m|$. Suppose we fix $T/|m|$ such that we have a Minkowski embedding and then increase $T/|m|$, say by holding $|m|$ fixed but increasing $T$. In the bulk the AdS-Schwarzschild horizon will grow and move toward the boundary, eventually encountering the D7-brane. The D7-brane solution then becomes a black hole embedding. Such a process involves a change in topology, so we have reason to expect that in general any observable associated with the flavor fields in the field theory will exhibit discontinuous behavior. Indeed, the bulk transition from Minkowski to black hole embedding appears in the field theory as a first-order phase transition \cite{Babington:2003vm,Kirsch:2004km,Ghoroku:2005tf,Mateos:2006nu,Albash:2006ew,Mateos:2007vn}.

Perhaps the most dramatic change in that transition occurs in the spectrum of D7-brane excitations, dual to the spectrum of mesons. For a Minkowski embedding the fluctuations of worldvolume fields are normal modes, \textit{i.e.} standing waves trapped between the $AdS_5$ boundary and the endpoint of the D7-brane. That translates into a field theory meson spectrum that is gapped and discrete \cite{Kruczenski:2003be}. For a black hole embedding the worldvolume fluctuations are quasi-normal modes, that is, the eigenfrequencies acquire an imaginary part. Physically, these fluctuations can leak energy into the AdS-Schwarzschild horizon and hence are damped. In the field theory the meson spectrum is gapless and continuous. The transition between the two is thus a kind of ``meson melting'' transition \cite{Hoyos:2006gb}. 

Introducing nonzero $B$, still keeping $\o=0$, qualitatively has the same effect as in the pure $AdS_5$ case: increasing $B$ pushes the D7-brane toward the boundary. If we start with a black hole embedding, for example,  and keep $T/|m|$ fixed while increasing $B/|m|^2$, eventually a transition occurs to a Minkowski embedding \cite{Filev:2007gb,Filev:2007qu,Albash:2007bk,Erdmenger:2007bn}.

Now consider nonzero $|m|$, $T$, $B$, and $\o$. The induced D7-brane metric is then
\beq
ds^2_{D7} = g_{tt}^{D7} dt^2 + 2 g_{tr}^{D7} dt dr + g_{rr}^{D7} dr^2 + \rho^2 \frac{\g^2}{2} H(\rho) d\vec{x}^2 + \frac{r^2}{\rho^2} ds^2_{S^3},
\eeq
where $g_{tr}^{D7}$ and $g_{rr}^{D7}$ are the same as in eq.~\eqref{eq:zerotempmetric} but
\beq
g_{tt}^{D7} = \rho^2 \left(- \frac{\g^2}{2} \frac{f(\rho)^2}{H(\rho)} + \frac{\o^2R^2}{(r^2 + R^2)^2} \right), \qquad f(\rho) = 1 - \frac{1}{\rho^4}, \qquad H(\rho) = 1 + \frac{1}{\rho^4},
\eeq
and we recall that $\rho^2 = r^2 + R^2$, the AdS-Schwarzschild horizon is at $\rho=1$, and the temperature is $T=\g/\pi$ in our conventions. The location of the D7-brane's worldvolume horizon $r_*$ is now given by
\beq
\label{eq:horT}
\frac{\g^2}{2} \frac{f(\rho_*)^2}{H(\rho_*)} - \frac{\o^2R(r_*)^2}{(r_*^2 + R(r_*)^2)^2} = 0,
\eeq
or equivalently
\beq
f(\rho_*)^2 = \frac{2 H(\rho_*)}{\g^2} \frac{\o^2R(r_*)^2}{(r_*^2 + R(r_*)^2)^2}.
\eeq
Clearly the D7-brane worldvolume horizon is always outside of the AdS-Schwarzschild horizon, since $f(\rho_*)>f(\rho=1)=0$. If $\o/\g  \propto \mu_5 /T \rightarrow 0$ then $\rho_* \rightarrow 1$ and the two horizons coincide. At fixed mass, we may think of the $\mu_5 /T \rightarrow 0$ limit either as small $\mu_5$ at fixed $T$ or large $T$ at fixed $\mu_5$.

The physical arguments of the last subsection for the reality of the action are unchanged. Applying those arguments to fix $\a$ and $\b$ we find
\beq
\label{eq:alphaT}
\a = -\N \frac{\g^4}{4}R(r_*) r_*^3 f(\rho_*) H(\rho_*) \sqrt{1 + \frac{4 B^2}{\g^4 (R(r_*)^2 + r_*^2)^2H(\rho_*)^2}} \ , \quad \b = -\N B \o \frac{r_*^4}{(R(r_*)^2 + r_*^2)^2}.
\eeq

In AdS-Schwarzschild our D7-brane solutions fall into three categories. The first two are the straightforward generalizations of the categories of the last subsection: Minkowski embeddings without worldvolume horizons and Minkowski embeddings with worldvolume horizons. The new category consists of black hole embeddings, which necessarily have worldvolume horizons, as explained above. As in the last subsection, solutions without a worldvolume horizon describe field theory states with no CME and no spontaneous breaking of CT, while solutions with a worldvolume horizon, whether Minkowski or black hole, describe field theory states with a CME and spontaneous breaking of CT.

The trivial solution $R(r)=0$ falls into the third category of embeddings, since it necessarily intersects the AdS-Schwarzschild horizon. From eq.~\eqref{eq:alphaT} we see that for the trivial solution $\a=0$ and $\b$ takes the value in eq.~\eqref{eq:trivialsolbeta}, so $\jz$ again takes the value in eq.~\eqref{CMEcurrent}, hence we see that the value of $\jz$ in the chirally-symmetric case, being fixed by the anomaly, is independent of temperature.

For Minkowski embeddings the procedure to generate numerical solutions is the same as in the last subsection. In particular, for solutions with a worldvolume horizon we first choose a point on the worldvolume horizon and then use the equation of motion to determine the first derivative. We use the latter procedure for black hole embeddings too, since these necessarily have a worldvolume horizon. Notice that when we impose boundary conditions at the worldvolume horizon, the behavior of $R(r)$ and its derivative at $r=0$ or at the AdS-Schwarzschild horizon, which determines whether the solution has a conical singularity, is an output of the calculation.

Figure~\ref{D7FiniteT} shows numerical solutions for $R(r)$ for various values of $|m|$, $\o$, $B$, and $T$. The results for Minkowski embeddings are similar to those of the last subsection. In particular, Minkowski embeddings with a worldvolume horizon have a nonzero $R'(0)$ and hence a conical singularity. The black hole embeddings, however, do not have such a conical singularity: as figure~\ref{D7FiniteT} suggests, and numerical analysis confirms, in the $(r,R)$ plane depicted the D7-brane ``hits'' the black hole horizon perpendicularly.

\FIGURE[t]{
	\includegraphics[width=6cm]{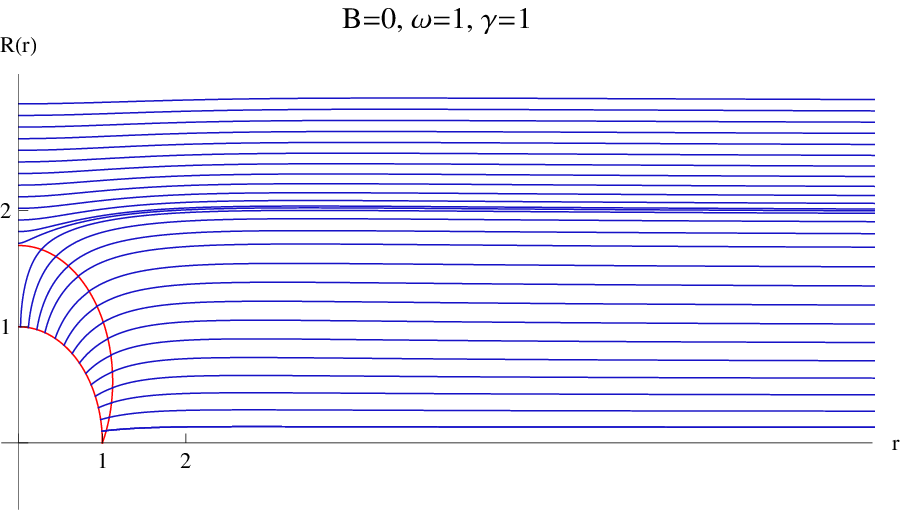}\qquad
	\includegraphics[width=6cm]{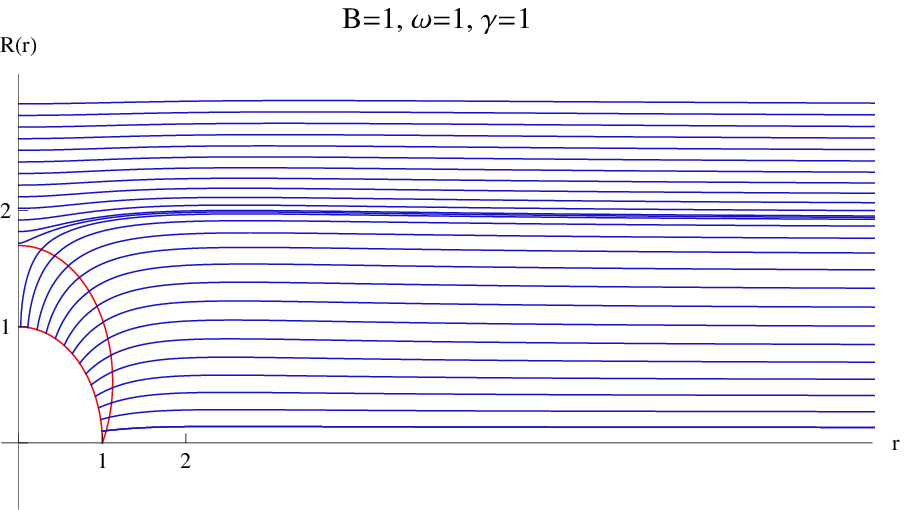}\\
	\vspace{0.4cm}
	\hspace{1cm} (a.) \hspace{6cm} (b.)\\
	\includegraphics[width=6cm]{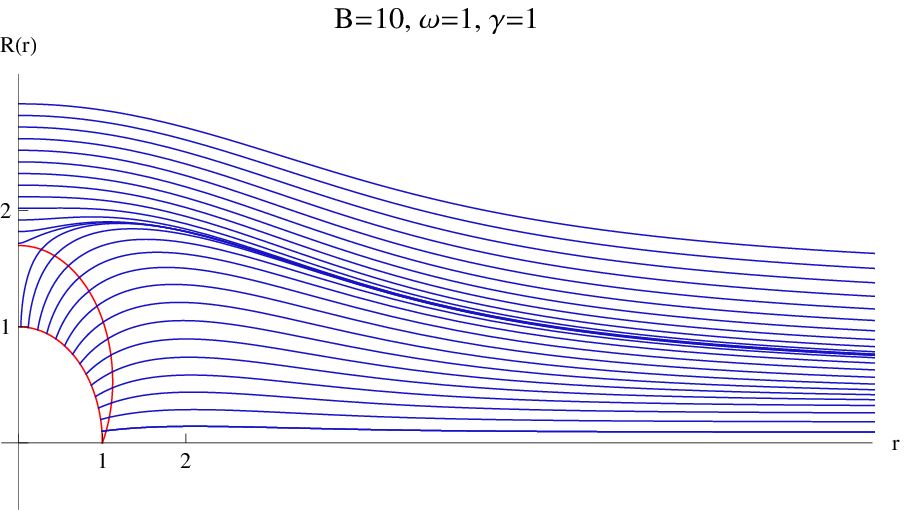}\qquad
	\includegraphics[width=6cm]{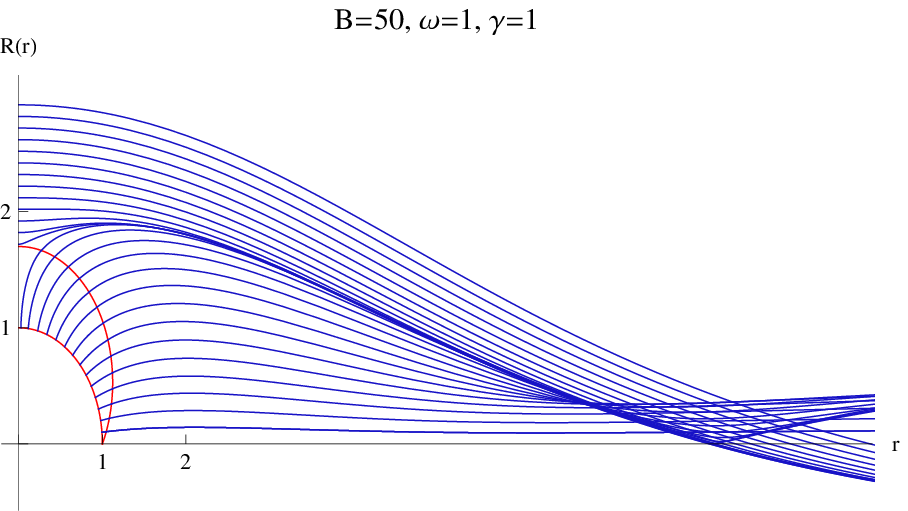}\\
	\vspace{0.1cm}
	\hspace{-0.4cm} (c.) \hspace{6cm} (d.)
	\caption{Numerical D7-brane embeddings $R(r)$ for $\o=1$ and $\g=1$, corresponding to a temperature $T=\g/\pi=1/\pi$, for various values of $B$, in units of the $AdS_5$ radius. The red quarter-circle represents the AdS-Schwarzschild horizon while the other red curve (the near-quarter-circle) denotes the worldvolume horizon of eq.~\eqref{eq:horT}. (a.) $B=0$. (b.) $B=1$. (c.) $B=10$. (d.) $B=50$. The different classes of solutions, and their behavior as functions of $B$, $|m|$, and $T$, are discussed in the accompanying text.}
	\label{D7FiniteT}
}

Many of the conclusions from our $T=0$ analysis remain valid at finite temperature. All solutions with nonzero $\a$ have nonzero $c_0 \propto |m|$. The worldvolume horizon eventually disappears as we increase $|m|$ or $B$: chiral symmetry breaking works against the CME in our system.

Fig.~\ref{Fig2} (a.) shows the chiral magnetic current $\jz$, normalized to the anomaly-determined value in eq.~\eqref{CMEcurrent}, versus $|m|/\mu_5$ for $B=1$ and several values of $T$. At higher $T$ the chiral magnetic current can persist to higher values of $|m|/\mu_5$ before dropping to zero. Fig.~\ref{Fig2} (b.) shows $\Opv/\N$ versus $|m|/\mu_5$ for $B=1$ and the same values of $T$ as in fig.~\ref{Fig2} (a.). The qualitative behavior of the pseudo-scalar condensate is similar to the $T=0$ case, increasing, reaching a maximum, and then dropping to zero as we increase $|m|/\mu_5$. At higher $T$, the maximum of the condensate is larger, and the condensate also persists to higher values of $|m|/\mu_5$.

As mentioned above, seeing $\Opv \to 0$ as $|m| \to 0$ is not enough to conclude that CT is restored in that limit. $\Op$ is $|m|$ times a dimension three operator, and the expectation value of that operator could remain finite as $|m| \to 0$. For the states we consider, we can argue that the expectation value of the dimension three operator vanishes as $|m| \to 0$ as follows. In the limit $|m| \to 0$ we expect the solution to approach the constant one $R(r) \approx c_0 \propto |m|$, and we expect the worldvolume horizon to approach the AdS-Schwarzschild horizon, so $\rho_* \approx 1$. Inserting these approximations into eq.~\eqref{eq:horT} we find $f(\rho_*) \simeq |m|\o/\g$, and then from eq.~\eqref{eq:alphaT} we find $\a \propto |m|^2$, so $\Opv$ vanishes as $|m|^2$ as $|m| \to 0$, indicating that the dimension-three operator $\Opv/|m|$ vanishes as $|m|$. We have confirmed that our numerical results for $\Opv$ in fig. \ref{Fig2} (b.) behave as $|m|^2$ as $|m| \to 0$.

For black hole embeddings we expect the spectrum of worldvolume excitations will be gapless and continuous, as in the $\o=0$ case. We expect the same for Minkowski embeddings with a worldvolume horizon: fluctuations of worldvolume fields will ``see'' the worldvolume horizon as a genuine horizon, and hence we expect them to behave in a fashion similar to those of black hole embeddings.\footnote{A gapless, continuous spectrum appears in the presence of a worldvolume electric field, which is closely related to rotation via T-duality as we argued above~\cite{Albash:2007bq,Mas:2009wf}.} More specifically, since we can associate a temperature with the worldvolume horizon, we expect the solutions for linearized fluctuations to translate into field theory two-point functions with a form characteristic of thermal diffusion at that temperature \cite{Das:2010yw}. Moreover, in the bulk the worldvolume and AdS-Schwarzschild horizons generally will not coincide, so the worldvolume temperature will generally be different from the $\N=4$ SYM plasma temperature. This provides another hint that the system is not in equilibrium.

Although we avoided an obvious instability by demanding reality of the D7-brane action, more subtle instabilities may exist in the spectrum of worldvolume fluctuations. In other words, some worldvolume fluctuations may still be tachyonic. Indeed, such instabilities have been found in very similar systems \cite{PremKumar:2011ag}. In appendix~\ref{ss:stability} we present a preliminary analysis of stability of linearized fluctuations. We find suggestive evidence that instabilities may occur for black hole embeddings with large enough magnetic field or chemical potential. We leave a complete analysis for the future.

\section{Comparing Holographic Models of the CME}\label{ss:anomaly}

Our goal in this section is to compare our system to the previous holographic studies of the CME in refs.~\cite{Rebhan:2009vc,Yee:2009vw,Gorsky:2010xu,Rubakov:2010qi,Gynther:2010ed,Brits:2010pw,Kalaydzhyan:2011vx}. The first issue we discuss is holographic realizations of the $U(1)_A$ anomaly.  For concreteness we focus on the Sakai-Sugimoto model \cite{Sakai:2004cn,Sakai:2005yt} because that involves probe flavor branes, just like our system, although much of our discussion applies for any theory in which the axial anomaly is realized holographically via a (4+1)-dimensional Chern-Simons term. The second issue we discuss is the regularity of bulk solutions describing the CME, which is intimately related to the question of whether the system is in equilibrium. Our states describing the CME are not in equilibrium, since axial charge and energy can leak to the adjoint sector. We compute the rates at which those occur in section~\ref{ss:equilcme}. We end the section with proposals for new holographic systems that should exhibit a CME.

\subsection{Anomalies and Definitions of Axial and Vector Currents}\label{ss:anomalydefinitions}

The Sakai-Sugimoto model is a holographic model of QCD that begins in type IIA string theory with the following non-supersymmetric intersection of $N_c$ D4-branes, $N_f$ D8-branes, and $N_f$ $\overline{\mbox{D8}}$-branes
$$
\begin{array}{c|cccccccccc}
 & x_0 & x_1 & x_2 & x_3 & (x_4) & x_5 & x_6 & x_7 & x_8 & x_9 \\ \hline
 \mbox{D4} & \times & \times & \times & \times & \times & & & & & \\
\mbox{D8}/\overline{\mbox{D8}} & \times & \times & \times & \times & & \times & \times & \times & \times & \times
\end{array}
$$
The low-energy theory living on the D4-branes is maximally-supersymmetric (4+1) dimensional $SU(N_c)$ YM theory. One spatial direction is compact, denoted $(x_4)$ above, with supersymmetry-breaking boundary conditions (fermions are anti-periodic in that direction). The holographic dual is type IIA string theory in the near-horizon D4-brane geometry with compact $x_4$ \cite{Witten:1998zw}, which consists of an $S^4$ with $N_c$ units of four-form flux $F_4$ and a six-dimensional space with (3+1)-dimensional Poincar\'e symmetry (at zero temperature), a radial/holographic direction, and the compact $x_4$ direction whose size goes smoothly to zero at a finite radial position in the bulk. The radial and $x_4$ directions thus form a ``cigar'' geometry.

The supergravity approximation is only reliable in the 't Hooft large-$N_c$ limit with an 't Hooft coupling that is large at the $x_4$ compactification scale. If the 't Hooft coupling were small at that scale, then the low-energy effective dynamics would be (3+1)-dimensional pure $SU(N_c)$ YM theory: with supersymmetry broken, the SYM scalars acquire a mass due to loop effects while the fermions acquire a mass at tree level. When the 't Hooft coupling is large at the $x_4$ compactification scale, the low-energy theory is $SU(N_c)$ YM theory plus additional degrees of freedom, namely a tower of Kaluza-Klein modes associated with $x_4$.

The probe D8- and $\overline{\mbox{D8}}$-branes localized in $x_4$ introduce flavor fields \cite{Sakai:2004cn,Sakai:2005yt}. At the (3+1)-dimensional intersection with D4 branes, the D8-branes introduce left-handed Weyl fermions while the $\overline{\mbox{D8}}$-branes introduce right-handed Weyl fermions. Since we choose the same number of each, we can package them into massless Dirac fermions. The field theory then has a non-Abelian global flavor symmetry $U(N_f)_L\times  U(N_f)_R$ of which the $U(1)_L\times U(1)_R$ subgroup is anomalous. In the near-horizon D4-brane geometry the D8/$\overline{\mbox{D8}}$-branes are extended along the $S^4$ and the (3+1)-dimensional field theory directions, and describe a curve on the cigar.

For the sake of comparison, we will consider also another holographic model of QCD, proposed in ref.~\cite{Kruczenski:2003uq}, which uses the same compactified D4-branes but probe D6-branes instead of D8/$\overline{\mbox{D8}}$-branes,
$$
\begin{array}{c|cccccccccc}
 & x_0 & x_1 & x_2 & x_3 & (x_4) & x_5 & x_6 & x_7 & x_8 & x_9 \\ \hline
\mbox{D4} & \times & \times & \times & \times & \times & & & & & \\
\mbox{D6} & \times & \times & \times & \times & & \times & \times & \times & &
\end{array}
$$
In fact, starting from our D3/D7 system, if we compactify $x_4$ and then T-dualize in that direction we arrive at the D4/D6 intersection above. As in D3/D7, for D4/D6 the flavor fields are $\N=2$ hypermultiplets localized at the (3+1)-dimensional intersection, where the hypermultiplet scalars acquire a mass due to loop effects but chiral symmetry prevents the fermions from acquiring a mass. In the near-horizon geometry the D6-branes are extended along an $S^2 \subset S^4$ and the (3+1)-dimensional field theory directions, and also describe a curve along the cigar. Most of the features of the D3/D7 and D4/D6 systems relevant for the CME are the same, so we will treat these two systems on an equal footing in the following. For example, the D4/D6 system has only an Abelian chiral symmetry, just like the D3/D7 system.

In the Sakai-Sugimoto and D4/D6 systems thermal equilibrium is realized holographically by Wick-rotating to Euclidean time, compactifying the Euclidean time direction with period $1/T$, and imposing anti-periodic boundary conditions for fermions in that direction. Two possible dual geometries are known, with the preferred one (dominating the bulk path integral) depending on $T$. For $T$ sufficiently small compared to the $x_4$ radius, the bulk solution has a compact time direction but no horizon, and is dual to a confined state with the center symmetry preserved. At sufficiently large $T$ a first-order phase transition occurs to a black brane geometry, dual to a deconfined state with spontaneously broken center symmetry \cite{Aharony:2006da}.

The main difference between the Sakai-Sugimoto and D3/D7 or D4/D6 systems that is relevant for the CME is the bulk realization of the $U(1)_A$ symmetry. In the Sakai-Sugimoto model, axial charge is carried by the flavor fields only and not by the adjoint fields. In the bulk that becomes the statement that the dual $U(1)_A$ gauge fields are associated with the D8/$\overline{\mbox{D8}}$-branes alone, and the axial anomaly is realized via Chern-Simons terms. More generally the $U(1)_L\times  U(1)_R$ symmetries are dual to gauge fields $L_M$ and $R_M$ on the D8- and $\overline{\mbox{D8}}$-branes, respectively. We define vector and axial fields as $V_M=(R_M+L_M)/2$, $A_M=(R_M-L_M)/2$. The $U(1)_A$ anomaly is then realized holographically via a WZ term in the D8-brane action
\beq
S_{WZ}^{D8}=\frac{1}{3!} (2\pi\alpha')^3 N_f T_{D8} \int_{D8} P[C_3]\wedge F\wedge F\wedge F,
\eeq
where $T_{D8}$ is the D8-brane tension and $P[C_3]$ is the pullback of the RR 3-form giving rise to the four-form, $F_4 = d C_3$. Integrating by parts, using $\int_{S^4} F_4= N_c$, and adding the contribution from the $\overline{\mbox{D8}}$-branes (corresponding to the right-handed Weyl fermions), we find\footnote{In this section, uppercase Latin letters $M,N,\ldots$ will denote (4+1)-dimensional bulk coordinates, including the holographic radial direction, while lowercase Greek letters $\mu,\nu,\ldots$ will denote (3+1)-dimensional boundary field theory directions.}
\beq
S_{WZ}^{D8}=-\kappa\int d^5 x \, \epsilon^{MNLPQ} L_M F^L_{NL} F^L_{PQ}+ \kappa\int d^5 x \, \epsilon^{MNLPQ} R_M F^R_{NL} F^R_{PQ},
\eeq
where $\kappa = \frac{1}{3!} (2\pi\alpha')^3 T_{D8} N_f N_c$. We thus obtain (4+1)-dimensional Chern-Simons terms, which are gauge invariant up to boundary terms. In gauge/gravity duality that suggests the associated field theory gauge transformations are anomalous, which is indeed the case here: both $U(1)_L$ and $U(1)_R$ are anomalous. In terms of the axial and vector fields the Chern-Simons terms become
\beq\label{eq:SSanomalies}
S_{WZ}^{D8}=\frac{\kappa}{2}\int d^5 x \, \epsilon^{MNLPQ} \left [ V_M F^V_{NL} F^A_{PQ} + \frac{1}{2} A_M F^V_{NL} F^V_{PQ} + \frac{1}{2} A_M F^A_{NL} F^A_{PQ} \right].
\eeq
Both the vector and the axial combinations are anomalous, although the vector anomaly vanishes when $F^A=0$, \ie when external axial field strengths vanish.\footnote{Throughout this paper we only consider the effects of the anomaly on one-point functions of currents, \ie on charge conservation equations. That is what we mean when we say an anomaly vanishes. The anomaly would still be visible in three-point functions of the current, even when external fields vanish, however.} In the field theory, the first and second terms correspond to $U(1)_AU(1)_V^2$ anomalies, while the the last term corresponds to a $U(1)_A^3$ anomaly. In all cases the coefficient is of the same order, $N_f N_c$, and survives the probe limit.

In contrast, in the D3/D7 and D4/D6 systems the $U(1)_A$ is realized holographically as rotations of the flavor branes in a transverse plane, which in the near-horizon limit becomes a transverse direction $\phi$ inside the internal space, $S^5$ or $S^4$. We have seen how that works for the D3/D7 system, so let us explain what happens in the D4/D6 system. The D6-brane action includes a WZ term
\beq
S_{WZ}^{D6}= \frac{1}{2} (2\pi\alpha')^2 N_f T_{D6} \int_{D6} P[C_3]\wedge F\wedge F,
\eeq
where $T_{D6}$ is the D6-brane tension and $F$ is the $U(1)$ field strength on the D6-branes, dual to the $U(1)_V$ current. Given the similarity with the D3/D7 WZ term, we expect similar physics. The WZ term should be associated with the $U(1)_AU(1)_V^2$ global anomaly.\footnote{Other WZ terms are associated with different anomalies, for instance the $U(1)_A SU(N_c)^2$ axial anomaly was derived using a WZ term of the form $\int_{D6} P[C_7]$ \cite{Barbon:2004dq}.} The pullback of $C_3$ will produce a factor of the derivative of $\phi$. In particular if we introduce $\partial_t \phi = \omega$ and $F_{xy} = B$, and if we demand a real on-shell action, then we expect a nontrivial $A_z$ and hence a CME in the dual field theory.

Unlike the Sakai-Sugimoto model, the bulk gauge fields dual to $U(1)_A$ are not associated with the flavor branes (D7- or D6-branes) alone. Instead the bulk $U(1)_A$ gauge field is a component of the metric that transforms as a gauge potential under diffeomorphisms of the form $\phi\to \phi+\xi(x)$. Notice that a dimensional reduction to five dimensions can produce a Chern-Simons term for these gauge fields. Moreover notice that in order to introduce an axial field strength we must deform the geometry, \ie we cannot use flavor brane worldvolume fields alone.\footnote{We cannot use the scalar $\phi$ on the flavor brane worldvolume to introduce an axial field strength: as explained in section~\ref{ss:sols}, the probe brane action will depend only on derivatives of $\phi$, \ie on the exact one-form $d\phi$, which produces only zero field strengths $d^2 \phi=0$.} Bulk solutions with (from our point of view) an external axial magnetic field were constructed in refs.~\cite{D'Hoker:2009mm,D'Hoker:2009bc}.

The shift $\phi\to\phi+a$, with constant $a$, corresponds to an R-symmetry transformation in the dual field theory: the $U(1)_A$ symmetry in the D3/D7 or D4/D6 systems is part of the R-symmetry, so in the field theory not only flavor fields but also adjoint fields are charged. This is a crucial difference with the Sakai-Sugimoto model, and has two major consequences. First, an axial charge in the flavor sector is not necessarily conserved, even in the absence of anomalies, since axial charge can leak into the adjoint sector.\footnote{For the D4/D6 system in the confined phase, the dual field theory spectrum includes glueballs charged under the R-symmetry, so R-charge introduced in the mesonic sector alone need not be conserved.} Second, the adjoint fields can contribute to the axial anomaly. In the 't Hooft and probe limits the anomaly will have contributions of order $N_c^2$ and $N_f N_c$. We have seen that for both the D3/D7 and D4/D6 systems, the $U(1)_A U(1)_V^2$ anomaly survives the probe limit, and is realized in the bulk via the probe flavor brane's WZ terms.

Given the different realizations of chiral symmetry in these systems, chiral symmetry breaking will also be realized in different ways. In the Sakai-Sugimoto model in the low-temperature confining phase the D8- and $\overline{\mbox{D8}}$-brane join, forming a U-shaped curve along the cigar, so that what appeared to be distinct $U(N_f)_L \times U(N_f)_R$ symmetries are broken to the diagonal. That is interpreted in the field theory as spontaneous chiral symmetry breaking.\footnote{A quark mass is possible in principle in the Sakai-Sugimoto model if the D8- and $\overline{\mbox{D8}}$-branes are localized at the same point of the $x_4$ circle, in which case a tachyon field should be part of the low-energy description even close to the boundary. The tachyon action is not known in general, nevertheless this approach has been pursued in models based on the Sakai-Sugimoto model \cite{Bergman:2007pm,Dhar:2007bz,Dhar:2008um} and in an AdS/QCD context \cite{Casero:2007ae,Iatrakis:2010zf,Iatrakis:2010jb}, where a scalar field typically describes chiral symmetry breaking \cite{Erlich:2005qh,Karch:2006pv}.} In the D3/D7 and D4/D6 systems, on the other hand, chiral symmetry breaking means breaking rotational invariance in a transverse plane. We can do that either by separating the flavor branes from the color branes, corresponding to a flavor mass in the field theory and hence an explicit breaking, or by the flavor brane bending, as occurs for example when we introduce the magnetic field $F_{xy} = B$ (recall section~\ref{ss:solszeroT}).

Let us now compare in more technical detail the $U(1)_A U(1)_V^2$ anomalies in both types of systems, with emphasis on the definition of gauge-invariant currents. We will schematically write the relevant parts of the probe brane actions as
\begin{align}\label{eq:actioncomparison1}
S_{D8} &= S_{YM}^{D8}(F^L,F^R)+\kappa\int d^5 x \epsilon^{MNLPQ} R_M F^R_{NL} F^R_{PQ}-\kappa\int d^5 x \epsilon^{MNLPQ} L_M F^L_{NL} F^L_{PQ},\\
\label{eq:actioncomparison2}
S_{D7} &= S_{YM}^{D7}(F^V)+\kappa\int d^5 x\, \Omega(r)\,\epsilon^{MNLPQ}\partial_M\phi F^V_{NL} F^V_{PQ}.
\end{align}
The terms $S_{YM}^{D8}$ and $S_{YM}^{D7}$ can be either the DBI action or a Maxwell's action obtained from an $\alpha'$ expansion to lowest nontrivial order, plus possible couplings to other fields. We actually don't care about the explicit form of these terms. We only care that they are gauge invariant. The factors of $\k$ for each system are straightforward to derive, being products of the brane tension, $N_f$ and numerical factors. We have written $S_{D7}$ in the second line above, although the same form appears for the D6-brane action in the D4/D6 system. We will use a radial coordinate $r$ for both systems, with the holographic boundary at $r \rightarrow \infty$. We have included a generic factor of the radial direction, $\Omega(r)$, in $S_{D7}$, since that may appear for general flavor brane embeddings in the D3/D7 and D4/D6 systems: for the D7-brane case see eq.~\eqref{actionden}.

Crucially, notice that the actions in eqs.~\eqref{eq:actioncomparison1} and \eqref{eq:actioncomparison2} are more general than the probe flavor systems we have been discussing. In fact the actions above are simply representatives of two categories of holographic systems describing the CME: those for which the $U(1)_V \times U(1)_A$ symmetry is realized via gauge fields (not necessarily arising from probe branes) with (4+1)-dimensional Chern-Simons terms and those for which the $U(1)_V$ symmetry is realized via a gauge field while the $U(1)_A$ symmetry is realized via rotation in $\phi$. Most of what follows is relevant for any holographic system in one of these categories, although we will continue to use the language of probe branes.

The equations of motion obtained from these actions are, in the Sakai-Sugimoto model
\begin{align}
&-\partial_N\frac{\delta S_{YM}^{D8}}{\delta \partial_N R_M}+3\kappa \epsilon^{MNLPQ} F^R_{NL} F^R_{PQ}=0,\\
&-\partial_N\frac{\delta S_{YM}^{D8}}{\delta \partial_N L_M}-3\kappa \epsilon^{MNLPQ} F^L_{NL} F^L_{PQ}=0,
\end{align}
and in the D3/D7 or D4/D6 systems,
\beq
-\partial_N\frac{\delta S_{YM}^{D7}}{\delta \partial_N V_M}+4\kappa \epsilon^{MNLPQ}\partial_N \Omega \partial_L\phi  F^V_{PQ}=0.
\eeq

For the sake of argument, we will consider two different definitions of the field theory currents. The first comes from the equations for the radial components of the gauge fields $R_r$, $L_r$ and $V_r$, which take the form of conservation equations for currents in four dimensions,
\begin{align}\label{eq:eomscurr1}
&\partial_\mu\left[-\frac{\delta S_{YM}^{D8}}{\delta \partial_\mu R_r}+6\kappa \epsilon^{r\mu\nu\sigma\rho} R_\nu F^R_{\sigma\rho}\right]=0,\\
\label{eq:eomscurr2}
&\partial_\mu\left[-\frac{\delta S_{YM}^{D8}}{\delta \partial_\mu L_r}-6\kappa \epsilon^{r\mu\nu\sigma\rho} L_{\nu} F^L_{\sigma\rho}\right]=0,
\end{align}
and
\beq\label{eq:d7eomscurr}
\partial_\mu\left[-\frac{\delta S_{YM}^{D7}}{\delta \partial_\mu V_r}+4\kappa \epsilon^{r\mu\nu\sigma\rho} \Omega \partial_\nu\phi  F^V_{\sigma\rho}\right]=0.
\eeq
We will refer to the near-boundary $r \to \infty$ limit of the quantities in brackets above as the ``conserved currents.'' In the $r \to \infty$ limit, the terms with explicit factors of the gauge fields $R_{\nu}$ and $L_{\nu}$ and derivatives in the field theory directions acquire their boundary values, so they become terms depending on the sources.

The second definition of the field theory currents is the standard holographic one, from variation of the on-shell action. We will call these the ``canonical currents.'' For the Sakai-Sugimoto model, varying $R_{\mu}$ and $L_{\mu}$ separately we find
\begin{align}
\delta_R S_{D8} &= \int d^5 x\, \left[-\frac{\delta S_{YM}^{D8}}{\delta \partial_\mu R_r} +4\kappa\epsilon^{r\mu\nu\sigma\rho} R_\nu F^R_{\sigma\rho}\right]\delta R_\mu, \\
\delta_L S_{D8} &= \int d^5 x\, \left[-\frac{\delta S_{YM}^{D8}}{\delta \partial_\mu L_r} -4\kappa\epsilon^{r\mu\nu\sigma\rho} L_\nu F^L_{\sigma\rho}\right]\delta L_\mu.
\end{align}
For the D3/D7 or D4/D6 systems we vary with respect to $V_{\mu}$ to find
\beq
\delta_V S_{D7} = \int d^5 x\, \left[-\frac{\delta S_{YM}^{D7}}{\delta \partial_\mu V_r} +4\kappa\epsilon^{r\mu\nu\sigma\rho} \Omega \partial_\nu\phi F^V_{\sigma\rho} \right]\delta V_\mu, 
\eeq
where we have used that $S_{YM}$ is gauge invariant, so for any of the gauge fields ${\cal A}_M$
\beq
\frac{\delta S_{YM}}{\delta \partial_r {\cal A}_\mu} = -\frac{\delta S_{YM}}{\delta \partial_\mu {\cal A}_r} .
\eeq
We identify the canonical currents as the $r \to \infty$ limit of the quantities in brackets above. Notice that for the D3/D7 and D4/D6 systems the two definitions produce identical currents. This is not true for the Sakai-Sugimoto model.

In the Sakai-Sugimoto model the $U(1)_L$ and $U(1)_R$ conserved currents are not gauge-invariant, while the canonical currents are neither conserved nor gauge-invariant. The same is then true for the $U(1)_V$ and $U(1)_A$ currents, constructed as linear combinations of the $U(1)_L$ and $U(1)_R$ currents. We can add counterterms at the boundary that modify the definitions of the currents, as discussed in ref.~\cite{Rebhan:2009vc}, which allow us to choose the vector current to be either conserved or gauge-invariant, but not both. This is the usual situation in field theory when the currents are anomalous. No matter how we choose to define the currents, the gauge-invariant part has an anomaly fixed by the equations of motion, so the vector current cannot be both conserved and gauge invariant. However, we can still make the vector current conserved and gauge-invariant under $U(1)_V$ transformations alone, \ie we can eliminate the explicit appearance of $V_M$ in the action in eq.~\eqref{eq:SSanomalies} by an integration by parts,
\begin{equation}
S_{D8}=\frac{\kappa}{4}\int d^5 x \epsilon^{MNLPQ} A_M\left( 3 F^V_{NL} F^V_{PQ}+ F^A_{NL} F^A_{PQ}\right).
\end{equation}
The integration by parts produces boundary terms that can be canceled by a counterterm, or, turning things around, we can add counterterms that make the vector current conserved and gauge-invariant under $U(1)_V$. Most important for the CME in the Sakai-Sugimoto model, with appropriate counterterms to make the $U(1)_V$ current conserved (though gauge-variant), the chiral magnetic current vanishes \cite{Rebhan:2009vc}.

In the D3/D7 or D4/D6 systems, in the probe limit we have no explicit axial gauge field and so cannot introduce an external axial field strength. The vector current is then obviously gauge-invariant and conserved: the conserved current defined in eq.~\eqref{eq:d7eomscurr} depends only on the field strength $F_{\mu\nu}^V$. No special counterterms are required.

The realization of the axial anomaly is somewhat more subtle than in the Sakai-Sugimoto model. The $U(1)_A$ symmetry is actually an R-symmetry, whose associated charge corresponds to angular momentum in $\phi$. For our probe flavor branes that is intimately related to the canonical momentum associated with the worldvolume scalar $\phi$, which we denote $\pi_\phi^M$,
\beq\label{eq:phicanonicalmomentum}
\pi_\phi^M\equiv\frac{\delta S_{D7}}{\delta (\partial_M \phi)}, \ \ \partial_M \pi_\phi^M=0.
\eeq
The second equation is just the $\phi$ equation of motion. The probe flavor contribution to the expectation value of the R-symmetry current in the field theory directions, $J_R^{\mu}$, is the integral of $\pi_\phi^{\mu}$ over the worldvolume of the brane.\footnote{Here and in section \ref{ss:equilcme} we ignore any potential divergences that may appear at the $r \to \infty$ endpoint of the integral, which can be cancelled with counterterms that do not affect our main results.}
\beq
\langle J_R^\mu \rangle=\int dr \,\pi_\phi^\mu.
\eeq
Again, in the probe limit we have no external axial field strengths to contribute to an axial anomaly. The vector field strengths still contribute, however. If we distinguish two contributions to the angular momentum current from the YM action and the other from the WZ term, then the $U(1)_AU(1)_V^2$ anomaly comes from the WZ contribution,
\beq
\left . \langle J_R^\mu \rangle \right|_{WZ}=-4\kappa\int dr\epsilon^{r \mu\nu\sigma\rho}\Omega(r)\,F^V_{r\nu} F^V_{\sigma\rho}.
\eeq
Upon integrating by parts, we obtain
\beq
\left . \langle J_R^\mu \rangle \right |_{WZ}=\left.-4\kappa\epsilon^{r \mu\nu\sigma\rho}\Omega(r)\,V_{\nu} F^V_{\sigma\rho}\right|_{r \to \infty}+4\kappa\int dr\epsilon^{r \mu\nu\sigma\rho}\partial_r\Omega(r) \,V_\nu F^V_{\sigma\rho}.
\eeq
Where we have dropped a term proportional to $\partial_\nu V_r$ because it is a total derivative in the field theory directions. The second term on the right-hand-side is invariant under $r$-independent gauge transformations $\delta V_\nu=\partial_\nu f(x_{\mu})$ since these also produce terms that are total derivatives in the field theory directions. Notice, however, the first term is not invariant under gauge transformations that are non-vanishing at the boundary. Adding the gauge-invariant YM contribution, we obtain the total probe flavor axial/R-symmetry current. The total current is not gauge invariant, due to the WZ contribution. Equivalently, we can define a gauge-invariant current that is not conserved, hence the axial/R-symmetry is anomalous.

\subsection{Thermal Equilibrium and the Holographic CME}\label{ss:equilcme}

In the Sakai-Sugimoto model, with appropriate counterterms the $U(1)_V$ current can be made gauge-invariant and conserved, but the chiral magnetic current is then zero \cite{Rebhan:2009vc}. For any holographic model of the CME exploiting (4+1)-dimensional Chern-Simons terms, a nonzero chiral magnetic current can be achieved if the axial chemical potential is carefully defined, as argued in ref.~\cite{Gynther:2010ed}. In a bulk geometry with a horizon at some position $r_H$, the gauge-invariant holographic definition of the field theory chemical potential is
\beq
\mu_5=\int_{r_H}^\infty dr F_{rt}^A = A_t(\infty)-A_t(r_H),
\eeq
where in the second equality we chose $A_r=0$ gauge. If $A_t(r_H)=0$ then we may identify $\mu_5=A_t(\infty)$. If $A_t(r_H)\neq0$, then $A_t(\infty)$ is an external source, which is physically distinct from the chemical potential. Indeed, part of the argument of ref.~\cite{Gynther:2010ed} was that a nonzero $A_t(r_H)$ is necessary to obtain a nonzero chiral magnetic current.

A serious issue arises with a nonzero $A_t(r_H)$, however. If the system is in thermal equilibrium, then in the field theory we may perform an analytic continuation to Euclidean signature, where time becomes a compact coordinate with periodic boundary conditions for bosons and anti-periodic boundary conditions for fermions. On the gravity side we can analytically continue the metric to a Euclidean black hole, where the time coordinate is also compact and smoothly shrinks to zero size at $r_H$. The issue is that solutions with nonzero $A_t(r_H)$ are not regular in the Euclidean geometry: put briefly, the Killing vector $\frac{d}{dt}$ vanishes at $r_H$, so the gauge field, which is a one-form, is regular only if $A_t(r_H)=0$ (see for example ref.~\cite{Kobayashi:2006sb}). We may thus question the starting assumption, that the system is in thermal equilibrium. More generally, we may question whether gauge-gravity duality can describe a CME in a thermal equilibrium state at all.

In fact, non-regular gauge field solutions appeared already in holographic studies of the hydrodynamics of charged fluids, including the effects of anomalies~\cite{Erdmenger:2008rm,Banerjee:2008th}.\footnote{We thank Andreas Karch for explaining the following to us.} In these cases the bulk solutions represent small perturbations about static black hole solutions, including a perturbation of a bulk gauge field. The boundary conditions on the gauge fields are regular only on the future horizon: strictly speaking, the solutions do not represent a system in equilibrium, then perturbed, then relaxing back to equilibrium, but rather solutions out of equilibrium in the infinite past that eventually settle into equilibrium in the infinite future. The key point is that non-regular boundary conditions on bulk gauge fields are a generic signal of out-of-equilibrium physics. 

Even when $A_t(r_H)=0$, some holographic descriptions of the CME require non-regular solutions, if a time-dependent scalar is involved. Consider for example the realization of the CME in the soft-wall AdS/QCD model \cite{Gorsky:2010xu,Brits:2010pw}, which is essentially AdS space with a certain profile for a dilaton-like scalar, representing the running of the field theory gauge coupling \cite{Karch:2006pv}. That model involves a (4+1)-dimensional Chern-Simons action, but also includes a term with the derivative of a scalar field, dual to a pseudo-scalar meson, analogous to the field $\phi$ on the flavor branes in the D3/D7 or D4/D6 systems (hence we use the same symbol, $\phi$),
\beq
S_{CS}^{AdS/QCD}= \int d^5 x\,(A+d\phi)\wedge F^V \wedge F^V,
\eeq
where $A$ is dual to the $U(1)_A$ current. As in the D3/D7 system, the CME appears when the time derivative of the boundary value of the scalar field is nonzero, $\partial_t\phi(\infty)\neq0$. Through a gauge transformation we can set $A_r=-\partial_r\phi$, in which case the axial chemical potential becomes
\beq
\mu_5=\int_{r_H}^\infty dr\, F_{rt} = A_t(\infty)-A_t(r_H)+\partial_t\phi(\infty)-\partial_t\phi(r_H).
\eeq
We can now set $A_t(r_H)=0$ and still obtain a nonzero chiral magnetic current, although regular Euclidean solutions should also have $\partial_t\phi(r_H)=0$ since $\partial_M \phi$ is a one-form.

The upshot is that all of the holographic models of the CME that we know involve non-equilibrium physics. For the D3/D7 and D4/D6 systems we can also easily see that the CME occurs only out of equilibrium. The simplest observation is that the worldvolume scalar $\phi$ depends on time explicitly: $\partial_t \phi = \o$. The equivalent field theory statement is that whenever $|m|$ is nonzero the action explicitly depends on time, as mentioned in section~\ref{ss:fieldtheory}.

More subtle non-equilibrium physics also occurs in the D3/D7 realization of the CME. Axial current conservation can be violated in three ways. The first way is explicitly via a nonzero $|m|$. The second way is due to anomalies. The third way is due to the fact that the axial symmetry is part of the R-symmetry, so that an axial charge density introduced in the flavor sector alone can leak into the adjoint sector.

From the field theory point of view, axial charge can be lost to the adjoint sector in two ways, depending on whether mesons are melted or not. The equivalent bulk statement is that D7-brane angular momentum can be lost in two ways, depending on whether the D7-brane has a worldvolume horizon or not.

If mesons are not melted, then the loss of R-charge occurs when mesons radiate R-charged glueballs. In the 't Hooft limit the interactions of color singlet mesons and glueballs are suppressed, so we expect the loss of R-charge to the adjoint sector not to be apparent in that limit. The dual statement is that a D7-brane with angular momentum but no worldvolume horizon can only lose angular momentum via radiation of closed strings, but the relevant interaction is proportional to the string coupling $g_s\sim 1/N_c$ and so is suppressed in the classical supergravity limit, where $g_s \to 0$.

If mesons are melted, then the spectrum in the flavor sector is no longer just delta-functions representing color-singlet mesons, but a continuum of modes whose interactions with the adjoint fields experience no large-$N_c$ suppression. The dual statement is that a D7-brane with angular momentum and a worldvolume horizon (whether a Minkowski or black hole embedding) can transfer angular momentum across the horizon even in the supergravity limit.

We will now confirm explicitly the above expectations by computing the time rate of change of axial/R-charge, both directly in the field theory and also from the D7-brane description, following refs.~\cite{Karch:2008uy,Das:2010yw}, with exact agreement between the two. Notice that when axial/R-charge leaks into the adjoint sector it should take energy with it, so we will also compute the time rate of change of the energy density, or more precisely the expectation value of the ``tt'' component of the stress-energy tensor (density), $\TT$.

We begin with the field theory calculations. For the rate of change of R-charge we just need to compute a Ward identity. Since an R-symmetry transformation is equivalent to a shift in the source, \ie a shift in the phase of the flavor mass $\phi\to\phi+\delta\phi$, we have, using the definition of $\Op$ in terms of a variation of minus the field theory action with respect to the phase of the flavor mass,\footnote{The variation of the path integral is $\int d^4 x \left\langle -\frac{1}{2}\partial_\mu\delta\phi \overline{\psi}\gamma^\mu \gamma^5 \psi-{\cal O}_\phi \delta\phi\right\rangle$. Integrating by parts and demanding that the variation vanish for any $\delta\phi$, we find the relation in eq.~\eqref{eq:wardcurr}.}
\beq\label{eq:wardcurr}
\partial_\mu \langle J_R^\mu\rangle =  \langle {\cal O}_\phi\rangle.
\eeq
In our case nothing is changing in space, so we obtain for the time rate of change of the R-charge density $\partial_t \langle J_R^t \rangle =  \Opv$. Recall that $\Op \propto |m|$, so a necessary condition for the rate of change to be nonzero is for $|m|$ to be nonzero. In that case the potential and therefore the Hamiltonian are time-dependent, so the energy density will also not be conserved. Recalling the definitions of the potential terms $V_q$ and $V_{\psi}$ from section~\ref{ss:fieldtheory}, the change in energy density is
\beq\label{eq:wardenergy}
\partial_t \TT  =  \partial_t \langle V_q+V_\psi \rangle= \Opv \partial_t \phi, 
\eeq
where the second equality follows from the chain rule. In our case $\partial_t \phi = \o$, and we just saw that $\Opv =  \partial_t \langle J^t_R \rangle$, so for our system
\beq
\partial_t \TT= \o \, \partial_t \langle J^t_R \rangle.
\eeq
The two rates of change are directly proportional. As expected, when R-charge leaks into the adjoint sector, it takes energy with it.

We now turn to the bulk calculation of the same rates of change. The R-charge density $\langle J_R^t \rangle$ is given by the angular momentum of the D7-brane,
\beq
\label{eq:totalRcharge}
\langle J_R^t \rangle= \int_{r_H}^{\infty} dr\, \pi_\phi^t = \int_{r_H}^{\infty} dr\, \frac{\delta S_{D7}}{\delta \partial_t \phi}.
\eeq
For concreteness we have written the lower endpoint of the $r$ integration as $r_H$, as appropriate for a black hole embedding. For a Minkowski embedding the lower endpoint is $r=0$. Taking $\partial_t$ of the charge density and using the $\phi$ equation of motion in eq.~\eqref{eq:phicanonicalmomentum}, we find
\beq
\partial_t \langle J^t_R \rangle = \int_{r_H}^{\infty} dr \, \partial_t \pi_\phi^t = -\int_{r_H}^{\infty} dr \, \partial_r \pi_{\phi}^r = -\left . \pi_{\phi}^r \right |^{\infty}_{r_H},
\eeq
where in the second equality we assumed homogeneity, so the derivative of $\pi_{\phi}^M$ in any field theory spatial direction vanishes. Recalling from eq.~\eqref{eq:pirho} that $\pi_{\phi}^r = \a$, which in our system is independent of $r$, the R-charge density appears to be constant in time, $\partial_t \langle J_R^t \rangle = 0$. That is indeed true since the states we study are stationary. The reason why is nontrivial, however: the two terms in the final equality above cancel one another. The contribution from the lower endpoint represents the angular momentum that the D7-brane is losing, or equivalently the R-charge that the flavors are dissipating into the adjoint sector, while the contribution from the upper endpoint represents angular momentum that we are pumping into the system by hand via a boundary condition on the D7-brane, since we are forcing the D7-brane to rotate at the boundary, or equivalently in the field theory R-charge that we are pumping into the system from an external source.\footnote{From a field theory point of view, the external source does not have a clear, intuitive interpretation, in contrast to, say, dragging strings  \cite{Gubser:2006bz,Herzog:2006gh,CasalderreySolana:2006rq}. In those cases a stationary string solution represents a heavy quark being pushed through a plasma by a simple external source, a $U(1)_V$ electric field \cite{Gubser:2006bz,Herzog:2006gh,CasalderreySolana:2006rq}. In our case we simply modify the potential by hand to produce the appropriate source.} Implicitly in our solutions we choose the latter precisely to cancel the former. The upshot is that the loss rate is the contribution from the lower endpoint,
\beq
\label{eq:bulklossrateresult}
\left . \partial_t \langle J^t_R \rangle \right |_{\rm loss} = \a = \Opv,
\eeq
in perfect agreement with the field theory Ward identity. Only D7-branes with a worldvolume horizon have nonzero $\a$, hence in the field theory only states with melted mesons have a nonzero rate of change for R-charge (in the 't Hooft limit), in conformity with our field theory intuition.

To compute the rate of change of the energy density, we need to compute the stress-energy tensor of the D7-brane. As explained in ref.~\cite{Karch:2008uy}, we can do that in two equivalent ways. The first way is directly, by variation of the D7-brane action with respect to the background metric. The second way is via a Noether procedure. Although we have used both methods, we will only present the latter, which is more efficient. Defining a Lagrangian via $S_{D7} = \int dr \lagr$, the stress-energy tensor \textit{density} of the D7-brane, in the $AdS_5$ directions, is
\beq
\Theta^M_{~N} = \lagr \, \delta^M_{~N} + 2 F_{LN} \frac{\delta \lagr}{\delta F_{ML}} - \partial_N \theta \frac{\delta \lagr}{\delta \partial_M \theta} - \partial_N \p \frac{\delta \lagr}{\delta \partial_M \p}.
\eeq
The D7-brane stress-energy tensor is then the integral of $\Theta^M_{~N}$ over $r$. For values of $M$ and $N$ in field theory directions, we may equate the D7-brane stress-energy tensor's components with the flavor fields' contribution to the expectation value of the field theory stress-energy tensor \cite{Karch:2008uy},
\beq
\langle T^{\mu}_{~\nu} \rangle = \int dr \, \Theta^{\mu}_{~\nu}.
\eeq
To compute the rate of change of energy density we will only need one component of the D7-brane stress-energy tensor density, $\Theta^r_{~t} = - \o \frac{\partial \lagr}{\partial \phi'} = -\o \a$. We proceed in a similar manner to the calculation of the R-charge rate of change. Using conservation of the D7-brane stress-energy tensor density, $\partial_M \Theta^M_{~N} = 0$, we find
\beq
\langle \partial_t T_{tt}\rangle=-\langle\partial_t T^t_{\ t}\rangle =  -\int^{\infty}_{r_H} dr \, \partial_t \Theta^t_{~t} =  \int^{\infty}_{r_H} dr \, \partial_r \Theta^r_{~t} =  \left . \Theta^r_{~t} \right|^{\infty}_{r_H}=\left.-\o \a\right|^\infty_{r_H}.
\eeq
The total energy is conserved, but the flux of energy at the boundary and the horizon is nonzero. The flux at the horizon corresponds to the rate of energy dissipation,
\begin{equation}\label{energyrate}
\left . \partial_t \TT \right |_{\rm loss} = \o\a=\omega\Opv,
\end{equation}
where $\a$ is given by \eqref{eq:alphaT} and is negative. More generally, the components $\Theta^r_{\ \mu}$ represent  ``external forces'' acting on the probe \cite{Gubser:2008vz}. The holographic calculation reproduces the relation between energy and charge loss rates,\footnote{For black hole embeddings the loss rates can also be extracted from (suitably regulated) divergences in the angular momentum $\frac{\delta S_{D7}}{\delta \o}$ and in the D7-brane stress-energy tensor at the AdS-Schwarzschild horizon, as explained in ref.~\cite{Karch:2008uy}.}
\beq
\left.\partial_t \TT \right|_{\rm loss}  = \left.\omega \partial_t \langle J_R^t\rangle\right|_{\rm loss} .
\eeq

The dissipation rate is only of order $\alpha\sim \lambda N_f N_c$. That means that the flavor sector will only transfer an order $N_c^2$ amount of charge into the adjoint sector over a time of the order $N_c/\lambda \propto 1/g_{YM}^2\gg 1$. For times parametrically shorter than $N_c/\lambda$, we can ignore the dissipation rate and treat the background as a reservoir, in which case the stationary solution in the probe limit is a reliable approximation to the actual solution. This is similar to what occurs with constant electric fields on the D7-brane, where both energy and momentum (but not angular momentum) are dissipated in the bulk \cite{Karch:2008uy}.

Recall that the above analysis is valid for Minkowski embedddings with worldvolume horizon  (or for black hole embeddings in the zero-temperature limit) simply by taking $r_H \to 0$. In section \ref{ss:sols} we saw that such D7-branes have a conical singularity at $r=0$. We can understand this singularity as a consequence of the angular momentum and energy flux along the brane. When the angular momentum and energy flowing along the brane reach the ``bottom'' at $r=0$, they must be dumped into some source, or really a sink. For black hole embeddings that source is hidden behind the AdS-Schwarzschild horizon, and the part of the D7-brane outside of that horizon is non-singular. In the absence of the AdS-Schwarzschild horizon, and neglecting the backreaction of the D7-brane, the source is manifested as the ``naked'' conical singularity of the embedding. Something very similar occurs for the holographic dual of $\N=4$ SYM formulated on a spatial $S^3$ with R-charge chemical potentials. Bulk solutions, known as ``superstars'' (dual to zero-temperature BPS states), exhibit naked singularities that have a sensible physical interpretation in terms of charged sources, namely giant gravitons \cite{Myers:2001aq}.

\subsection{Other Possible Holographic Models of the CME}

Finally, we will comment briefly on other possible holographic realizations of the CME, using probe flavor branes. The main objective is to devise systems where the probe brane WZ terms produce a coupling of the form  $d \phi \wedge F \wedge F$, although not necessarily from rotation in a transverse plane. We emphasize that these models are speculative, \ie we have not actually constructed supergravity solutions for them.

Consider D4-branes compactified on a spatial circle, as described in section~\ref{ss:anomalydefinitions}. Instead of rotating probe branes, we can consider a black hole with angular momentum in the internal space, or different fluxes in the background geometry. For instance, in the high-temperature black hole geometry we can introduce a background RR two-form $F_2=d C_1$, where $C_1 = h(r) dt$ and $h(r)$ approaches a constant, $h_0$, as $r \to \infty$. The meaning of this $C_1$ in the field theory dual can be understood by introducing a probe D4-brane parallel to the color D4-branes. When the probe D4-brane is taken to infinity, the coupling to the $C_1$ form becomes
\beq
\int_{D4} P[C_1]\wedge F \wedge F \to h_0  \int_{xyz} A\wedge F,
\eeq
where on the right-hand-side we have integrated over $r$. We can regard this as a Chern-Simons term in the effective three-dimensional theory that lives on the compactified D4-branes at high temperatures. We can thus associate $C_1$ with a topological charge. Probe D6-branes couple to the $C_1$ flux via a WZ term of the form $\int P[C_1]\wedge F \wedge F \wedge F$. The D6-branes wrap an $S^2\subset S^4$, so we can see that a CME may be possible in the presence of magnetic flux\footnote{A flux on the $S^2$ implies that the D6-branes are a magnetic source of RR four-form flux, which can be interpreted as a bound state of D6-branes with D4-branes in the directions transverse to the sphere. Similar statements apply for the D8-, D7- and D9-branes we discuss next: flux on the internal space represents a bound state with D4- or D3-branes, respectively.} $\int_{S^2} F \neq 0$. Similarly, probe D8-branes will have a WZ coupling of the form $\int P[C_1] \wedge F \wedge F \wedge F \wedge F$, and hence may exhibit a CME if the worldvolume instanton number on the $S^4$ is nonzero, $\int_{S^4} F\wedge F \neq 0$. In the chirally-symmetric phase, the D8- and $\overline{\mbox{D8}}$-branes are disconnected and cross the black hole horizon at different points \cite{Aharony:2006da}. In that case the instanton number on the D8-branes ($n_R$) and $\overline{\mbox{D8}}$-branes ($n_L$) can be different. The chiral magnetic currents for $U(1)_V$ and $U(1)_A$ will then be proportional to (with both parallel to the magnetic field),
\beq
J_V \propto  (n_R-n_L) B, \qquad J_A \propto (n_R+n_L) B.
\eeq

The type IIB versions of the D4/D6 and D4/D8 intersections with nontrivial $C_1$ are D3/D7 and D3/D9 intersections in the presence of a time-dependent axion $C_0$. The relevant WZ term for the D7-brane is $\int P[dC_0] \wedge A \wedge F \wedge F \wedge F$. A CME may be possible if we introduce non-zero flux on the internal three-sphere $\int_{S^3} A\wedge F \neq 0$. The relevant D9-brane WZ coupling is $\int P[dC_0] \wedge A \wedge F \wedge F \wedge F \wedge F$, and a CME may be possible with nonzero flux on the five-sphere $\int_{S^5} A\wedge F \wedge F\neq 0$. Notice that in these cases the boundary terms have been chosen in such a way that the probe brane actions are not gauge-invariant, so whether the $U(1)_V$ current is anomalous would need to be checked.

\section{Summary and Discussion}\label{ss:discuss}

We used AdS/CFT to study the CME in large-$N_c$, strongly-coupled $\N=4$ SYM theory coupled to a number $N_f \ll N_c$ of $\N=2$ supersymmetric flavor hypermultiplets. We introduced a time-dependent phase for the hypermultiplet mass, which for the hypermultiplet fermions is equivalent to an axial chemical potential, and we introduced an external, non-dynamical $U(1)_V$ magnetic field. When the magnitude of the hypermultiplet mass $|m|$ was zero, we found at both zero and finite temperature that the chiral magnetic current $\jz$ coincided with the weak-coupling result in the chirally-symmetric limit, eq.~\eqref{CMEcurrent}. When $|m|$ was nonzero we found that $\jz$ had a smaller value than eq.~\eqref{CMEcurrent}, and also that the $U(1)_V$-invariant and CT-odd pseudo-scalar operator $\Op$ acquired a nonzero expectation value. Indeed, for sufficiently large $|m|$ or $B$, compared to the axial chemical potential or the temperature, both $\jz$ and $\Opv$ dropped to zero (recall fig.~\ref{Fig2}). In these cases we interpret the appearance of a chiral magnetic current as the conversion of the pseudo-scalar condensate to a vector condensate via the magnetic field.

We compared several holographic models of the CME, and in particular highlighted the fact that all of them describe a CME only in non-equilibrium states (in contrast to lattice QCD analyses \cite{Buividovich:2009zzb,Buividovich:2009zj,Abramczyk:2009gb,Yamamoto:2011gk}). Our holographic system also describes the CME only in non-equilibrium states. Whenever $|m|$ is nonzero, the scalars in the hypermultiplet have masses with time-dependent phases, hence the Hamiltonian has explicit time dependence and energy is not conserved. 
Moreover, in our system the axial symmetry is part of the R-symmetry, so axial charge in the flavor sector can leak into the adjoint sector, also taking energy with it. We computed the associated loss rates, which we found to be proportional to $\Opv$. When the CME occurs in our system at nonzero $|m|$, the flavor fields are losing axial charge and energy to the adjoint sector. In the probe limit these loss rates are negligible, however, and our states describing a CME were in fact stationary.\footnote{The loss rates are of order $\Opv \propto \lambda N_c$, and so can be neglected for times shorter than $N_c/\lambda$. Taking into account the change in angular momentum and energy, \ie computing the back-reaction of the D7-branes, would probably lead to an expanding horizon \cite{Das:2010yw}. Gravity solutions exhibiting precisely that behavior have been constructed for external electric fields in ref.~\cite{Sahoo:2010sp}.} 

The supergravity description of the above was a number $N_f$ of probe D7-branes extended along $AdS_5 \times S^3$ inside $AdS_5 \times S^5$, rotating on the $S^5$ and with worldvolume gauge fields that encode the magnetic field and chiral magnetic current. AdS space is effectively a gravitational potential well in which the local speed of light decreases as we move away from the boundary. A D7-brane rotating sufficiently quickly may at some point be rotating faster than the local speed of light and hence may develop a worldvolume horizon. We saw that indeed D7-brane solutions thus split into two categories, those that rotate quickly enough to develop a worldvolume horizon and those that don't. For the former the D7-brane becomes imaginary, signaling the presence of a tachyon, unless we introduce certain worldvolume fields and adjust their integration constants to maintain reality of the action. Via the holographic dictionary these integration constants were precisely the values of $\jz$ and $\Opv$. In the bulk the loss of axial charge and energy appear as the flow of angular momentum and energy across the worldvolume horizon. We found numerically that the angular momentum and energy flux produces a conical singularity in Minkowski embeddings with worldvolume horizon at the point where the $S^3$ collapses to zero size.

Although we focused on D7-branes, the CME can be realized in many similar flavor brane systems. The basic ingredients are a holographic spacetime with probe flavor D-branes satisfying two conditions: they describe (3+1)-dimensional flavor fields, and they have at least two transverse directions in which to rotate. An axial chemical potential, implemented as a time-dependent fermion mass, will be realized via rotation in a transverse plane, and the axial anomaly will be realized via a WZ coupling to RR flux in the internal space. A model relevant for applications to QCD is that of ref.~\cite{Kruczenski:2003uq}, with flavor D6-branes in the near-horizon geometry of D4-branes.

From a phenomenological point of view, models of the D3/D7 or D4/D6 type have advantages and disadvantages when compared to the Sakai-Sugimoto or AdS/QCD models. Consider first the disadvantages. D3/D7-type models are typically less similar to large-$N_c$ QCD, for example, non-Abelian chiral symmetries will generically be explicitly broken by (super)potential terms. Given that the general objective of holography is to uncover universal physics, this disadvantage may not be fatal. Beyond that we suspect that the problems with D7-branes will be generic: $U(1)_A$ charge may leak into the adjoint sector, Minkowski embeddings describing a CME will likely be singular, and many solutions describing the CME may in fact be perturbatively unstable. On the other hand, the fact that Sakai-Sugimoto or AdS/QCD models require a source at the horizon to describe the CME suggests that the same (or similar) problems may appear in those models as well, once the meaning and effects of the source are clarified. Perhaps the principal advantage of D3/D7-type models is that the $U(1)_V$ current is conserved and gauge invariant under $U(1)_V$ transformations by construction, so among other things comparison with weak coupling calculations is more straightforward. Another advantage is that quark masses are easy to introduce and the effects of chiral symmetry breaking are easy to study.

Although our focus was on the CME, our spinning D7-branes without worldvolume horizon may have useful applications as well. When the solution describes massless flavors, the field theory is in a state with a finite charge density of fermions with $U(1)_A$ spontaneously broken, \ie superfluid states \cite{O'Bannon:2008bz}. These solutions have no loss of axial charge or energy and no obvious instabilities, at least to leading order in the $1/N_c$ expansion, and so deserve further study as models for strongly-coupled, many-body fermion physics.\footnote{These solutions may in fact also describe a CME, beyond leading order in the large-$N_c$ approximation. In the presence of a black hole, Hawking radiation can induce thermal excitations on the probe brane which are dual to a gas of mesons in the field theory. The magnetic field can polarize pseudo-scalar mesons, producing an excess of vector mesons through the anomaly, hence a CME can occur. From the perspective of the gravitational theory, this is a quantum process, and as such is not captured by our classical calculation. From the perspective of the field theory, the meson gas produces only a $O(1/N_c^2)$ contribution to the total free energy of the system.}

An important task for the future is a complete analysis of linearized fluctuations of worldvolume fields, to determine whether our solutions are stable. In particular, in our analysis we assumed homogeneity of the ground state, but in QCD with $U(1)_A$ or $U(1)_V$ chemical potentials and a strong magnetic field, in a state with chiral symmetry broken, the ground state may be the inhomogeneous ``chiral magnetic spiral'' \cite{Basar:2010zd}. Such a phase was indeed detected in the Sakai-Sugimoto model via analysis of linearized fluctuations \cite{Kim:2010pu}, and a similar analysis should be done for the D3/D7 model.

\section*{Acknowledgments}

We thank O.~Bergman, J.~Charbonneau, S.~Das, M.~Kaminski, A.~Karch, K.-Y.~Kim, P.~Kumar, K.~Landsteiner, A.~Rebhan, S.~Ryu, D.T.~Son, S.~Sugimoto, Y.~Tachikawa, T.~Takayanagi, L.~Yaffe, N.~Yamamoto, A.~Yarom, A.~Zayakin, and A.~Zhitnitsky for useful conversations. We also thank the Galileo Galilei Institute for Theoretical Physics for hospitality and the INFN for partial support during the completion of this work. C.H. also wants to thank the Erwin Schr\"{o}dinger Institute in Vienna and T.N. would like to thank the IPMU in Tokyo for hospitality during the completion of this work. The work of T.N. was supported in part by the US NSF under Grants No.\,PHY-0844827 and PHY-0756966. The work of A.O'B. was supported by the European Research Council grant ``Properties and Applications of the Gauge/Gravity Correspondence.'' This work was supported in part by DOE grant DE-FG02-96ER40956.

\appendix

\section{Holographic Renormalization}\label{holorg}

In this appendix we compute holographically the one-point functions of the operators $\Om$, $\Op$ and $J^z$ defined in section~\ref{ss:fieldtheory}.

The AdS/CFT dictionary equates minus the on-shell bulk action with the CFT generating functional. The on-shell action is generically divergent, however, due to the integration over the infinite volume of AdS space. In the field theory these are UV divergences which require renormalization: we introduce a regulator, add counterterms, take variational derivatives of the generating functional to obtain regulated correlators, and then remove the regulator to obtain finite, renormalized correlators. The analogous procedure in AdS/CFT is called holographic renormalization (see for example refs.~\cite{deHaro:2000xn,Skenderis:2002wp}). Holographic renormalization for probe D-brane worldvolume fields has been studied in various places. We will follow refs.~\cite{Karch:2005ms,Karch:2006bv,O'Bannon:2008bz}.

We begin by writing the metric of $AdS_5 \times S^5$ (in $L\equiv 1$ units) as
\beq
\label{eq:feffermangrahamcoords}
ds^2 = \frac{1}{u^2} \left( du^2 - dt^2 + d\vec{x}^2 \right) + d\theta^2 + \sin^2 \theta d\p^2 + \cos^2 \theta ds^2_{S^3},
\eeq
where $u$ is the radial coordinate with the boundary at $u=0$. These coordinates are related to those of eq.~\eqref{eq:metric} as
\beq\label{eq:coordtranslate}
u^2 = \frac{1}{\rho^2} = \frac{1}{r^2 + R^2}, \qquad \theta = \arctan \frac{R}{r}.
\eeq
The ansatz for the worldvolume fields described in section~\ref{ss:sols} is now
\beq
\theta(u), \, A_z(u), \, A_y(x) = B x, \phi(t,u) = \omega \, t + \varphi(u).
\eeq
The metric and ansatz are identical in form to those of section \ref{ss:sols}, so formally we may use all of the equations of that section, with the replacements $r \rightarrow u$ and $R(r) \rightarrow \theta(u)$.
In particular, we may derive $\theta(u)$'s equation of motion from eq.~\eqref{eq:shathat}, with $r\rightarrow u$ and $R(r)\rightarrow \theta(u)$,
\bea
\hat{\hat{S}}_{D7} & = & -\N \int du \sqrt{g_{uu} + g_{\th\th} \th'^2} \nn \\
& \times & \sqrt{g_{xx}^3 g_{SS}^3 \left(|g_{tt}|-g_{\p\p} \dot{\p}^2\right)\left(1 + \frac{B^2}{g_{xx}^2}- \frac{\a^2}{\N^2 |g_{tt}|g_{\p\p} g_{xx}^3g_{SS}^3}\right)- g_{xx} \left( \frac{\b}{\N} + B \o g_{SS}^2\right)^2}. \nn
\eea
From the equation of motion we find $\theta(u)$'s asymptotic expansion,
\beq\label{eq:thetaexpansion}
\theta(u) = c_0 \, u + \tilde{c}_2 \, u^3 - \frac{c_0 \o^2}{2} u^3 \log u + O\left(u^5 \log u\right).
\eeq
A straightforward exercise using eq.~\eqref{eq:coordtranslate} shows that the coefficient of the leading term, $c_0$, is identical to that of $R(r)$'s expansion in eq.~\eqref{eq:Rasymptoticexpansion}, hence we use the same symbol. The coefficients of the sub-leading terms in eqs.~\eqref{eq:Rasymptoticexpansion} and \eqref{eq:thetaexpansion} are related as $\tilde{c}_2= c_2 + \frac{1}{6}c_0^3$. For the rest of this appendix we use only $c_0$ and $\tilde{c}_2$ unless stated otherwise.

From eqs.~\eqref{phisol} and \eqref{azsol}, with $r\rightarrow u$ and $R(r) \rightarrow \theta(u)$, we obtain the asymptotic expansions of the other fields
\bea
A_z(u) & = & c_z + \frac{1}{2} \left( \frac{\b}{\N} + B \o\right) u^2 + \frac{1}{2} c_0^2 \, \frac{\b}{\N} \, u^4 + O(u^6), \nonumber \\
\phi(t,u) & = & \omega t + \frac{\a}{\N} \left( \frac{1}{2c_0^2} u^2 + \left( \frac{7}{12} - \frac{1}{2}\frac{\tilde{c}_2}{c_0^3}-\frac{1}{16}\frac{\o^2}{c_0^2}\right) u^4 +\frac{1}{4}\frac{\o^2}{c_0^2} u^4 \log u \right) + O\left(u^5 \log u\right). \nonumber
\eea

We now plug $\theta(u)$, $\phi(t,u)$ and $A_z(u)$ into $S_{D7}$ to find the divergences of the on-shell action. To regulate the divergences we integrate only to $u=\epsilon$, producing the regulated action
\bea
S_{D7}^{reg} & = & -\N \int_{_\epsilon} du \left[ \frac{1}{u^5} - \frac{c_0^2}{u^3} + \frac{B^2 - 2 \o^2 c_0^2}{2 u} + \ldots \right] \\
\label{eq:regaction}
& = & +\N \left [ -\frac{1}{4} \frac{1}{\epsilon^4} + \frac{c_0^2}{2} \frac{1}{\e^2}  - \o^2 c_0^2 \log \e + \frac{1}{2} B^2 \log \e + O\left(\e^2 \log \e\right) \right] + \ldots,
\eea
where we have presented only the terms that diverge as $\epsilon \rightarrow 0$, with the $\ldots$ representing all other terms.

We next add counterterms at the $u=\epsilon$ surface to remove the $\epsilon \rightarrow 0$ divergences. The counterterms we need are those written in refs.~\cite{Karch:2005ms,Karch:2006bv} for the field $\theta(u)$, but modified to account for $\phi(t,u)$ as follows. Roughly speaking, $\theta(u)$ is dual to the \textit{magnitude} of the mass operator, while $\phi$ is dual to its \textit{phase}. The precise operators are written in section~\ref{ss:fieldtheory}. When both scalars are active, the natural thing to do is thus to package these two fields into a single \textit{complex} scalar field $\Theta(t,u) = \theta(u) e^{i \phi(t,u)}$, and rewrite the counterterms of refs.~\cite{Karch:2005ms,Karch:2006bv} in terms of $\Theta(t,u)$:
\beq
L_1 = + \N \, \frac{1}{4} \, \sqrt{-\g}, \qquad L_2 = - \N \, \frac{1}{2} \, \sqrt{-\g} \, |\Theta|^2, \qquad L_3 = + \N \, \frac{5}{12} \, \sqrt{-\g} \, |\Theta|^4,
\eeq
\beq
L_4 = + \N \, \frac{1}{2} \, \sqrt{-\g} \, \log |\Theta| \, \Theta^* \Box_{\g} \Theta, \qquad L_5 = + \N \, \frac{1}{4} \, \sqrt{-\g} \, \Theta^* \Box_{\g} \Theta,
\eeq
\beq
L_6 = -\N \, \frac{1}{4} \, \sqrt{-\g} \, \log \e \, F^{\mu\nu} F_{\mu\nu},
\eeq
where $\g_{\mu\nu} = \e^{-2} \eta_{\mu\nu}$ is the induced metric at $u=\e$, with $\g = -\e^{-8}$ its determinant, and $\Box_{\g}$ is the scalar Laplacian associated with $\g_{\mu\nu}$. Inserting the asymptotic expansions for the fields, we find
\bea
L_1 & = & + \N \, \frac{1}{4} \, \frac{1}{\e^4}, \\
L_2 & = & + \N \, \left( - \frac{c_0^2}{2} \frac{1}{\e^2} - c_0 \tilde{c}_2 + \frac{1}{2} \o^2 \, c_0^2 \log \e \right) + O\left( \e^2 \log \e\right),\\
L_3 & = & + \N \, \frac{5}{12} \, c_0^4 +O\left( \e^2 \log \e\right), \\
L_4 & = & + \N \, \left( +\frac{1}{2} \o^2 \, c_0^2 \log \e + \frac{1}{2} \o^2 c_0^2 \log c_0 \right ) + O\left(\e^2 \log \e\right),\\
L_5 & = & + \N \, \frac{1}{4} \, \o^2 c_0^2 + O\left(\e^2 \log \e\right),\\
L_6 & = & +\N \, \left( -\frac{1}{2} B^2 \log \e \right).
\eea
Comparing the above equations for the counterterms with the regulated action in eq.~\eqref{eq:regaction}, we see that all divergences will cancel.

Notice that $L_3$ and $L_5$ are finite in the $\epsilon \rightarrow 0$ limit. We could also introduce a third finite counterterm, identical to $L_6$ but without the factor of $\log \epsilon$. Fixing the coefficients of these finite counterterms corresponds to a choice of renormalization scheme in the field theory. The coefficient of the $|\Theta|^4$ counterterm is fixed by demanding that the one-point function $\Omv$, which we will compute shortly, vanishes for the solution that represents a constant $\N=2$-supersymmetric flavor mass at zero temperature with $B=0$ and $\o=0$ \cite{Karch:2005ms}. The coefficient of the $\Theta^* \Box_{\g} \Theta$ counterterm is fixed by demanding that $\Omv$ vanish in the large-mass limit $|m| \propto c_0 \rightarrow \infty$, as demonstrated in the appendix of ref.~\cite{O'Bannon:2008bz}. For the third finite counterterm, proportional to $B^2$, we follow the convention ref.~\cite{Jensen:2010vd} and set the coefficient to zero.

Now let us compute renormalized one-point functions.  To do so we define a ``subtracted action'' as
\beq
S_{D7}^{sub} \equiv S_{D7}^{reg} + \sum_{i=1}^6 L_i.
\eeq
The finite, renormalized on-shell action is then
\beq
S_{D7}^{ren} = \lim_{\e\rightarrow0} S_{D7}^{sub}.
\eeq
The AdS/CFT dictionary then equates $-S_{D7}^{ren}$ with the generating functional of the field theory. For a (pseudo)scalar field $\Phi(u)$ dual to an operator $\mathcal{O}$ of dimension $\Delta$, the renormalized one-point function is
\beq
\langle \mathcal{O} \rangle = -\lim_{\e \rightarrow 0} \frac{1}{\e^{\Delta}} \frac{1}{\sqrt{-\g}} \frac{\delta S_{D7}^{sub}}{\delta \Phi(\e)}.
\eeq
Our scalar $\theta(u)$ is dual to the dimension-three mass operator of the flavor fields, $\mathcal{O}_m$. More precisely, $\mathcal{O}_m$ is given by taking the variational derivative of the generating functional with respect to $|m|$. Recalling from the discussion below eq.~\eqref{eq:azasymptoticexpansion} that $|m| = \frac{c_0}{2\pi \a'}$, we find
\beq
\Omv = -(2\pi\a')\lim_{\e \rightarrow 0} \, \frac{1}{\e^3} \frac{1}{\sqrt{-\g}} \frac{\delta S_{D7}^{sub}}{\delta \theta(\e)}.
\eeq
The calculation of $\Omv$ is straightforward, so we present only the result,
\beq
\Omv =  (2\pi\a')\N \left( -2 \tilde{c}_2 + \frac{1}{3} c_0^3 - \frac{1}{2} \o^2 c_0 - \frac{1}{2} \o^2 c_0 \ln c_0^2 \right).
\eeq
Upon using $\tilde{c}_2 = c_2 + \frac{1}{6} c_0^3$ we obtain eq.~\eqref{eq:omvev}, which is identical to the result for $\Omv$ in ref.~\cite{O'Bannon:2008bz}. The pseudoscalar $\phi(t,u) = \o t + \varphi(u)$ is dual to the dimension-four operator $\Op$ written in eq.~\eqref{vevphiFT}, so for $\Opv$ we have
\beq
\Opv = -\lim_{\e \rightarrow 0} \frac{1}{\e^4} \frac{1}{\sqrt{-\g}} \frac{\delta S_{D7}^{sub}}{\delta \varphi(\e)}.
\eeq
Notice that $\frac{1}{\e^4} \frac{1}{\sqrt{-\g}} = 1$ so we only need to compute the variation,
\beq
\delta S_{D7}^{sub} = \int_\epsilon du \, \frac{\delta S_{D7}^{sub}}{\delta \partial_u \varphi} \partial_u \delta \varphi = \a \int_\epsilon du \, \partial_u \delta \varphi = -\a \, \delta \varphi(u=\e),
\eeq
where we used the fact that on-shell $\frac{\delta S_{D7}^{sub}}{\delta \partial_u \phi} = \a$ is independent of $u$, and we demanded that the fluctuation be fixed at the endpoint in the bulk of AdS space so only the $u=\e$ endpoint of the $u$ integral contributes. We thus have
\beq\label{vevphi}
\Opv = \a.
\eeq
Arguments very similar to those for $\phi$ and $\Opv$ apply also for $A_z$ and $\jz$. Recalling that we absorbed a factor of $2\pi\a'$ into the vector field above eq.~\eqref{eq:prefactordefinition}, we have
\beq
\jz = (2\pi\a')\lim_{\e \rightarrow 0} \frac{1}{\e^4} \frac{1}{\sqrt{-\g}} \frac{\delta S_{D7}^{sub}}{\delta A_z(\e)}.
\eeq
The variation of the action is
\beq
\delta S_{D7}^{sub} = \int_\e du \, \frac{\delta S_{D7}^{sub}}{\delta \partial_u A_z} \, \partial_u \delta A_z = \frac{\delta S_{D7}^{sub}}{\delta \partial_u A_z} \int_\e du \, \partial_u \delta A_z = -\b \delta A_z(u=\e).
\eeq
where in the second equality we used the fact that on-shell $\frac{\delta S_{D7}^{sub}}{\delta \partial_u A_z}$ is independent of $u$ and in the final equality we demanded that the fluctuation be fixed at the endpoint in the bulk of AdS space. Notice that when evaluated on a solution $\frac{\delta S_{D7}^{sub}}{\delta \partial_u A_z}$ must be identical to the variational derivative of the Legendre transform of the action defined in eq~\eqref{eq:legendretransformphi}: $\frac{\delta S_{D7}^{sub}}{\delta \partial_u A_z} = \frac{\delta \hat{S}_{D7}^{sub}}{\delta \partial_u A_z}$. The latter is the integration constant $\b$ in eq.~\eqref{eq:azeq}. Assembling these ingredients, we find
\beq
\jz = - (2\pi\a') \b.
\eeq

\section{Comments on Stability}\label{ss:stability}

In this appendix we do a partial analysis of fluctuations of spinning D7-branes to look for indications of instabilities. Since Minkowski embeddings with worldvolume horizons are not regular, we focus only on black hole embeddings. We will not compute the full spectrum of quasi-normal modes: we will consider only zero-momentum fluctuations, and study only their behavior near the AdS-Schwarzschild horizon.

To our knowledge the only holographic calculation of quasi-normal modes for spinning probe D-branes appears in ref.~\cite{PremKumar:2011ag}, for a different system. In ref.~\cite{PremKumar:2011ag} the field theory was $\N=4$ SYM on a spatial three-sphere $S^3$ at zero temperature with massless $\N=2$ hypermultiplets and zero magnetic field but finite axial chemical potential. Instabilities did indeed appear there which most likely survive the infinite-volume limit in which the $S^3$ decompactifies. Our analysis is complementary to that of ref.~\cite{PremKumar:2011ag}, in the decompactified limit, in that we work with nonzero mass, temperature and magnetic field.

We will use the radial coordinate $\rho$ defined in eq.~\eqref{eq:metric} and our worldvolume scalar dual to $\Om$ will be $\theta(\rho)$ as in appendix~\ref{holorg}, which is related to $r$ and $R(r)$ as in eq.~\eqref{eq:coordtranslate}. The boundary is at $\rho \rightarrow \infty$ and the worldvolume horizon is at $\rho_*$ defined in eq.~\eqref{eq:horT}. The equation of motion for $\theta(\rho)$ comes from eq.~\eqref{eq:shathat}, with $r\rightarrow \rho$ and $R(r)\rightarrow \theta(\rho)$.

To study small fluctuations we will use the full nonlinear equations of motion for the embedding that we derived in section \ref{ss:sols} and expand around a background solution to linear order. For illustrative purposes we will begin with the trivial background solution, $\theta(\rho)=0$, which describes massless flavor fields.

Close to the horizon $\rho_*=1$, linearized fluctuations $\delta\theta(t,r)=e^{-i k_0 t} \vartheta(\rho)$ obey an equation of motion independent of the value of the magnetic field,
\beq
\vartheta''+\frac{1}{\rho-1} \vartheta' +\frac{\omega^2}{8(\rho-1)^2}\vartheta = -k_0^2 \vartheta.
\eeq
We can transform this into a Schr\"{o}dinger form by defining $\vartheta(\rho)=(\rho-1)^{-1/2}\psi(\rho)$,
\beq\label{eq:trivialfluceq}
-\psi''-\frac{\frac{1}{4}+\frac{\omega^2}{16}}{(\rho-1)^2}\psi = k_0^2 \psi.
\eeq
For any $\omega\neq 0$, this equation allows for an infinite set of negative energy $k_0^2<0$ solutions representing bound states. This is the same situation as for a field in AdS with a mass below the Breitenlohner-Freedman bound, which we take as strong evidence of an instability of fluctuations in $\theta$, based on previous experience in similar holographic models \cite{PremKumar:2011ag,Gubser:2008px}. However, a full analysis of the equations of motion is required to demonstrate the existence of an unstable mode.

We can do a similar analysis for nontrivial black hole embeddings $\theta(\rho)\neq 0$, although the expressions are much more involved. Setting $k_0=0$ and expanding close to the worldvolume horizon $\rho=\rho_*$, we can again derive a Schr\"{o}dinger-type equation,
\beq\label{eq:nontrivialfluceq}
-\psi''+\frac{M-\frac{1}{4}}{(\rho-\rho_*)^2}\psi = 0.
\eeq
Here $\psi$ is related to $\vartheta$ by a factor that is a power of $(\rho-\rho_*)$, where the power is a complicated function of the parameters $\o$, $B$, $\rho_*$, etc. Similarly $M$ is a complicated function of the parameters. The condition to avoid negative-energy solutions to the Schr\"{o}dinger equation is $M\geq 0$. We plot $M$ as a function of $\rho_*$ for various values of $B$ and several values of positive $\o$ in fig.~\ref{FigMass1} and several values of negative $\o$ in fig.~\ref{FigMass2}. With our conventions physical embeddings are those with $\omega >0$, where the body of the D7-brane drags \textit{behind} the part at the boundary. For $\o>0$, we find embeddings that satisfy $M>0$, for sufficiently small $\omega$ and $B$, while for larger values of $\o$, even solutions close to the trivial one (which has $\rho_*=1$) have $M<0$. As the magnetic field increases, the region where $M>0$ shrinks, and can disappear completely if the magnetic field is large enough.

Notice that the behavior of fluctuations does not connect smoothly with the chirally-symmetric case $\rho_*=1$. We can see non-analytic behavior at $\rho_*=1$ by comparing the $\omega>0$ curves in Figure \ref{FigMass1} with the $\omega<0$ curves in Figure \ref{FigMass2}: the limiting value of $M$ at $\rho_*=1$ is different for positive and negative $\o$, despite the fact that the equation of motion for fluctuations about the trivial solution, eq.~\eqref{eq:trivialfluceq}, is independent of the sign of $\o$.

We should stress again that we cannot decisively conclude whether embeddings are stable or not without a full analysis of the fluctuations. We should also mention that we have not considered other possible sources of instabilities that can appear for more generic fluctuations, for example fluctuations with nonzero momentum.
 

\FIGURE[t]{
	\includegraphics[width=6cm]{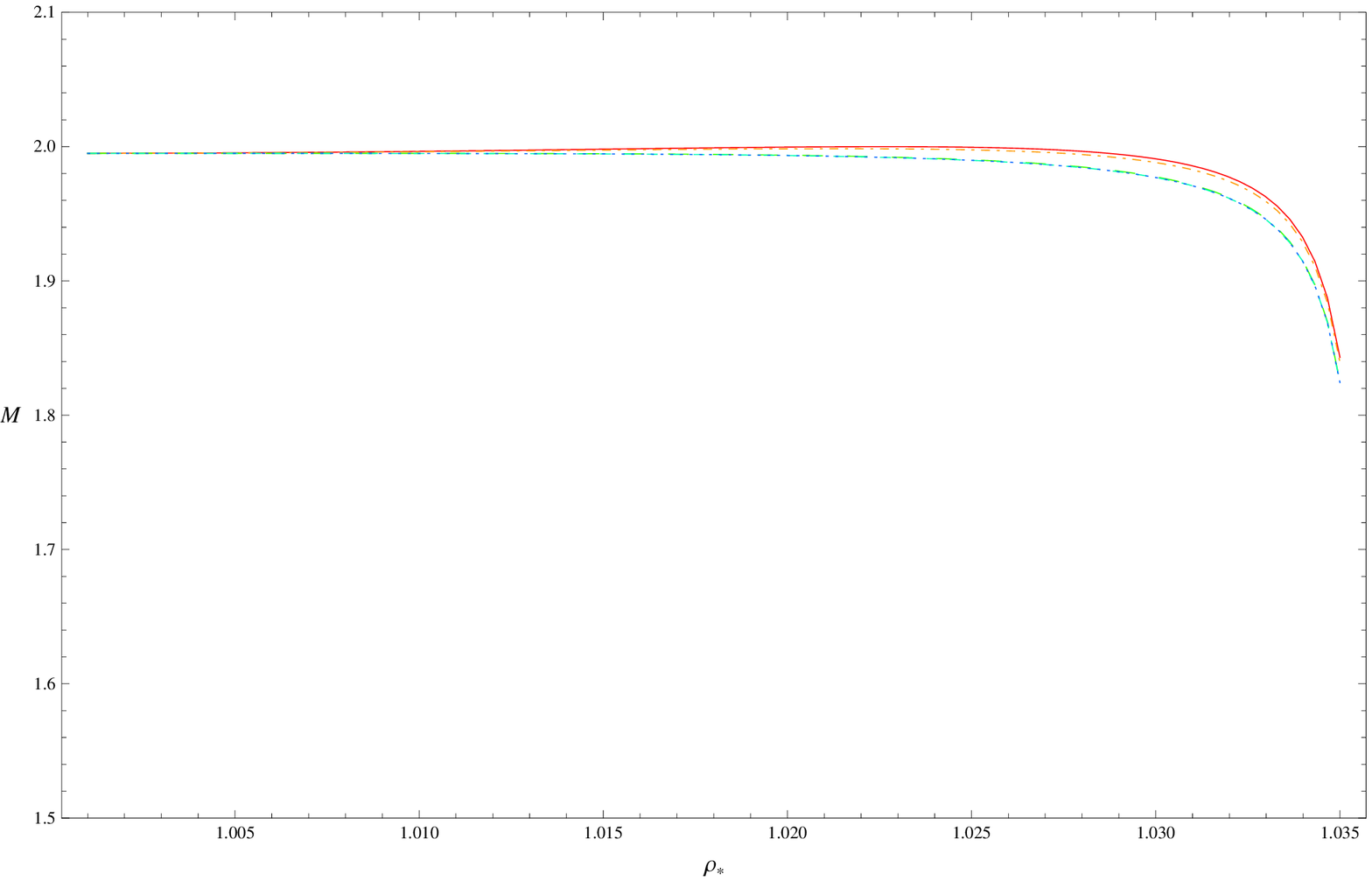}\qquad
	\includegraphics[width=6cm]{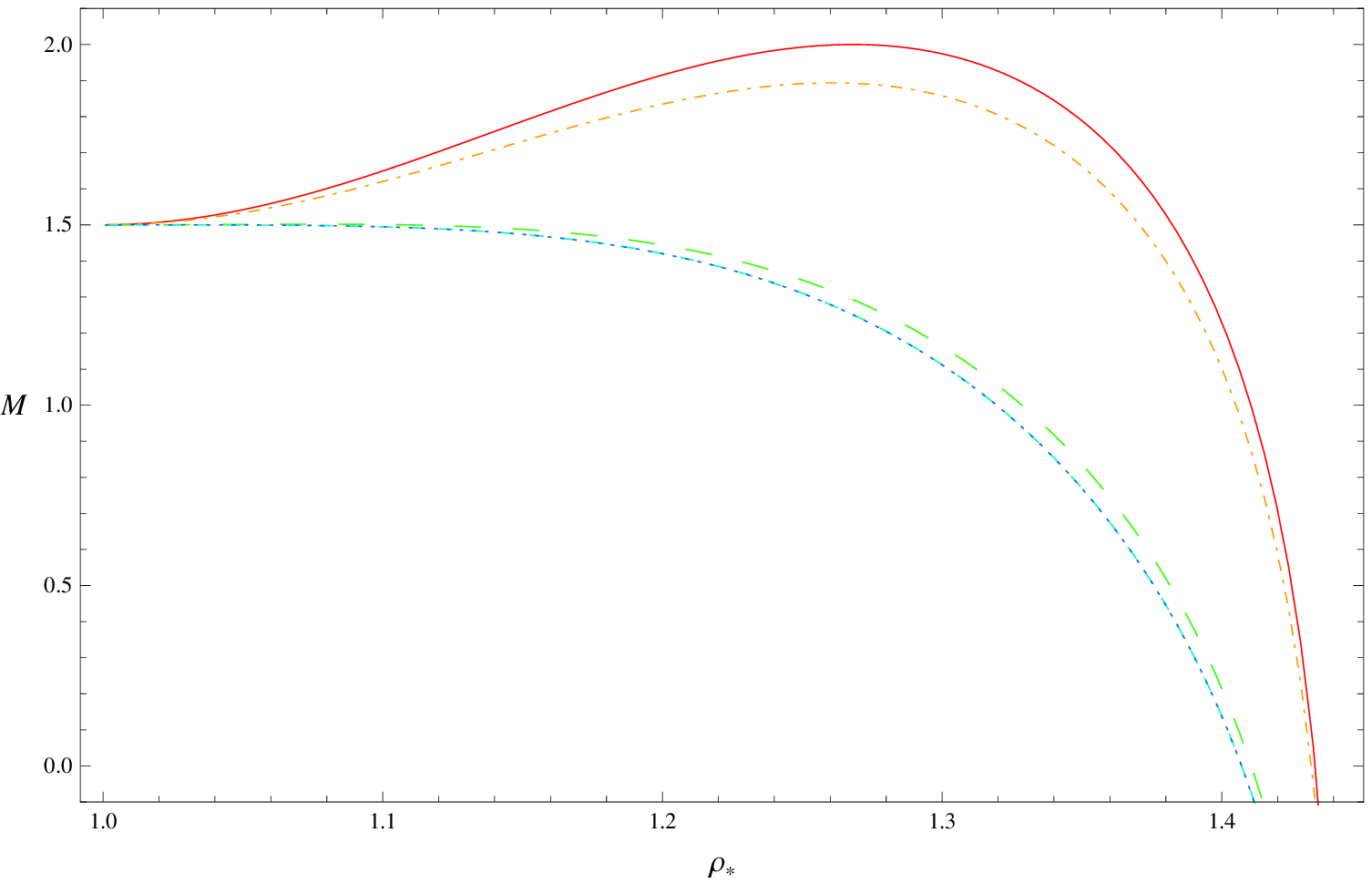}\\
	\vspace{0.5cm}
	\hspace{1.5cm} (a.) \hspace{6cm} (b.)\\
	\includegraphics[width=6cm]{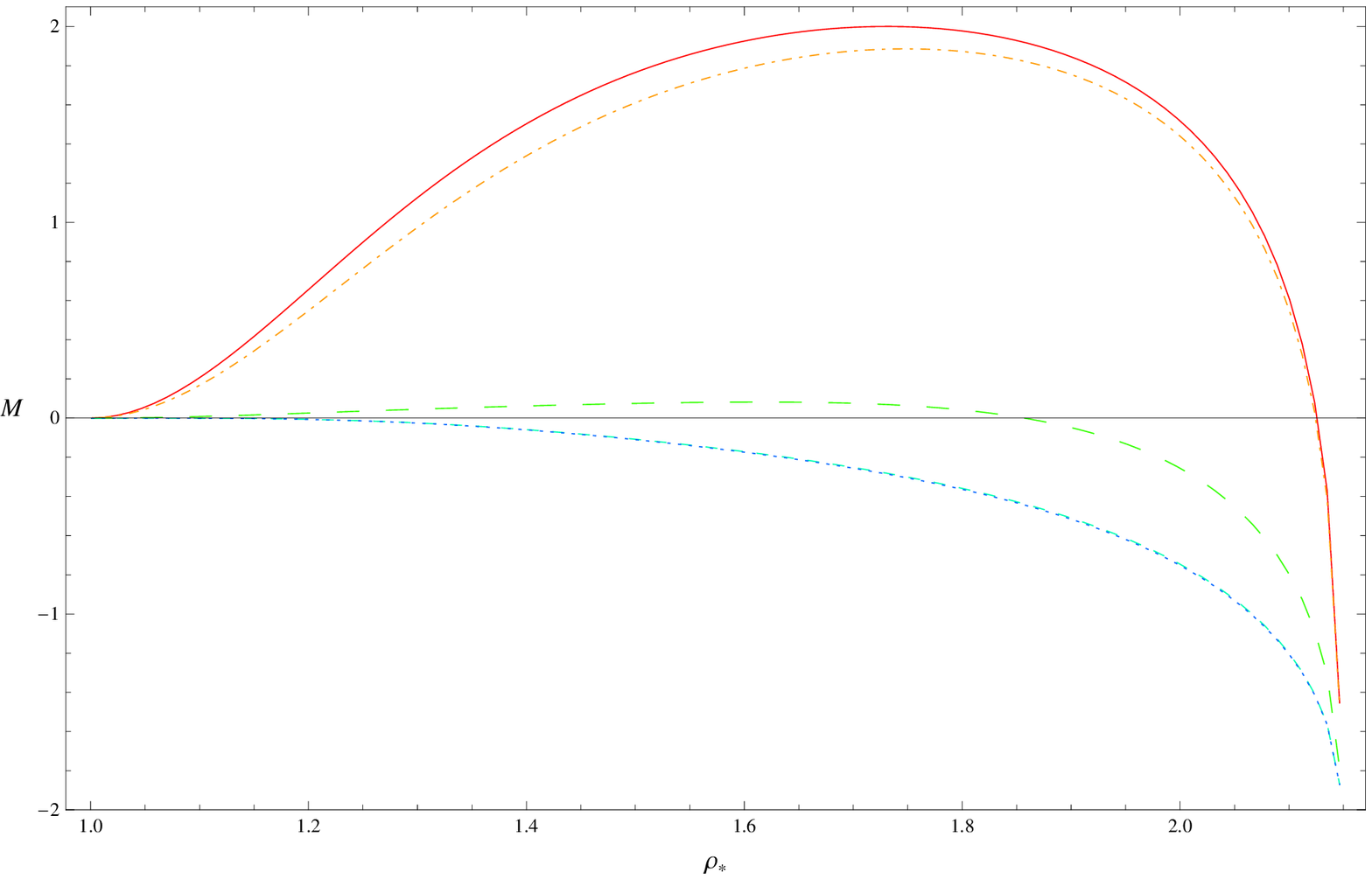}\qquad
	\includegraphics[width=6cm]{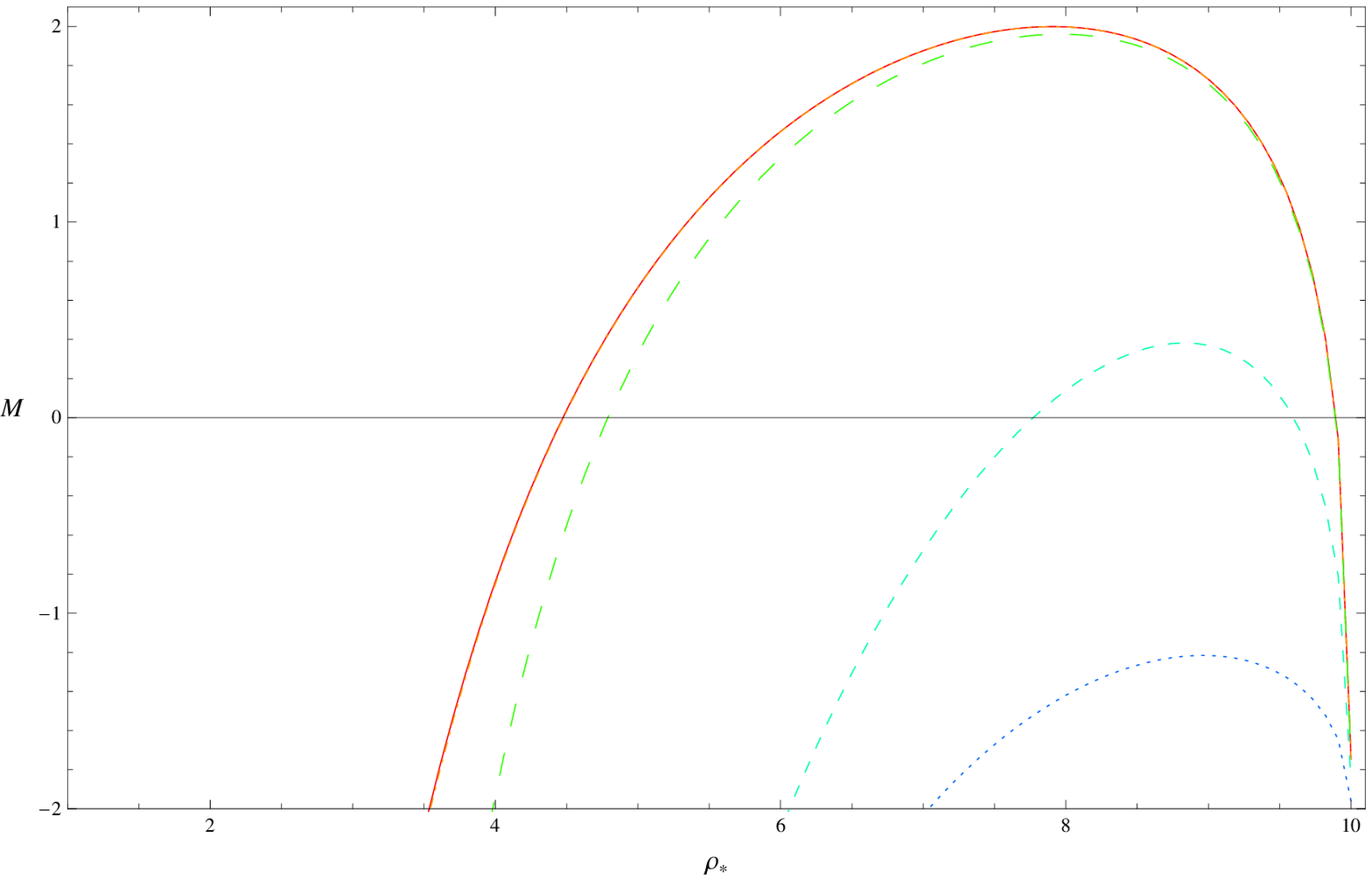}\\
	\vspace{0.05cm}
	\hspace*{0.2cm} (c.) \hspace{6cm} (d.)
	\caption{The value of $M$ in eq.~\eqref{eq:nontrivialfluceq} as a function of $\rho_*$, the position of the D7-brane worldvolume horizon. (a.) $\omega=0.1$, (b.) $\o=1$, (c.) $\o=2$, and (d.) $\o=10$, in units of the AdS radius. The red (solid), orange (dot-dashed), green (large dashed), cyan (medium dashed) and blue (dotted) curves correspond to $B/\gamma^2=0.005, 0.5, 5, 50$ and $500$, where the temperature of the AdS-Schwarzschild black hole is $T=\g/\pi$. A value $M>0$ is highly suggestive that the D7-brane embedding is stable, while a value $M<0$ is strong evidence that the embedding is unstable.}
	\label{FigMass1}
}

\FIGURE[t]{
	\includegraphics[width=6cm]{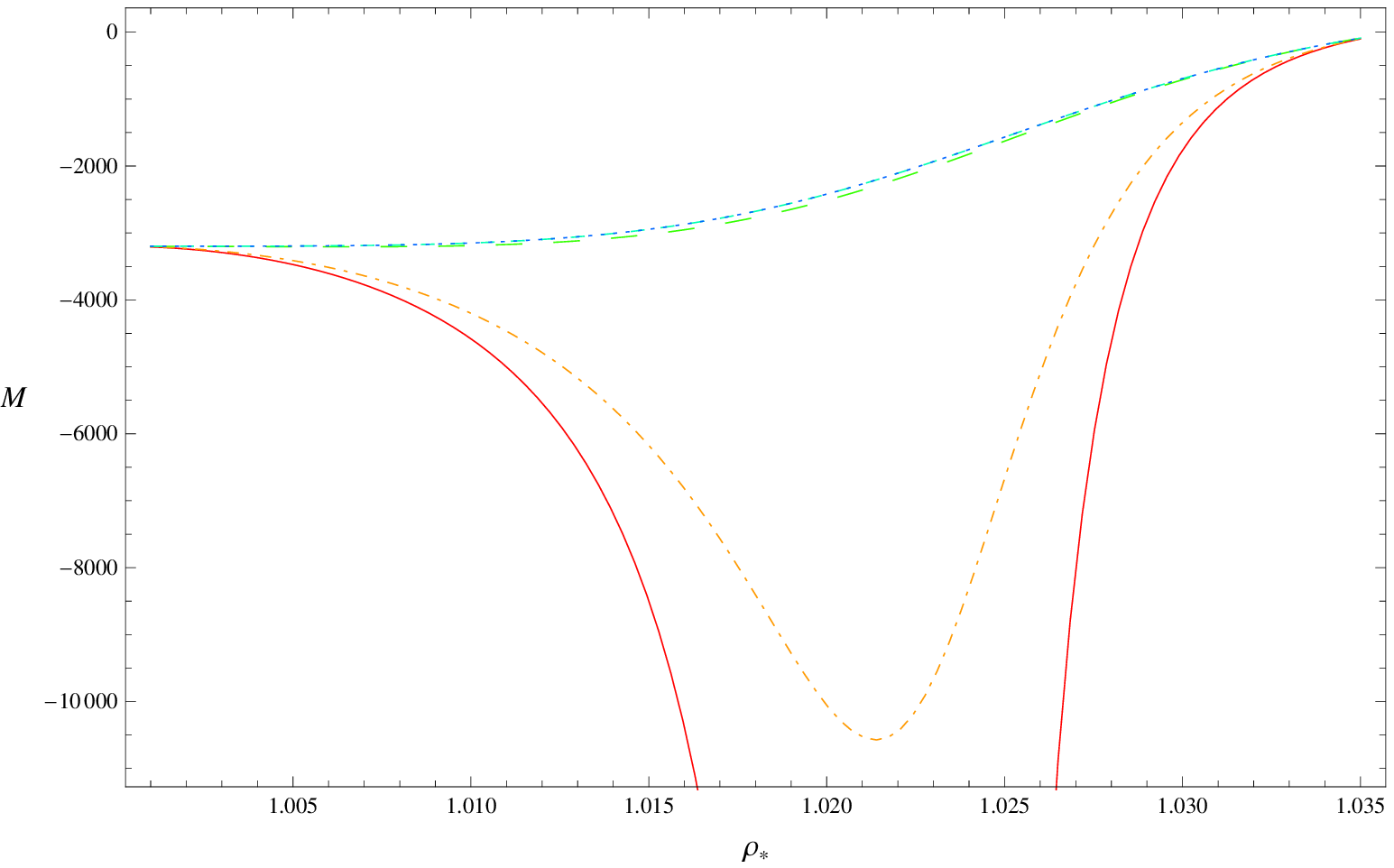}\qquad
	\includegraphics[width=6cm]{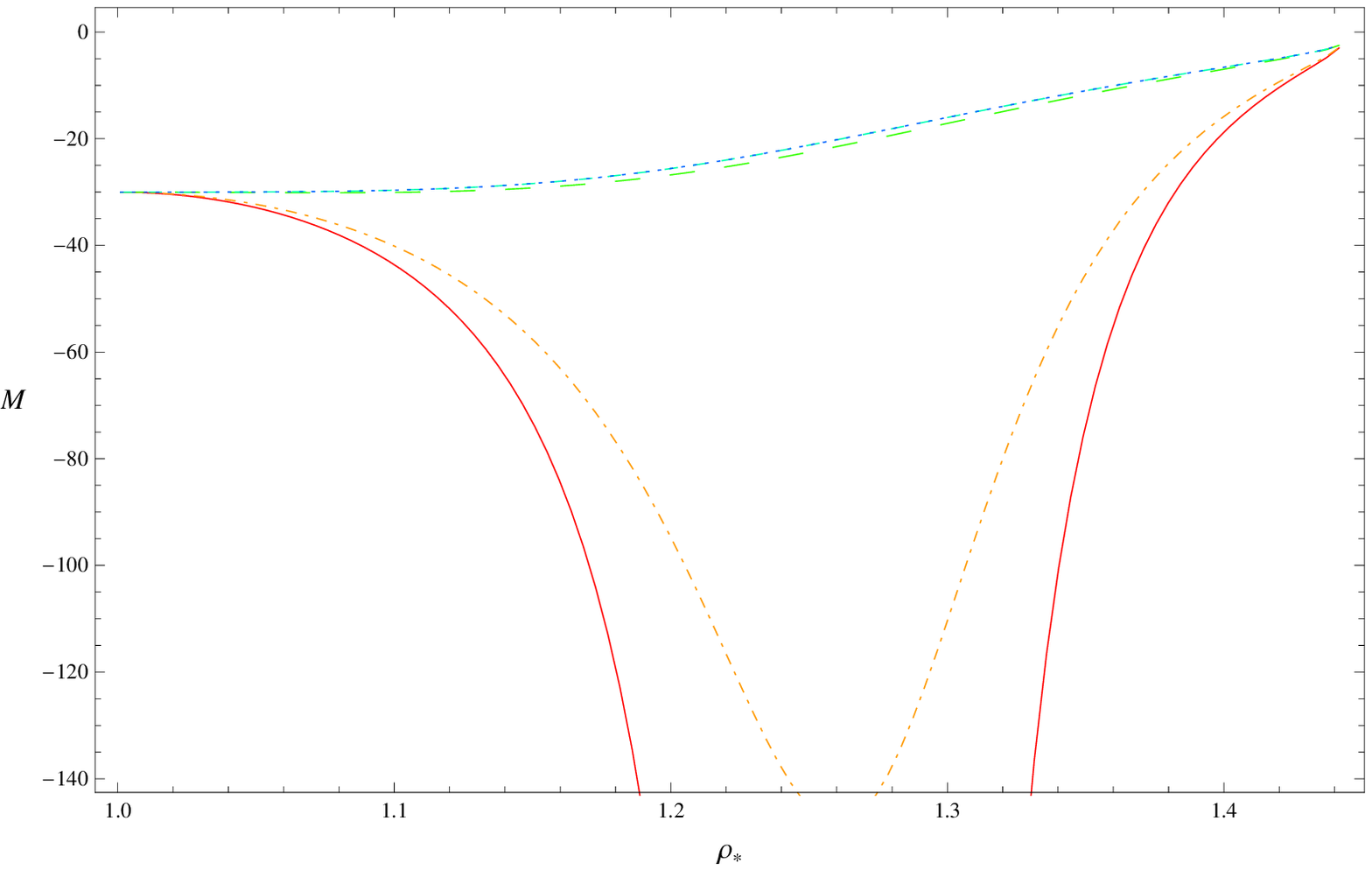}\\
	\vspace{0.05cm}
	\hspace*{1cm} (a.) \hspace{5.8cm} (b.)
	\caption{The value of $M$ in eq.~\eqref{eq:nontrivialfluceq} as a function of $\rho_*$, the position of the D7-brane worldvolume horizon. (a.) $\omega=-0.1$, (b.) $\o=-1$, in units of the AdS radius. The red (solid), orange (dot-dashed), green (large dashed), cyan (medium dashed) and blue (dotted) curves correspond to $B/\gamma^2=0.005, 0.5, 5, 50$ and $500$, where the temperature of the AdS-Schwarzschild black hole is $T=\g/\pi$. Notice that the values of $M$ at $\rho_*=1$ are very different from those in fig.~\ref{FigMass1}, indicating some non-analyticity there.}
	\label{FigMass2}
}


\section{Effective temperature, entropy and heat capacity of spinning branes}\label{ss:heatcap}

As mentioned in section~\ref{ss:solszeroT}, we can associate a temperature with the worldvolume horizon of spinning D7-branes. Recall the induced metric of eq.~\eqref{eq:zerotempinducedmetric},
\beq
ds_{D7}^2 = g_{tt}^{D7}dt^2 + 2g_{tr}^{D7}dtdr + g_{rr}^{D7}dr^2 + g_{xx}d{\vec x}^2 + g_{SS}ds_{S^3}^2.
\eeq
Since the induced metric depends only on the radial coordinate $r$, we can diagonalize it with the coordinate transformation 
\beq
d\hat t = dt + \frac{g_{tr}^{D7}}{g_{tt}^{D7}}dr,
\eeq
which gives us
\beq
ds_{D7}^2 = g_{\hat t \hat t}^{D7} d\hat t^2 + g_{\hat r \hat r}^{D7}dr^2 + g_{xx}d{\vec x}^2 + g_{SS}ds_{S^3}^2,
\eeq
where 
\beq
g_{\hat t \hat t}^{D7} = g_{t t}^{D7} \ , \qquad g_{\hat r \hat r}^{D7} = g_{rr}^{D7} - \frac{(g_{tr}^{D7})^2}{g_{t t}^{D7} }.
\eeq
If $g_{\hat t \hat t}^{D7}$ vanishes at some point $r = r_*$, then we can define a Hawking temperature $T_H$ by demanding regularity of the Euclidean geometry. Defining the Euclidean time $\tau \equiv i\hat t$, we have
\beq\label{indtemp}
\tau \sim \tau + \frac{1}{T_H} \ , \qquad T_H = \frac{|g_{\hat t \hat t}^{D7}|'}{4\pi \sqrt{|g_{\hat t \hat t}^{D7}| g_{\hat r \hat r}^{D7}}}\Bigg|_{r=r_*}.
\eeq
$T_H$ is the Unruh temperature that an accelerating observer on the brane feels. In the field theory, the flavor degrees of freedom effectively feel the temperature $T_H$ while the adjoint degrees of freedom remain at zero temperature. We expect two-point functions of flavor operators to take a form characteristic of thermal diffusion with temperature $T_H$, even if the actual temperature is zero \cite{Das:2010yw}. The fact that $T_H$ can differ from the actual temperature is another indication that the system is not in thermal equilibrium.

As mentioned in section~\ref{ss:fieldtheory}, in the flavor sector only the fermions carry $U(1)_A$ charge. The scalars are neutral. Our D7-brane solutions with a worldvolume horizon have nonzero angular momentum, corresponding to states with a nonzero $U(1)_A$ charge, \ie states that na\"ively should have a density of strongly-coupled fermions. An obvious question is whether we can detect a Fermi surface. The most direct way to do so would be to compute two-point functions of fermionic operators, looking for a peak at some finite (Fermi) momentum \cite{Liu:2009dm,Cubrovic:2009ye,Faulkner:2009wj,Ammon:2010pg}. Another approach is to look for the de Haas-van Alphen effect, in which the magnetization, magnetic susceptibility, conductivity, and other observables exhibit oscillations as a function of the magnetic field whose period is inversely proportional to the cross-sectional area of the Fermi surface. Detecting such oscillations may require computing $1/N_c$ effects, as in ref.~\cite{Denef:2009yy}. Both of these approaches are beyond the scope of this paper.\footnote{Our system, and more generally any system with a CME, has an interpretation in condensed matter physics not only in terms of Fermi physics but also in terms of topological insulators. An action of the form $a(t,\vec{x}) F \wedge F$ has been proposed not only as an effective action for the CME \cite{Kharzeev:2009fn} but also as an effective action for (3+1)-dimensional topological insulators \cite{Qi:2008ew}. While $a(t,\vec{x})=\mu_5 t$ produces a CME, an $a(t,\vec{x})$ independent of $t$ but varying in space as a step-function from zero to $\pi$ produces a $T$-invariant topological insulator. An early derivation of the CME in the condensed matter context appears in ref. \cite{1998PhRvL..81.3503A}, and an analysis of the CME in the topological insulator context appears in section IV B of ref. \cite{Qi:2008ew}. Indeed, if the QGP created in heavy-ion collisions does in fact exhibit a CME, then the QGP may be a kind of topological insulator. A holographic analysis of $T$-invariant topological insulators using D7-branes in $AdS_5 \times S^5$ appears in ref.~\cite{HoyosBadajoz:2010ac}.}

We will follow ref.~\cite{Karch:2008fa} and characterize our system by asking how the heat capacity $c_V$ scales with temperature $T$ at low temperatures: a Fermi liquid will have $c_V \propto T$ while a Bose liquid will have $c_V \propto T^d$ in $d$ spatial dimensions. One result of ref.~\cite{Karch:2008fa} was that for D7-branes describing a $U(1)_V$ density at low temperatures the heat capacity scaled as $c_V \sim T^6$, which looks like neither a Fermi nor Bose liquid, and hence may be a whole new kind of quantum liquid. Here we are asking what happens with a $U(1)_A$ density.

The problem in our case, of course, is that our system is not in equilibrium, so how can we define a heat capacity, or any thermodynamic quantity? If the degrees of freedom on the D7-brane were gravitational, \ie if they included spin two fields rather than just spin zero and one, then we would have a natural definition of entropy as the area of the worldvolume horizon, which would provide a starting point for a thermodynamic analysis. Lacking that, we must look for a substitute. The obvious candidate is the on-shell D7-brane action, which in equilibrium indeed represents the flavor contribution to the free energy. In our non-equilibrium setting, we will define an ``effective free energy'' as minus the on-shell D7-brane action, and extract an entropy and a heat capacity by taking derivatives with respect to the effective temperature. Whether these quantities satisfy thermodynamic laws is not at all clear. In particular, notice that $\omega$, which is equivalent to the axial chemical potential, and the temperature $T_H$ are not independent variables for a fixed $|m|$.

The Lorentzian D7-brane action (not just an action density) is
\beq
S_{D7} = - {\cal N} \int d\hat t \, dr \, d^3 x \ \sqrt{-g_{\hat t \hat t}^{D7} g_{\hat r \hat r}^{D7} g_{xx}^3 \left( 1+ \frac{B^2}{g_{xx}^2}\right)}  + \frac{\cal N}{2} B\omega \int d\hat t \, dr \, d^3 x \ A_z'(r).
\eeq
Here and below we do not write the counterterms of appendix~\ref{holorg}, although we include them in our calculations in order to obtain a finite, renormalized on-shell action. We can obtain the Euclidean action $S_{D7}^E$ by taking
\beq
S_{D7}^{E} = -i S_{D7}(\hat t \to -i \tau) = -S_{D7}(\hat t \to \tau) \ .
\eeq
Since the D7-brane Lagrangian is time-independent in our case, the replacement $\hat t \to \tau$ acts only on the measure of the integral. Accordingly, the equation of motion in Euclidean signature is the same as in Lorentzian signature. Notice that we do not Wick rotate $\o$, so the worldvolume scalar $\phi$ becomes imaginary. From a field theory point of view we could equivalently begin in Euclidean signature with an imaginary axial chemical potential.\footnote{For a probe brane system where an imaginary worldvolume electric field was used to guarantee that the Euclidean equations of motion were identical to the Lorentzian ones see ref.~\cite{Bergman:2008sg}.}

We define the effective free energy density as $f = T_H S_{D7}^{E}/\textrm{vol}_{\mathbb{R}^3}$. The effective entropy density $s$ and the heat capacity $c_V$ are then
\beq
s = - \frac{\partial f}{\partial T_H}\Big|_{B,c_0}, \qquad c_V = T_H\frac{\partial s}{\partial T_H}\Big|_{B,c_0} .
\eeq
In fig.~\ref{FigCV} we plot $s$ and $c_V$ as functions of $T_H$ for $c_0 = 10^{-3}$. We find that $c_V \propto T_H$, similar to a Fermi liquid. We also find that $s$ does not vanish in the $T_H \to 0$ limit. Such residual zero-temperature entropy, \ie a large degeneracy of states, is another signal that our solutions may not be stable, since a generically any perturbation will break the degeneracy and the system will settle in a new, presumably non-degenerate ground state. Notice also that our entropy density $s$ is not proportional to the area of the induced horizon, which goes as $\omega^3 \theta_\ast^3 \sim T_H^3$ for small $T_H$ and hence vanishes at $T_H=0$.


\FIGURE[h]{
	\includegraphics[width=7cm]{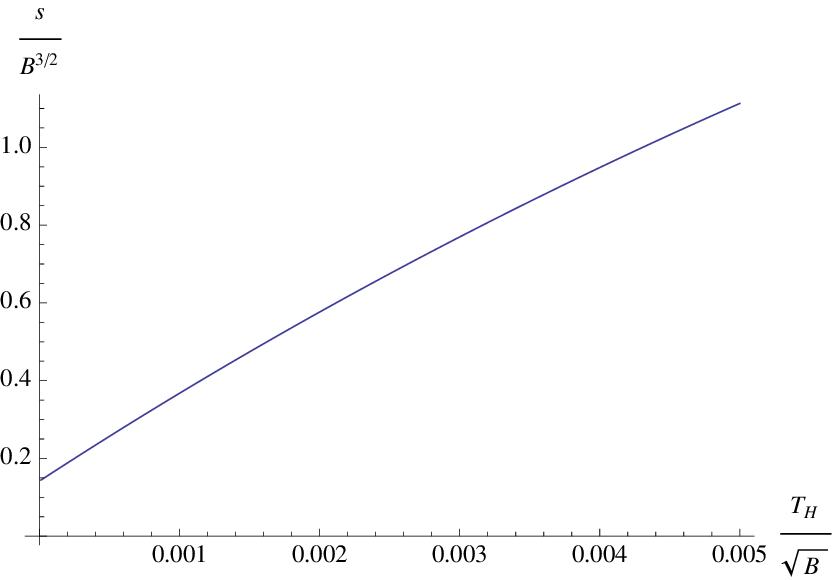}\qquad
	\includegraphics[width=7cm]{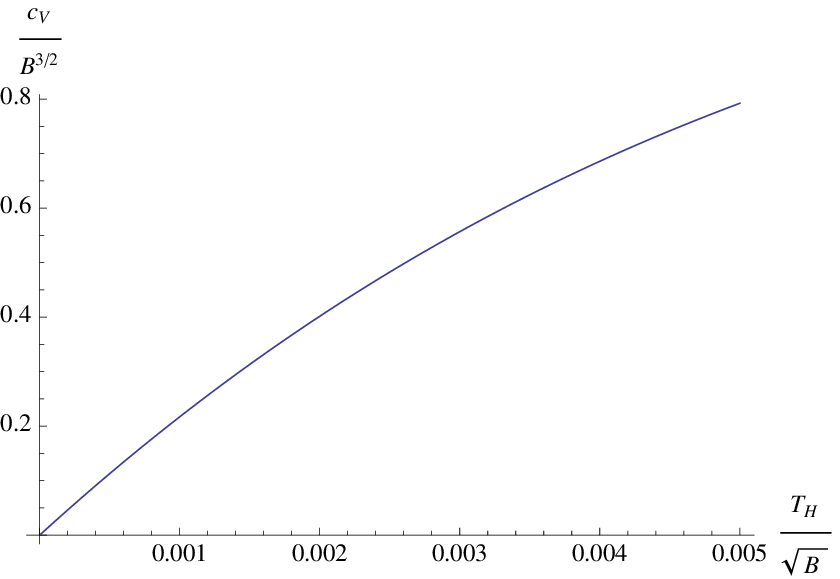}
	\vspace{0.05cm}
	\hspace*{1cm} (a.) \hspace{6.8cm} (b.)
	\caption{(a.) The effective entropy density $s$ versus the effective temperature $T_H$, both in units of the magnetic field $B$, for $c_0 = 10^{-3}$, where the flavor mass is $|m| = \sqrt{\lambda} \frac{c_0}{2\pi}$ in units of the $AdS_5$ radius. (b.) The heat capacity $c_V$ versus $T_H$, both in units of $B$, for the same value of $c_0$ as in (a.).}
	\label{FigCV}
}


Calculations similar to the above were done for spinning probe D3-branes without a magnetic field in ref.~\cite{Das:2010yw}. In that case the on-shell action was not renormalized, and divergent terms appeared in the free energy and entropy density proportional to powers of a UV cutoff. The authors of ref.~\cite{Das:2010yw} proposed that the entropy be regarded as an entanglement entropy between the flavor sector and the ${\cal N} = 4$ SYM sector, in which case the UV divergences are the usual ones, with the leading divergence having a coefficient proportional to the area of boundary of the entangled region~\cite{Bombelli:1986rw,Srednicki:1993im}. The UV divergences can also be understood from the holographic prescription for entanglement entropy~\cite{Ryu:2006bv,Ryu:2006ef,Nishioka:2009un}. We renormalize the action using $\omega$-dependent counterterms, so from the perspective of ref.~\cite{Das:2010yw}, we are extracting some part of the finite contributions to the entanglement entropy.

\bibliographystyle{JHEP}
\bibliography{d7cme}

\end{document}